\documentclass[reprint,
superscriptaddress,nofootinbib,
amsmath,
amssymb,
aps,
prb,
floatfix,
english
]{revtex4-2}

\usepackage{bibunits}
\usepackage[unicode=true, colorlinks=true, citecolor={blue!80!black}, urlcolor={blue!50!black}, linkcolor = {blue!80!black}]{hyperref}
\defaultbibliography{references}
\defaultbibliographystyle{apsrev4-2}
\usepackage{graphicx}% Include figure files
\usepackage{svg}
\usepackage{dcolumn}% Align table columns on decimal point
\usepackage{siunitx}
\usepackage[T1]{fontenc}
\usepackage[utf8]{inputenc}
\usepackage{bbm}
\usepackage{babel}

\newcommand{\wout}{\omega_\mathrm{\rm out}}
\newcommand{\win}{\omega_\mathrm{\rm in}}
\newcommand{\wq}{\omega_\mathrm{q}}
\newcommand{\wres}{\omega_\mathrm{res}}

\newcommand{\nbar}{\bar{n}}
\newcommand{\nzpf}{N_{\rm zpf}}
\newcommand{\ghz}{\:\mathrm{GHz}}
\newcommand{\mhz}{\:\mathrm{MHz}}

\begin{document}
\widetext

\title{Full characterization of measurement-induced transitions of a superconducting qubit}
\author{Thomas~Connolly}
\thanks{These two authors contributed equally.\\
tom.connolly@yale.edu, pavel.kurilovich@yale.edu}
\affiliation{Departments of Applied Physics and Physics, Yale University, New Haven, Connecticut 06520, USA}
\author{Pavel~D.~Kurilovich}
\thanks{These two authors contributed equally.\\
tom.connolly@yale.edu, pavel.kurilovich@yale.edu}
\affiliation{Departments of Applied Physics and Physics, Yale University, New Haven, Connecticut 06520, USA}
\author{Vladislav~D.~Kurilovich}\thanks{{Present address: Google Quantum AI, 301 Mentor Dr, Goleta, CA93111, USA}}
\affiliation{Departments of Applied Physics and Physics, Yale University, New Haven, Connecticut 06520, USA}

\author{Charlotte~G.~L.~B\o ttcher}
\thanks{{Present address: Department of Applied Physics, Stanford University, Stanford, California 94305, USA}}
\affiliation{Departments of Applied Physics and Physics, Yale University, New Haven, Connecticut 06520, USA}

\author{Sumeru~Hazra}
\affiliation{Departments of Applied Physics and Physics, Yale University, New Haven, Connecticut 06520, USA}
\author{Wei~Dai}
\affiliation{Departments of Applied Physics and Physics, Yale University, New Haven, Connecticut 06520, USA}

\author{Andy~Z.~Ding}
\affiliation{Departments of Applied Physics and Physics, Yale University, New Haven, Connecticut 06520, USA}

\author{Vidul~R.~Joshi}
\thanks{{Present address: Microsoft Quantum}}
\affiliation{Departments of Applied Physics and Physics, Yale University, New Haven, Connecticut 06520, USA}

\author{Heekun~Nho}
\affiliation{Departments of Applied Physics and Physics, Yale University, New Haven, Connecticut 06520, USA}

\author{Spencer~Diamond}
\affiliation{Departments of Applied Physics and Physics, Yale University, New Haven, Connecticut 06520, USA}

\author{Daniel~K.~Weiss}
\thanks{{Present address: Quantum Circuits, Inc., New Haven, CT, USA}}
\affiliation{Departments of Applied Physics and Physics, Yale University, New Haven, Connecticut 06520, USA}
\affiliation{Yale Quantum Institute, Yale University, New Haven, Connecticut 06511, USA}

\author{Valla~Fatemi}
\affiliation{Departments of Applied Physics and Physics, Yale University, New Haven, Connecticut 06520, USA}
\affiliation{School of Applied and Engineering Physics, Cornell University, Ithaca, New York 14853, USA}
\author{Luigi Frunzio}
\affiliation{Departments of Applied Physics and Physics, Yale University, New Haven, Connecticut 06520, USA}
\author{Leonid~I.~Glazman}
\affiliation{Departments of Applied Physics and Physics, Yale University, New Haven, Connecticut 06520, USA}
\affiliation{Yale Quantum Institute, Yale University, New Haven, Connecticut 06511, USA}
\author{Michel~H.~Devoret}\thanks{michel.devoret@yale.edu\\
{Present address: Physics Dept., U.C. Santa Barbara, Santa Barbara, California 93106, USA and Google Quantum AI, 301 Mentor Dr, Goleta, California 93111, USA}}
\affiliation{Departments of Applied Physics and Physics, Yale University, New Haven, Connecticut 06520, USA}
\begin{abstract}
Repeated quantum non-demolition measurement is a cornerstone of quantum error correction protocols \cite{kitaev_fault-tolerant_2003, kitaev_anyons_2006}. In superconducting qubits, the speed of dispersive state readout can be enhanced by increasing the power of the readout tone. However, such an increase has been found to result in additional qubit state transitions that violate the desired quantum non-demolition character of the measurement \cite{sank_measurement-induced_2016, khezri_measurement-induced_2023}. Recently, the readout of a transmon superconducting qubit was improved by using a tone with frequency much larger than the qubit frequency \cite{kurilovich_high-frequency_2025}. Here, we experimentally identify the mechanisms of readout-induced transitions in this regime. In the dominant mechanism, the energy of an incoming readout photon is partially absorbed by the transmon and partially returned to the transmission line as a photon with lower frequency. Other mechanisms involve the excitation of unwanted package modes, decay via material defects \cite{thorbeck_readout-induced_2024}, and, at higher qubit frequencies, the activation of undesired resonances in the transmon spectrum \cite{sank_measurement-induced_2016,dumas_measurement-induced_2024}. Our work provides a comprehensive characterization of superconducting qubit state transitions caused by a strong drive.
\end{abstract}

\maketitle

\section{Introduction}
\begin{figure}[t]
  \begin{center}
    \includegraphics[scale = 1.0]{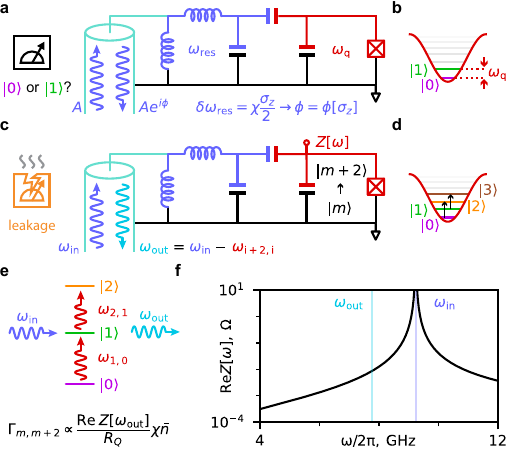}
\caption{Transmon state transitions caused by inelastic scattering of drive photons. (a) Dispersive readout of a transmon qubit is achieved by elastically scattering microwave photons off the readout resonator coupled to the qubit. Since the resonator frequency depends on the qubit state, the phase of the reflected signal can be used to infer whether the state is $|0\rangle$ or $|1\rangle$. Dispersive shift $\chi$ quantifies the difference of resonator frequencies for the two computational states. (b) Level diagram of the transmon. Colored horizontal lines show computational states while grey lines show non-computational states. (c) Due to the transmon non-linearity, readout photons with frequency $\win$ can scatter inelastically by giving off part of their energy to the qubit and producing a photon at a lower frequency $\wout$. This process leads to the leakage
error where transmon prepared in a computational state $|0\rangle$ excites to state $|2\rangle$. Similar process leads to excitation $|1\rangle\rightarrow |3\rangle$. (d) Schematic of the leakage process mediated by inelastic scattering. (e) Inelastic scattering is governed by a four-wave mixing non-linearity of the transmon. The rate of inelastic scattering is proportional to the drive power, quantified by the number of photons in the resonator $\bar{n}$. It is also proportional to the dissipative part of the impedance $Z[\omega]$ of the transmon island evaluated at the frequency $\wout$. Here, $R_Q = h/e^2\approx 25.8\:\mathrm{k}\Omega$ is the resistance quantum. (f) Dissipative part of the impedance $Z[\omega]$ computed withing lumped-element model of panel (a).
}\label{fig:intro}
  \end{center}
\end{figure}

Measurement is a central operation in quantum computation, forming the basis of quantum error correction protocols \cite{kitaev_fault-tolerant_2003, kitaev_anyons_2006, ofek_extending_2016, google_quantum_ai_suppressing_2023, sivak_real-time_2023, google_quantum_ai_quantum_2024}. In these protocols, a high-fidelity measurement must be applied to the same qubit repeatedly, for many subsequent cycles of error correction. Therefore, in addition to being precise, the measurement should also leave the observed state intact, i.e., present a quantum non-demolition (QND) character. In circuit quantum electrodynamics, an approximation to QND measurement is achieved by dispersive readout \cite{clerk_introduction_2010,blais_cavity_2004, wallraff_strong_2004, mallet_single-shot_2009, reed_high-fidelity_2010, jeffrey_fast_2014, walter_rapid_2017, dassonneville_fast_2020, swiadek_enhancing_2024, spring_fast_2024}, see Figure~\ref{fig:intro}(a,b). In this readout scheme, the qubit state is inferred from the non-resonant microwave tone elastically scattered off the qubit through the readout resonator.

{Dispersive readout has to be performed much faster than the qubit relaxation time~\cite{gambetta_protocols_2007}. Shortening the duration of the measurement necessitates increasing the power of the probe tone to maintain the same signal-to-noise ratio. However, increasing the power of the} tone has been found to induce unwanted transitions between the qubit states that diminish the QND character of the measurement. These drive-induced transitions can occur from one computational state to another \cite{zhang_engineering_2019,  petrescu_lifetime_2020, hanai_intrinsic_2021, thorbeck_readout-induced_2024, bista_readout-induced_2025} or from a computational state to a non-computational state \cite{reed_high-fidelity_2010, sank_measurement-induced_2016, walter_rapid_2017, lescanne_escape_2019,
khezri_measurement-induced_2023, hazra_benchmarking_2024, bista_readout-induced_2025}. The latter type of transitions -- ``leakage'' from the computational subspace -- is especially problematic. Leakage can lead to the loss of control over the qubit state and thus decapitate quantum error correction protocols \cite{aliferis_fault-tolerant_2007, fowler_coping_2013, ghosh_understanding_2013, suchara_leakage_2015, magnard_fast_2018, bultink_protecting_2020, varbanov_leakage_2020, mcewen_et_al_removing_2021, miao_overcoming_2023}. 
For transmon qubits, one of the dominant mechanisms of leakage is the activation of accidental multi-excitation resonances in the transmon spectrum involving non-computational states \cite{sank_measurement-induced_2016, shillito_dynamics_2022, xiao_diagrammatic_2023, cohen_reminiscence_2023, dumas_measurement-induced_2024, nesterov_measurement-induced_2024}.

Raising the frequency of the readout tone to well beyond the frequency of the transmon allows one to avoid activating multi-excitation resonances. The advantages of such an off-resonant readout were demonstrated in our previous work \cite{kurilovich_high-frequency_2025}. We achieved a QND fidelity above 99.9$\%$ by using a readout resonator with a frequency 12 times higher than that of the transmon. However, the performance of such a high-frequency readout could not be further improved by increasing the power. This raises the question of the mechanisms that limit the QND fidelity of the readout.

\begin{figure*}[t]
  \begin{center}
    \includegraphics[scale = 1]{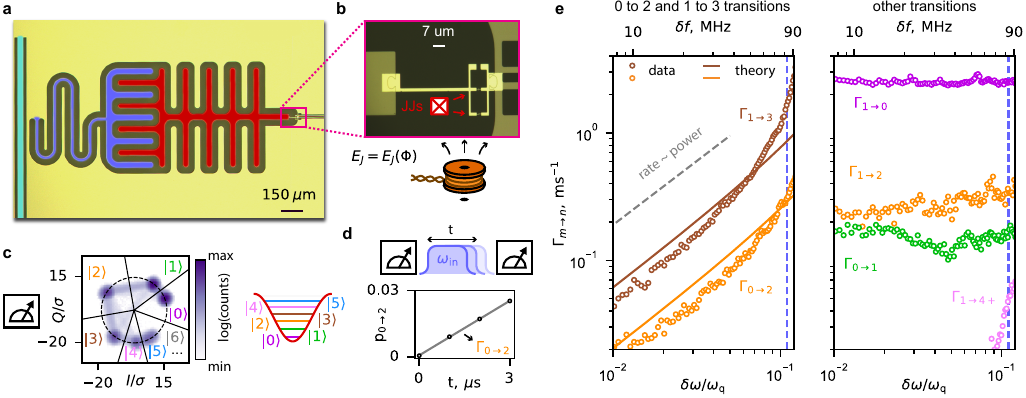}
    \caption{Transition rates of the transmon in the presence of the drive. The qubit is tuned to frequency $\wq/2\pi = 758\:\mathrm{MHz}$ (working point of Ref.~\cite{kurilovich_high-frequency_2025}). The drive frequency $\win/2\pi = 9280\mhz$ is close to that of the readout resonator, $\wres/2\pi = 9227\mhz$. (a) False-color microscope image of the device. Quarter-wavelength readout resonator (blue) is capacitively coupled to the transmon island (red). The resonator is inductively coupled to the transmission line (cyan). (b) Zoom-in on the region of the device containing the Josephson junctions. Two Josephson junctions are arranged in a loop. This allows us to tune the frequency of the device by threading magnetic flux through the loop. (c) Single-shot histogram of resonator measurements after intentionally scrambling the state of the transmon with a $\pi/2$-pulse. The measurement can resolve computational states $|0\rangle$ and $|1\rangle$ as well as the non-computational states $|2\rangle$, $|3\rangle$, $|4\rangle$. All non-computational states higher than $|4\rangle$ are lumped into a single distribution. The population of non-computational states in the histogram stems from transitions caused by inelastic scattering of readout photons. (d) Pulse sequence for rate measurement consists of a pair of measurements separated by a drive pulse of a variable duration and amplitude. We infer the rate by comparing the outcomes for different pulse durations. (e) Transition rates as a function of drive power. The power is quantified by the absolute value of the AC Stark shift experienced by the transmon, $\delta \omega$ (and $\delta f = \delta\omega/2\pi$). The shown range of powers is determined by relevance for qubit readout (see Ref.~\cite{kurilovich_high-frequency_2025}). Vertical dashed lines show the maximum power level reached by the optimal readout pulse in Ref.~\cite{kurilovich_high-frequency_2025}. Left panel: the rate of transitions $|0\rangle\rightarrow |2\rangle$ and $|1\rangle \rightarrow |3\rangle$ linearly increases with power. We attribute this to inelastic scattering of readout photons. Solid lines show the prediction of a parameter-free theory, see Eq.~\eqref{eq:gamma_m_m+2}. Right panel: transition rates $\Gamma_{0\rightarrow 1},\Gamma_{1\rightarrow 0}, \Gamma_{1\rightarrow 2}$ are roughly power independent. Transitions from $|1\rangle$ to $|4\rangle$ and higher states are strongly suppressed for the range of powers relevant for readout. They appear at highest powers, {$\delta\omega/\wq\sim 0.1$}, but their rate remains small compared to that of other transition channels. The rate of transitions from $|0\rangle$ to states $|4\rangle$ and higher is outside of the plotted range.}
    \label{fig:one_flux_point}
  \end{center}
\end{figure*}

In this work, we experimentally identify the mechanisms that violate the QND nature of the high-frequency readout, where $\wres \gg \wq$ ($\wres$ and $\wq$ are frequencies of the readout resonator and the qubit, respectively). The leading leakage mechanism is similar to the well-known Raman scattering of light: a readout photon incoming from the transmission line is ``split’’ between a qubit excitation and an outgoing photon at a lower frequency, see Figure~\ref{fig:intro}(c,d). We identify this inelastic scattering process by measuring the rates of unwanted transmon state transitions, and comparing these rates to the parameter-free theory which we develop.

According to the theory, when the drive photon scatters inelastically off the transmon, the latter excites to a non-computational state, see Figure~\ref{fig:intro}(e,f). We show that the strongest processes of this type are $|0\rangle \rightarrow |2\rangle$ and $|1\rangle \rightarrow |3\rangle$, where $|m\rangle$ denotes the $m$-th eigenstate of the transmon.
We independently measure the rates of such transitions and show that they are in agreement with the theory.
As expected, the rate of unwanted transitions is proportional to the power of the readout tone. The proportionality coefficient is sensitive to the electromagnetic environment of the transmon at the frequency of outgoing photons $\wout$. This frequency is different from both the qubit frequency $\wq$ and the readout frequency $\wres$. Undesired transitions can thus be suppressed by engineering the readout channel to reduce dissipation at frequency $\wout$.

The described above inelastic scattering process, tied to the readout channel itself, is only one of the possible mechanisms that violate the QND character of the measurement. By changing the qubit frequency, we resolve additional demolishing processes. They correspond to accidental resonances with spurious modes either in the electromagnetic environment or the materials of the device \cite{thorbeck_readout-induced_2024}. Due to these resonances, the rate of undesired state transitions becomes elevated at a discrete set of qubit frequencies. At the highest achievable qubit frequencies, we also witness the onset of transitions caused by the multi-excitation resonances between the computational and non-computational states.

Our experimental and theoretical results paint a complete picture of measurement-induced state transitions caused by the dispersive readout of a superconducting qubit. The mechanisms of state transitions that we uncover are also relevant for other contexts where a strong off-resonant drive is applied to the transmon. These include parametric gates \cite{gao_programmable_2018, chapman_high--off-ratio_2023, lu_high-fidelity_2023} and quantum control of linear oscillators \cite{eickbusch_fast_2022, sivak_real-time_2023}.

\section{Measuring the transition rates}
We begin by describing the details of our experimental system. It consists of a transmon qubit and a readout resonator implemented in a 2D architecture, see Figure~\ref{fig:one_flux_point}(a,b). The transmon has charging energy $E_C / h = 36\mhz$; its frequency can be tuned between $\wq/2\pi = 0.520\ghz$ and $\wq/2\pi = 1.53\ghz$ with magnetic flux through a SQUID loop. Our recent readout result \cite{kurilovich_high-frequency_2025} was obtained with this device tuned to a frequency $\wq/2\pi = 0.758\ghz$ (corresponding to magnetic flux $\Phi \approx 0.42\Phi_0$, where $\Phi_0$ is the flux quantum). A quarter-wave readout resonator has frequency between $\wres/2\pi = 9.223\ghz$ and $\wres/2\pi = 9.240\ghz$ depending on the transmon flux bias. The resonator linewidth is $\kappa/2\pi = 1.80\mhz$. The linewidth is determined by the inductive coupling of the resonator to the readout transmission line. At the bias point of Ref.~\cite{kurilovich_high-frequency_2025}, the strength of capacitive coupling between the qubit and the resonator is $g/2\pi=515\mhz$. The presence of the transmon results in the state-dependent pull of the cavity frequency. For example, at $\wq/2\pi=0.758\:\mathrm{GHz}$, the difference between the frequency shifts with transmon being in its ground or excited state is $\chi/2\pi = 0.90\mhz$, allowing us to readout the transmon state.

To probe the the unwanted transitions caused by the drive,
we initialize the transmon in either its ground state $|0\rangle$ or in its excited state $|1\rangle$. We then apply a pulsed tone at a frequency $\win$ close to that of the readout resonator. We detune the drive by many resonator linewidths to suppresses the measurement-induced broadening of qubit transitions (by 40 to 55 MHz depending on the flux bias). This simplifies the analysis of drive-induced transitions without changing the physical picture. After the pulse, we measure the resulting transmon state $|m\rangle$; the measurement can resolve states from $|0\rangle$ to $|4\rangle$, see Figure~\ref{fig:one_flux_point}(c,d). Repeating this experiment many times, we find the transition probability between initial state $|m\rangle$ and final state $|n\rangle$. Changing the duration of the pulse yields the transition rates $\Gamma_{m\rightarrow n}$. We then repeat this experiment at different powers of the drive tone. The resulting power-dependence of $\Gamma_{m\rightarrow n}$ is shown in Figure~\ref{fig:one_flux_point}(e). The power is quantified by the the absolute value of the independently-calibrated AC Stark shift experienced by the transmon, $\delta\omega = \chi\nbar$. See supplementary materials for the details of the Stark shift calibration.

At zero drive power, the transitions of the transmon are caused by its coupling to the equilibrium environment. The only non-zero transition rates are $\Gamma_{0\rightarrow 1}$, $\Gamma_{1\rightarrow 0}$, and $\Gamma_{1\rightarrow 2}$. Relaxation rate $\Gamma_{1\rightarrow 0}$ exceeds the excitation rate $\Gamma_{0\rightarrow 1}$ roughly by a factor of 10. This is consistent with the detailed balance relation, $\Gamma_{1\rightarrow 0} = e^{\hbar\wq / k_B T}\Gamma_{0\rightarrow 1}$, where $T = 16\:\mathrm{mK}$ is close to the base temperature of our dilution refrigerator.
%The measured excitation rate, $\Gamma_{1\rightarrow 2}$, exceeds $\Gamma_{0\rightarrow 1}$ by a factor of two, consistent with the expectation for a weakly anharmonic oscillator.
Rates $\Gamma_{0\rightarrow 1}$, $\Gamma_{1\rightarrow 0}$, and $\Gamma_{1\rightarrow 2}$ remain roughly unchanged upon application of a high-power pulse. 
%Isolated peaks in the transition rates $\Gamma_{0\rightarrow 1}$, $\Gamma_{1\rightarrow 0}$, and $\Gamma_{1\rightarrow 2}$ that appear at finite drive power occur when the AC-Stark-shifted qubit frequency matches that of the lossy environment modes \cite{thorbeck_readout-induced_2024}.

Finite drive power excites the transmon to non-computational states. The dominant drive-activated processes that we observe are $|0\rangle\rightarrow |2\rangle$ and $|1\rangle \rightarrow |3\rangle$. The rates of these processes, $\Gamma_{0 \rightarrow 2}$ and $\Gamma_{1 \rightarrow 3}$, grow linearly with the power of the drive, as long as the power is sufficiently low. The growth becomes faster than linear at higher powers. This drive-induced excitation was the dominant leakage process limiting QND character of our high-frequency readout introduced in Ref.~\cite{kurilovich_high-frequency_2025}. Direct excitation to states $|4\rangle$ and higher starts to emerge at high drive powers. However, the rate of such excitation remains small compared to $\Gamma_{0\rightarrow 2}$ and $\Gamma_{1\rightarrow 3}$ in the range of powers relevant for readout. %We attribute the appearance of such transitions at the highest powers to the higher-order nonlinear processes.

\section{Inelastic scattering theory}
\label{sec:inelastic}
Previously, readout-induced leakage transitions were attributed to the drive activating accidental multi-excitation resonances in the transmon spectrum \cite{sank_measurement-induced_2016, xiao_diagrammatic_2023, khezri_measurement-induced_2023, dumas_measurement-induced_2024}. In this process, several readout photons are simultaneously absorbed, promoting the transmon to a non-computational state. This explanation, however, is inadequate for our data. The transitions that we observe do not have a resonant character. Moreover, since $\wres\gg\wq$, non-linear resonances would excite transmon to much higher non-computational states than $|2\rangle$ or $|3\rangle$. As is shown in Fig.~\ref{fig:one_flux_point}(e), excitation to levels $|4\rangle$ and higher does not happen in the range of powers relevant for readout. Thus, an explanation different from transmon multi-excitation resonances is needed for the observed undesired transitions.

We attribute the undesired transitions caused by the readout tone to inelastic scattering of readout photons. As mentioned in the introduction, in the process of inelastic scattering, a readout photon with frequency $\win$ splits between qubit excitation $|m\rangle \rightarrow |m + 2\rangle$ and a photon reflected at a smaller frequency $\wout = \win - \omega_{m+2, m}$ [where $\omega_{m+2, m} = (\epsilon_{m+2} - \epsilon_m)/\hbar$, with $\epsilon_m$ being the energy of state $|m\rangle$ in the computational basis]. We evaluate the rate of inelastic scattering via Fermi's Golden rule in which we treat the non-linearity as a perturbation. We set out by outlining the outcome of the calculation. We then provide the details of how the calculation is carried out.

According to our calculation, the rate of qubit excitation $|m\rangle \rightarrow |m + 2\rangle$ due to the inelastic scattering of readout photons is
\begin{equation}
\label{eq:gamma_m_m+2}
\Gamma_{m\rightarrow m+2} = (m+1)(m+2) \frac{\wq}{\wout}\frac{2\pi\mathrm{Re}Z[\wout]}{R_Q} \delta\omega.
\end{equation}
Here, $\delta\omega$ is the previously-introduced AC Stark shift of the qubit; the transition rate scales linearly with this parameter. Quantity $\mathrm{Re}Z[\omega]$ is the dissipative part of the impedance measured between the transmon island and the ground, see Fig.~\ref{fig:intro}(c,f). It is evaluated at the frequency $\wout = \win - \omega_{m+2, m}$ of the inelastically scattered photons. The dissipation is enhanced near the frequency of the readout resonator due to the Purcell effect. Detuning $\wout$ from this frequency suppresses the rate of undesired transitions. Constant $R_Q = h/e^2 \approx 25.8\:\mathrm{k}\Omega$ is the resistance quantum.

Next, we compare the calculated rate with our measurement. To this end, we evaluate the impedance $Z[\omega]$ within a lumped model of our circuit depicted in Fig.~\ref{fig:intro}(c) (see supplement \cite{noauthor_see_nodate} for details) and substitute independently measured $\delta\omega$, and $\wq$ in Eq.~\eqref{eq:gamma_m_m+2}.
%\footnote{When calculating $\wout$, at high powers we account for the AC Stark shift of the transmon transitions.} $\wout = \win - \omega_{m+2, m}$. We calculate the impedance $Z[\omega]$ within a lumped model of our circuit depicted in Fig.~\ref{fig:intro}(f), see supplement \cite{noauthor_see_nodate}.
The result of this calculation is shown with solid lines in Fig.~\ref{fig:one_flux_point}(e). Our parameter-free theory is in a good agreement with the measurement. We attribute the discrepancy of about $25\%$ between the theory and the experiment to our imperfect knowledge of $Z[\omega]$ away from the resonator frequency, see Section~\ref{sec:inelastic_flux} for further discussion.
At the highest powers, the theory starts to underestimate the measured rate. We believe that this happens due to the activation of the higher-order nonlinear processes. These processes stem from high-frequency modes in the qubit environment. The effects of these modes are explored in Section~\ref{sec:full}.

To derive Eq.~\eqref{eq:gamma_m_m+2}, we use the Hamiltonian $H = H_{t} + H_{b} + V$, modeling the transmon coupled to a transmission line through a readout resonator. The term $H_{t} = 4E_C (N - n_g)^2 - E_J \cos \varphi$ describes the transmon; $N$ is the Cooper pair number on the island and $\varphi$ is the superconducting phase. Term $H_{b} = \sum_{k} \hbar\omega_k (b^\dagger_k b_k + 1/2)$ describes the transmission line modes labeled by index $k$; $\omega_k$ and $b_k$ are the corresponding frequencies and annihilation operators, respectively. The term $V$ describes the coupling between the transmon and the transmission line mediated by the readout resonator, $V=\sum_{k}N(\lambda_k b^\dagger_k + \lambda_k^\star b_k)$. In what follows, we link the coupling constants $\lambda_k$ to the impedance $Z[\omega]$ introduced above.

The core idea of our calculation is to treat transmon non-linearity as a perturbation. First, we diagonalize the Hamiltonian $H$ neglecting the non-linearity~\cite{nigg_black-box_2012}. This defines a set of normal modes; the phase operator can be expressed through the normal modes as $\varphi = \varphi_{\rm zpf}(ia - i a^\dagger) + \sum_k (i\mu_k^\star a_k - i\mu_k a_k^\dagger)$. 
Here, operator $a$ corresponds to the transmon mode dressed by its coupling to the environment; zero-point fluctuations of the transmon phase are given by $\varphi_{\rm zpf} = (\hbar\wq/2E_J)^{1/2}$. 
In turn, operators $a_k$ correspond to the dressed environment modes.
Their participation on the transmon island is described by coefficients $\mu_k = 2\hbar^{-1}\omega_k \lambda_k/ (\omega_k^2 - \wq^2)$.
The nonlinearity couples the normal modes to each other. When $E_J/E_C \gg 1$, the coupling can be captured with a term $-E_J \varphi^4/4!$ obtained by Taylor-expanding the Josephson potential to the fourth order. Applying Fermi's Golden rule with respect to this term, we compute the desired transition rates.

%Then, one can compute the transitions between these normal. tdiagonalize the \textit{linearized} Hamiltonian of the system including both the transmon and the environment modes. This gives us a set of normal modes which are independent as long as the non-linearity is neglected. Then, we decompose the operator of the phase $\varphi$ at the Josephson junction in terms of the normal modes. Again, the decomposition includes both the transmon mode and the environment modes. Then, we use the obtained decomposition to compute the matrix element of the $T$-matrix between the scattering states. The $T$-matrix is computed in Born's approximation, \textit{i.e.}, to the lowest order in transmon non-linearity. Third, we employ the resulting matrix element in Fermi's golden rule for the transition rate. This yields the transition rate in terms of parameters characterizing the drive power and coupling between the transmon and the environment. Finally, we relate parameters entering the expression for the rate to measurable quantities, such as AC Stark shift experienced by the qubit and the real part of the impedance $Z[\omega]$ of the transmon island. We assume that one of the transmon pads is grounded such that $Z[\omega]$ is a single number (as opposed to it being a matrix).

Specifically, we assume that the system is initialized in a state $|\Psi_{i}\rangle \propto (a_{\rm in}^\dagger)^{n_{\rm in}} |\Omega\rangle |m\rangle$, where $|\Omega\rangle$ is the vacuum state of the transmission line and $|m\rangle$ is the state of the transmon; the photon occupation number $n_{\rm in}$ and the mode with frequency $\omega_{\rm in}$ characterize the applied drive (index $k=\mathrm{in}$ is used to denote this mode). The final state, where one of the drive photons is inelastically scattered, is $|\Psi_{f}\rangle\propto (a_{\rm in}^\dagger)^{n_{\rm in}-1}a_{\rm out}^\dagger|\Omega\rangle |m+2\rangle$ (index $k=\mathrm{out}$ denotes the mode to which the drive photon is scattered). The transition rate can be written as
\begin{equation}
\label{eq:fgr}
    \Gamma_{m\rightarrow m+2} = \frac{2\pi}{\hbar}\nu_{\rm out} |\mathcal{M}_{fi}|^2,\quad \mathcal{M}_{fi} = \langle \Psi_f | E_J\frac{\varphi^4}{4!}| \Psi_i \rangle.
\end{equation}
Here, $\nu_{\rm out}$ is the photon density of states in the transmission line at the frequency $\omega_{\rm out}$.
%In harmonic approximation the $T$-matrix is 0. To the leading order in non-linearity, it can be expressed as $T = - \frac{1}{4!}E_J \varphi^4$; this equation is obtained by taking the dominant non-linear term in the expansion of the cosine potential of the transmon.

\begin{figure}[t!]
  \begin{center}
    \includegraphics[scale = 1.0]{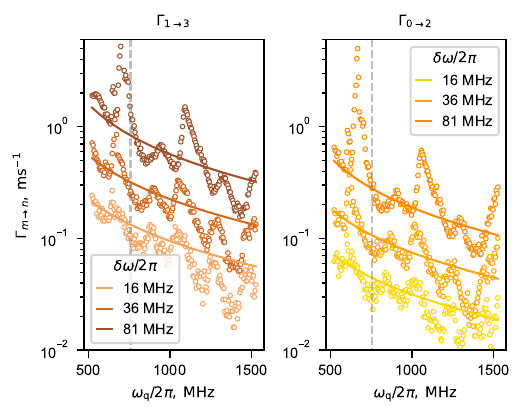}
\caption{Transition rates $\Gamma_{0\rightarrow 2}$ and $\Gamma_{1\rightarrow 3}$ as a function of qubit frequency controlled with flux bias. Both plots show rates measured at three different values of AC Stark shift $\delta\omega$. Grey dashed line indicates the working point of Ref.~\cite{kurilovich_high-frequency_2025}. Solid lines show the prediction of the inelastic scattering theory, Eq.~\eqref{eq:gamma_m_m+2}, where $Z[\omega]$ is computed within the lumped element model of the device. The deviations between the theory and experiment stem from two sources. The first is our imperfect knowledge of $Z[\omega]$ away from the resonator frequency (impedance mismatches). The second is higher-order non-linear processes involving high-frequency modes, see Section \ref{sec:full} for details.}\label{fig:raman_vs_flux}
  \end{center}
\end{figure}
\begin{figure*}[t]
  \begin{center}
    \includegraphics[scale = 1]{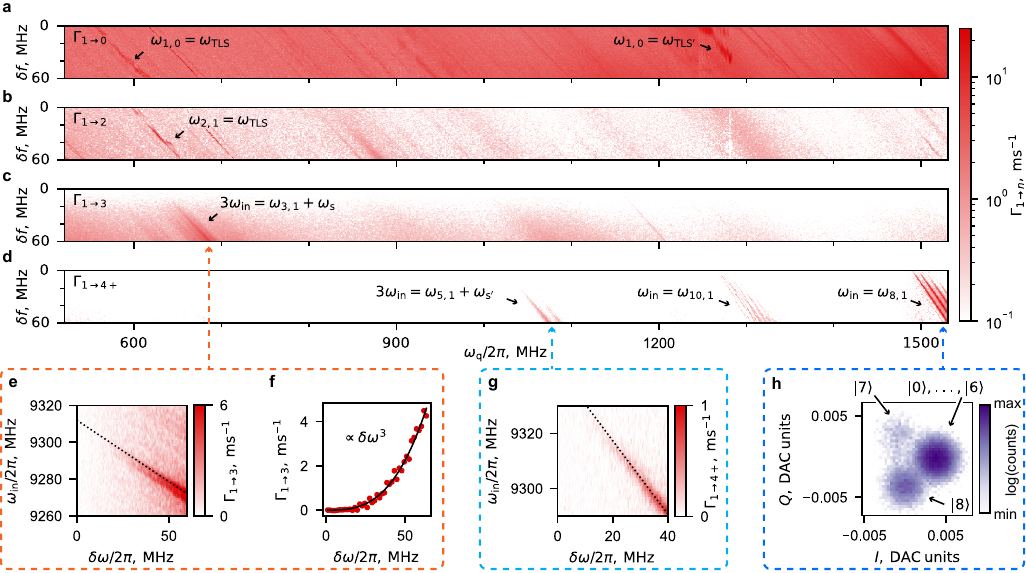}
    \caption{Full characterization of undesired drive-induced transitions of a transmon qubit. (a-d) Transition rates from $|1\rangle$ to various final states plotted as a function of qubit frequency (controlled with magnetic flux) and drive power (quantified by AC Stark shift $\delta f = \delta\omega/2\pi$). The drive frequency $\win/2\pi=9280\mhz$ is close to that of the readout resonator ($\wres/2\pi=9223-9240\mhz$ depending on the flux). (a,b) Rates $\Gamma_{1\rightarrow 0}$ and $\Gamma_{1\rightarrow 2}$. These transitions do not involve the drive photons. The stripey structure is determined by resonances between the AC-Stark shifted qubit frequency and modes in the environment. The frequency shift between the stripes in (a) and (b) is related to difference between $\omega_{1,0}$ and $\omega_{2,1}$ of about $2\pi \cdot 40\mhz$. (c) Rate $\Gamma_{1\rightarrow 3}$. Smooth background corresponds to the inelastic scattering process described in Fig.~1. Sharp features corresponds to the higher-order inelastic processes [see panels (e) and (f) for details]. (d) Rate $\Gamma_{1\rightarrow 4+}$ of transitions to $|4\rangle$ and higher states ($|5\rangle$, $|6\rangle$ and higher). These transitions stem either from higher order inelastic processes [see panel (g)] or from activation of multi-excitation resonances in the transmon spectrum [see panel (h)]. The splitting of the transition lines is explained in the main text. (e) The behavior of the sharp feature in $\Gamma_{1\rightarrow 3}$ at $\wq/2\pi=670\mhz$ as a function of the drive frequency $\win$ and power. Black dashed line corresponds to the condition $3\win = \omega_{3,1}[\delta\omega] + \omega_m$ describing a six-wave mixing process involving a mode with frequency $\omega_{m}/2\pi=26.72\ghz$ in the qubit environment. (f) The power-dependence of $\Gamma_{1\rightarrow 3}$ associated with the feature of panel (e). Consistently with the described six-wave mixing process, the rate scales as the cube of power. (g) The behavior of the sharp feature in $\Gamma_{1\rightarrow 4+}$ at $\wq/2\pi = 1070\mhz$ as a function of drive-frequency and power. The position of the feature is consistent with condition $3\win = \omega_{5,1}[\delta\omega] + \omega_{m^\prime}$, where $\omega_{m^\prime} = 24.150\ghz$.
    (h) Reading out the resonator at a lowered frequency allows us to resolve higher excited states of the transmon. The readout histogram shows that the qubit transitions to state $|8\rangle$ after exciting the sharp feature in $\Gamma_{1\rightarrow 4+}$ at $\wq/2\pi = 1500\mhz$.}
    \label{fig:big_flux_sweep}
  \end{center}
\end{figure*}
We compute $\mathcal{M}_{fi}$ by substituting the normal-mode decomposition of the phase operator $\varphi$ {into} Eq.~\eqref{eq:fgr}. %decompose $\varphi$ entering the $T$-matrix as a sum over the normal modes in the absence of non-linearity. Assuming that the coupling to the environment is weak, the decomposition reads $\varphi = \varphi_{\rm zpf}(ia - i a^\dagger) + \sum_k (i\mu_k a_k - i\mu_k^\star a_k^\dagger)$,
%where $a$ is the transmon annihilation operator, $\varphi_{\rm zpf} = (\hbar\wq/2E_J)^{1/2}$, and coefficient $\mu_k = 2\hbar^{-1}\omega_k \lambda_k/ (\omega_k^2 - \wq^2)$. Substituting the normal mode decomposition into the $T$-matrix, we obtain the transition matrix element
We then find
\begin{equation}
\label{eq:matr_elem}
    \mathcal{M}_{fi} = \sqrt{m+1}\sqrt{m+2} E_C \nzpf^2 \sqrt{n_{\rm in}} \mu_{\rm in} \mu_{\rm out}.
\end{equation}
To express the result in terms of measurable quantities, we note that $\mu_{\rm in}$ can be related to the Stark shift as $\delta\omega = \frac{1}{2}\wq|\mu_{\rm in}|^2 n_{\rm in}$ \cite{noauthor_see_nodate}. The constant $\mu_{\rm out}$ can be linked the impedance of the transmon island, $\mathrm{Re}Z[\omega_{\rm out}]/R_Q = \hbar\omega_{\rm out}\nu_{\rm out}|\mu_{\rm out}|^2 / 8$. Employing these substitutions in Eqs.~\eqref{eq:fgr} and \eqref{eq:matr_elem} we arrive at Eq.~\eqref{eq:gamma_m_m+2}.

In our derivation, we assumed that the qubit nonlinearity is weak. In the supplementary materials \cite{noauthor_see_nodate}, we show that weak non-linearity assumption can be lifted if the coupling between the transmon and the environment is small. We also develop the theory for other possible inelastic scattering processes. For example, we consider a process where two drive photons and a qubit excitation are converted into an excitation of an environment mode.

\section{Inelastic scattering rate vs qubit frequency}
\label{sec:inelastic_flux}
Next, to further validate the theory, we measure the rates $\Gamma_{0\rightarrow 2}$ and $\Gamma_{1 \rightarrow 3}$ at different qubit frequencies. To this end, we repeat the measurement described above at different values of the magnetic flux threaded through the SQUID loop of our device. At each value of the flux we independently calibrate the qubit frequency, the resonator frequency, the relation between the Stark shift and drive power, and the readout thresholds used for state assignment, see supplementary materials \cite{noauthor_see_nodate}.

The results of this measurement for several different drive powers are shown in Fig.~\ref{fig:raman_vs_flux}. At small powers, the theoretical prediction for the rates is close to the data at all qubit frequencies. We attribute the deviations to the impedance mismatches in the transmission line that create an uncertainty in our estimate of $\mathrm{Re}Z$ in Eq.~\eqref{eq:fgr}. Notably, for a fixed value of the AC Stark shift $\delta \omega$, the rate of unwanted transitions is enhanced at smaller qubit frequencies. The origin of this enhancement is the Purcell effect. Indeed, due to the Purcell effect, $\mathrm{Re}Z[\omega]\propto 1 / (\omega - \wres)^2$. Therefore, $\mathrm{Re}\:Z[\wout]\propto 1/\wq^2$ since $\omega_\mathrm{\rm out} \approx \win - 2\wq$ and $\win$ is close to $\wres$. Combined with the prefactor $\propto\wq$ in Eq.~\eqref{eq:fgr}, this results in $\Gamma_{m\rightarrow m + 2} \propto \delta \omega / \wq$.

At higher drive power additional peaks appear in the measured rates. These stem from the activation of higher-order nonlinear processes involving the spurious device modes. We explore these processes in the next section.

\section{Full characterization of transitions}
\label{sec:full}
The inelastic scattering process that we describe above is only one of the possible mechanisms by which the readout drive can cause unwanted transmon state transitions. In this section, we describe the measurement that reveals the other transition mechanisms. Specifically, we sweep a range of qubit frequencies and drive powers. At each point in this sweep, we measure the transition rates $\Gamma_{m\rightarrow n}$ from the computational states $|m\rangle$ with $m=0,1$ to the states $|n\rangle$ with $n = 0,1,2,3,4+$. Here, $4+$ label any excited non-computational state higher than 3, \textit{i.e.}, states $|4\rangle$, $|5\rangle$, $|6\rangle$ and higher.

The results of this measurement for $m=1$ are shown in Fig.~\ref{fig:big_flux_sweep} (similar data for $m=0$ is presented in the supplementary materials). The plots demonstrate two distinct behaviors for the transition rates. First, rates $\Gamma_{1 \rightarrow 0}$, $\Gamma_{1 \rightarrow 2}$, and $\Gamma_{1 \rightarrow 3}$ have a smooth background present at all qubit frequencies. The background in $\Gamma_{1 \rightarrow 0}$ and $\Gamma_{1 \rightarrow 2}$ is roughly independent of the drive power. Second, there is a set of resonant features in the measured transition rates. These resonances manifest as peaks in the transition rates at certain combinations of qubit frequency and drive power.

For the rate $\Gamma_{1 \rightarrow 0}$, the smooth background corresponds to the direct emission of the qubit excitation into the environment. This process is responsible for the relaxation of the undriven qubit. The rate of qubit decay increases with qubit frequency \cite{noauthor_see_nodate}. The smooth background in $\Gamma_{1\rightarrow 2}$ corresponds to thermal excitation by the same environment that is responsible for $\Gamma_{1\rightarrow 0}$.
%The ratio between rates $\Gamma_{1\rightarrow 2}$ and $\Gamma_{1\rightarrow 0}$ is consistent with the thermal occupation of the environment modes (at the base temperature of the refridgerator, $T=16\:\mathrm{mK}$) and the matrix elements for the respective transitions.
%are related to each other via $\Gamma_{1\rightarrow 2} = 2 \exp\left(-\hbar\wq/k_B T\right)\Gamma_{1\rightarrow 0}$, where $T\approx 16\:\mathrm{mK}$ is the base temperature of the refrigerator.
Finally, the smooth background in $\Gamma_{1\rightarrow 3}$ stems from the inelastic scattering process described in Section \ref{sec:inelastic} and \ref{sec:inelastic_flux}. The rate of this process is zero in the absence of the drive.

Resonant features in the rates $\Gamma_{1\rightarrow 0}$ (or $\Gamma_{1\rightarrow 2}$) correspond to resonances in the qubit environment at some frequencies $\omega_\mathrm{TLS}$. Most likely, these resonances originate from excitation (or relaxation) of material defects in the device that are strongly coupled to the qubit degree of freedom \cite{thorbeck_readout-induced_2024}.  The peaks in $\Gamma_{1\rightarrow 0}$ occur when a frequency matching condition is fullfilled, $\omega_{1,0}[\delta\omega] = \omega_{\mathrm{TLS}}$. Due to the AC-Stark shift the transition frequency $\omega_{1,0}$ depends on the drive power, $\omega_{1,0}[\delta\omega] = \omega_{1,0}[0] - \delta\omega$. This explains the slope of the resonant features in Fig.~\ref{fig:big_flux_sweep}(a). The peaks in $\Gamma_{1\rightarrow 2}$ happen when $\omega_{2,1}[\delta\omega] = \omega_{\mathrm{TLS}}$. Frequency $\omega_{2,1}$ differs from $\omega_{1,0}$ due to the qubit anharmonicity. Therefore, the resonant features are shifted by $\sim 40\:\mathrm{MHz}$ between Fig.~\ref{fig:big_flux_sweep}(a) and Fig.~\ref{fig:big_flux_sweep}(b).
%As expected, for a fixed value of the AC Stark shift $\delta\omega$, the frequency of the resonant features does not depend on the drive frequency, see Fig.~\ref{fig:big_map}(e).

The resonant feature in $\Gamma_{1\rightarrow 3}$ around $670\:\mathrm{MHz}$ stems from a six-wave process mixing three readout photons with two qubit excitation quanta and a spurious device mode~\cite{singh_impact_2025}. To determine the number of the involved drive photons, we vary the frequency of the drive and measure the power dependence of the transition rate (the magnetic flux is kept fixed such that in the absence of the drive the qubit frequency is $670\:\mathrm{MHz}$). The result of this measurement is shown in Fig.~\ref{fig:big_flux_sweep}(e). For a fixed amount of AC Stark shift, the position of the resonance depends on the frequency of the drive. The dependence is well described by a resonant condition $3\win = \omega_{3,1}[\delta\omega] + \omega_s$ that corresponds to the six-wave mixing process: absorption of three drive photons with frequency $\win$ results in the qubit transition $|1\rangle\rightarrow|3\rangle$ and excitation of a spurious mode with frequency $\omega_s/2\pi = 26.72\:\mathrm{GHz}$ (likely $3\lambda/4$ mode of the readout resonator). The transition frequency $\omega_{3,1}$ experiences AC Stark shift, $\omega_{3,1}[\delta\omega] = \omega_{3,1}[0] - 2\delta\omega$; this explains the slope in Fig. 4(e). Theoretically, the rate of the described six wave-mixing process is proportional to the cube of the drive power. As is shown in Fig.~\ref{fig:big_flux_sweep}(f), this is indeed the case in our data.

As the qubit frequency is increased, resonant features appear in the rate of excitation to higher excited states of the transmon, $\Gamma_{1\rightarrow 4+}$. The most prominent feature occurs around $\wq/2\pi = 1500\:\mathrm{MHz}$. This ``bear claw'' feature stems from the activation of a multi-excitation resonance in the transmon spectrum. The drive becomes resonant with the transition between states $|1\rangle$ and $|8\rangle$, \textit{i.e.}, condition $\win = \omega_{8,1}$ is fulfilled. As demonstrated in Fig.~\ref{fig:big_flux_sweep}(h), the drive does indeed excite the transmon to state $|8\rangle$. Notably, the spectroscopic line in Fig.~\ref{fig:big_flux_sweep}(d) is followed by several parallel ``replicas''. These replicas stem from the detuning of about 40 MHz between the drive frequency and the frequency of the readout resonator. The replicas correspond to a process in which $n$ drive quanta are converted into a qubit excitation from $|1\rangle$ to $|8\rangle$ and $n-1$ resonator quanta. They occur when the condition $n \win = \omega_{8,1}[\delta\omega] + (n-1)\omega_\mathrm{r}$ is fulfilled for some integer $n$.

Some of the other resonant features in Fig.~\ref{fig:big_flux_sweep} are labeled according to their character. They stem from one of the following mechanisms: (i) direct interaction with isolated lossy modes, (ii) nonlinear processes involving spurious degrees of freedom, (iii) activation of multi-excitation resonances in the transmon spectrum.
\section{Discussion and conclusion}
Applications of superconducting circuits to quantum information tasks require coupling them to microwave tones. In particular, projective measurement of a superconducting qubit is achieved by off-resonantly driving the qubit through a readout resonator. The speed and thus the performance of the measurement can be improved by increasing the drive power. {However, the increase of the drive power also leads to the appearance of unwanted state transitions.} These drive-activated transitions limit the measurement fidelity. The origin of these transitions is a subject of the present investigation.

In our work, we fully characterize the mechanisms of state transitions in an off-resonantly driven transmon qubit. We focus on the case of a dispersive qubit readout achieved with a readout resonator with a high-frequency, $\wres/\wq\gg 1$ \cite{kurilovich_high-frequency_2025}. By changing the qubit frequency using magnetic flux, we resolve two types of the unwanted drive-activated transitions.
Transitions of the first type are ubiquitously present at all qubit frequencies. In their course, the qubit is excited to a non-computational state with a rate proportional to the drive power. These transitions occur due to the inelastic scattering of drive photons back into the readout channel, a process allowed by the transmon non-linearity. Specifically, a drive photon with frequency $\win$ splits into a qubit excitation $|0\rangle\rightarrow |2\rangle$ [or $|1\rangle \rightarrow |3\rangle$] and an outgoing photon at a smaller frequency $\wout$, see Fig.~\ref{fig:intro}. The measured rate of inelastic scattering is in a good agreement with our parameter-free theory, Eq.~\eqref{eq:gamma_m_m+2}, see Fig.~\ref{fig:raman_vs_flux}. The inelastic scattering process is the dominant source of leakage error in our recent readout result \cite{kurilovich_high-frequency_2025}.

Transitions of the second type have a resonant character, i.e., they appear at a discrete set of qubit frequencies. The causes of these resonances are excitation of degrees of freedom associated with material defects or spurious electromagnetic modes in the qubit environment. With the increase of transmon frequency, transmon multi-excitation resonances also start to interfere with the readout process. In these resonances, the drive frequency matches that of a transition between a computational and a non-computational state. This leads to leakage from the computational basis.

The understanding developed in our work gives a comprehensive roadmap to mitigating the unwanted drive-induced state transitions in superconducting circuits. Multi-excitation resonances can be suppressed by increasing the drive frequency compared to that of a transmon \cite{kurilovich_high-frequency_2025}. Material defects contributing to the spectrum of excitations in the microwave frequency range can potentially be suppressed by improved fabrication techniques. The transitions caused by inelastic scattering of drive photons can be mitigated by means of microwave engineering. For example, the Raman-like process described above can be alleviated by filtering the transmission line at the frequency of inelastically scattered photons $\wout$.

While detuning of the readout tone from the qubit frequency allows one to increase the readout speed, this route eventually faces limitations due to higher-order inelastic scattering processes. For example, as described in Section \ref{sec:full}, we observe a strong process where three drive photons convert into two qubit excitations, $|1\rangle \rightarrow |3\rangle$, and an excitation of an electromagnetic mode at $\omega_m/2\pi\approx 27\:\mathrm{GHz}$. This shows the importance of microwave hygiene even at frequencies far exceeding that of the qubit or of the readout drive. 

Finally, we note that inelastic processes involving Bogoliubov quasiparticles can also degrade the performance of superconducting qubit operations carried out with microwave tones. This type of inelastic processes was recently analyzed in Refs.~\cite{kishmar_quasiparticle-induced_2025, chowdhury_theory_2025}.

\textit{Note:} Recently, two relevant papers appeared~\cite{fechant_offset_2025, wang_probing_2025} that focus on drive-induced transitions caused by multi-excitation resonances.

\section{Acknowledgements}
We thank Christian K. Andersen, Arno Bargerbos, Marta Pita-Vidal, Lukas Splitthoff, Jaap Wesdorp, and Daniel Sank for discussions. We thank Alessandro Miano and Elifnaz Önder for help with the measurement setup. Finally, we thank Y. Sun, K. Woods, L. McCabe, and M. Rooks for their assistance and guidance in the device fabrication processes.

{This research was sponsored by the Army Research Office (ARO) under grants no.~W911NF-22-1-0053 and W911NF-23-1-0051, by DARPA under grant no.~HR0011-24-2-0346, by the U.S.~Department of Energy (DoE), Office of Science, National Quantum Information Science Research Centers, Co-design Center for Quantum Advantage (C2QA) under contract number DE-SC0012704, and by the Air Force Office of Scientific Research (AFOSR) under award number FA9550-21-1-0209. The views and conclusions contained in this document are those of the authors and should not be interpreted as representing the official policies, either expressed or implied, of the ARO, DARPA, DoE, AFOSR or the US Government. The US Government is authorized to reproduce and distribute reprints for Government purposes notwithstanding any copyright notation herein. Fabrication facilities use was supported by the Yale Institute for Nanoscience and Quantum Engineering (YINQE) and the Yale Univeristy Cleanroom. L.F. is a founder and shareholder of Quantum Circuits Inc. (QCI).}
\bibliography{references.bib}

\begin{thebibliography}{52}%
\makeatletter
\providecommand \@ifxundefined [1]{%
 \@ifx{#1\undefined}
}%
\providecommand \@ifnum [1]{%
 \ifnum #1\expandafter \@firstoftwo
 \else \expandafter \@secondoftwo
 \fi
}%
\providecommand \@ifx [1]{%
 \ifx #1\expandafter \@firstoftwo
 \else \expandafter \@secondoftwo
 \fi
}%
\providecommand \natexlab [1]{#1}%
\providecommand \enquote  [1]{``#1''}%
\providecommand \bibnamefont  [1]{#1}%
\providecommand \bibfnamefont [1]{#1}%
\providecommand \citenamefont [1]{#1}%
\providecommand \href@noop [0]{\@secondoftwo}%
\providecommand \href [0]{\begingroup \@sanitize@url \@href}%
\providecommand \@href[1]{\@@startlink{#1}\@@href}%
\providecommand \@@href[1]{\endgroup#1\@@endlink}%
\providecommand \@sanitize@url [0]{\catcode `\\12\catcode `\$12\catcode `\&12\catcode `\#12\catcode `\^12\catcode `\_12\catcode `\%12\relax}%
\providecommand \@@startlink[1]{}%
\providecommand \@@endlink[0]{}%
\providecommand \url  [0]{\begingroup\@sanitize@url \@url }%
\providecommand \@url [1]{\endgroup\@href {#1}{\urlprefix }}%
\providecommand \urlprefix  [0]{URL }%
\providecommand \Eprint [0]{\href }%
\providecommand \doibase [0]{https://doi.org/}%
\providecommand \selectlanguage [0]{\@gobble}%
\providecommand \bibinfo  [0]{\@secondoftwo}%
\providecommand \bibfield  [0]{\@secondoftwo}%
\providecommand \translation [1]{[#1]}%
\providecommand \BibitemOpen [0]{}%
\providecommand \bibitemStop [0]{}%
\providecommand \bibitemNoStop [0]{.\EOS\space}%
\providecommand \EOS [0]{\spacefactor3000\relax}%
\providecommand \BibitemShut  [1]{\csname bibitem#1\endcsname}%
\let\auto@bib@innerbib\@empty
%</preamble>
\bibitem [{\citenamefont {Kitaev}(2003)}]{kitaev_fault-tolerant_2003}%
  \BibitemOpen
  \bibfield  {author} {\bibinfo {author} {\bibfnamefont {A.~Y.}\ \bibnamefont {Kitaev}},\ }\bibfield  {title} {\bibinfo {title} {Fault-tolerant quantum computation by anyons},\ }\href {https://doi.org/10.1016/S0003-4916(02)00018-0} {\bibfield  {journal} {\bibinfo  {journal} {Annals of Physics}\ }\textbf {\bibinfo {volume} {303}},\ \bibinfo {pages} {2} (\bibinfo {year} {2003})}\BibitemShut {NoStop}%
\bibitem [{\citenamefont {Kitaev}(2006)}]{kitaev_anyons_2006}%
  \BibitemOpen
  \bibfield  {author} {\bibinfo {author} {\bibfnamefont {A.}~\bibnamefont {Kitaev}},\ }\bibfield  {title} {\bibinfo {title} {Anyons in an exactly solved model and beyond},\ }\href {https://doi.org/10.1016/j.aop.2005.10.005} {\bibfield  {journal} {\bibinfo  {journal} {Annals of Physics}\ }\bibinfo {series} {January {Special} {Issue}},\ \textbf {\bibinfo {volume} {321}},\ \bibinfo {pages} {2} (\bibinfo {year} {2006})}\BibitemShut {NoStop}%
\bibitem [{\citenamefont {Sank}\ \emph {et~al.}(2016)\citenamefont {Sank}, \citenamefont {Chen},\ and\ \citenamefont {Khezri~et al.}}]{sank_measurement-induced_2016}%
  \BibitemOpen
  \bibfield  {author} {\bibinfo {author} {\bibfnamefont {D.}~\bibnamefont {Sank}}, \bibinfo {author} {\bibfnamefont {Z.}~\bibnamefont {Chen}},\ and\ \bibinfo {author} {\bibfnamefont {M.}~\bibnamefont {Khezri~et al.}},\ }\bibfield  {title} {\bibinfo {title} {Measurement-{Induced} {State} {Transitions} in a {Superconducting} {Qubit}: {Beyond} the {Rotating} {Wave} {Approximation}},\ }\href {https://doi.org/10.1103/PhysRevLett.117.190503} {\bibfield  {journal} {\bibinfo  {journal} {Physical Review Letters}\ }\textbf {\bibinfo {volume} {117}},\ \bibinfo {pages} {190503} (\bibinfo {year} {2016})}\BibitemShut {NoStop}%
\bibitem [{\citenamefont {Khezri}\ and\ \citenamefont {Opremcak~et al.}(2023)}]{khezri_measurement-induced_2023}%
  \BibitemOpen
  \bibfield  {author} {\bibinfo {author} {\bibfnamefont {M.}~\bibnamefont {Khezri}}\ and\ \bibinfo {author} {\bibfnamefont {A.}~\bibnamefont {Opremcak~et al.}},\ }\bibfield  {title} {\bibinfo {title} {Measurement-induced state transitions in a superconducting qubit: {Within} the rotating-wave approximation},\ }\href {https://doi.org/10.1103/PhysRevApplied.20.054008} {\bibfield  {journal} {\bibinfo  {journal} {Physical Review Applied}\ }\textbf {\bibinfo {volume} {20}},\ \bibinfo {pages} {054008} (\bibinfo {year} {2023})}\BibitemShut {NoStop}%
\bibitem [{\citenamefont {Kurilovich}\ \emph {et~al.}(2025)\citenamefont {Kurilovich}, \citenamefont {Connolly}, \citenamefont {Bøttcher}, \citenamefont {Weiss}, \citenamefont {Hazra}, \citenamefont {Joshi}, \citenamefont {Ding}, \citenamefont {Nho}, \citenamefont {Diamond}, \citenamefont {Kurilovich}, \citenamefont {Dai}, \citenamefont {Fatemi}, \citenamefont {Frunzio}, \citenamefont {Glazman},\ and\ \citenamefont {Devoret}}]{kurilovich_high-frequency_2025}%
  \BibitemOpen
  \bibfield  {author} {\bibinfo {author} {\bibfnamefont {P.~D.}\ \bibnamefont {Kurilovich}}, \bibinfo {author} {\bibfnamefont {T.}~\bibnamefont {Connolly}}, \bibinfo {author} {\bibfnamefont {C.~G.~L.}\ \bibnamefont {Bøttcher}}, \bibinfo {author} {\bibfnamefont {D.~K.}\ \bibnamefont {Weiss}}, \bibinfo {author} {\bibfnamefont {S.}~\bibnamefont {Hazra}}, \bibinfo {author} {\bibfnamefont {V.~R.}\ \bibnamefont {Joshi}}, \bibinfo {author} {\bibfnamefont {A.~Z.}\ \bibnamefont {Ding}}, \bibinfo {author} {\bibfnamefont {H.}~\bibnamefont {Nho}}, \bibinfo {author} {\bibfnamefont {S.}~\bibnamefont {Diamond}}, \bibinfo {author} {\bibfnamefont {V.~D.}\ \bibnamefont {Kurilovich}}, \bibinfo {author} {\bibfnamefont {W.}~\bibnamefont {Dai}}, \bibinfo {author} {\bibfnamefont {V.}~\bibnamefont {Fatemi}}, \bibinfo {author} {\bibfnamefont {L.}~\bibnamefont {Frunzio}}, \bibinfo {author} {\bibfnamefont {L.~I.}\ \bibnamefont {Glazman}},\ and\ \bibinfo {author} {\bibfnamefont {M.~H.}\ \bibnamefont {Devoret}},\ }\href
  {https://doi.org/10.48550/arXiv.2501.09161} {\bibinfo {title} {High-frequency readout free from transmon multi-excitation resonances}} (\bibinfo {year} {2025}),\ \bibinfo {note} {arXiv:2501.09161 [quant-ph]}\BibitemShut {NoStop}%
\bibitem [{\citenamefont {Thorbeck}\ \emph {et~al.}(2024)\citenamefont {Thorbeck}, \citenamefont {Xiao}, \citenamefont {Kamal},\ and\ \citenamefont {Govia}}]{thorbeck_readout-induced_2024}%
  \BibitemOpen
  \bibfield  {author} {\bibinfo {author} {\bibfnamefont {T.}~\bibnamefont {Thorbeck}}, \bibinfo {author} {\bibfnamefont {Z.}~\bibnamefont {Xiao}}, \bibinfo {author} {\bibfnamefont {A.}~\bibnamefont {Kamal}},\ and\ \bibinfo {author} {\bibfnamefont {L.~C.}\ \bibnamefont {Govia}},\ }\bibfield  {title} {\bibinfo {title} {Readout-{Induced} {Suppression} and {Enhancement} of {Superconducting} {Qubit} {Lifetimes}},\ }\href {https://doi.org/10.1103/PhysRevLett.132.090602} {\bibfield  {journal} {\bibinfo  {journal} {Physical Review Letters}\ }\textbf {\bibinfo {volume} {132}},\ \bibinfo {pages} {090602} (\bibinfo {year} {2024})}\BibitemShut {NoStop}%
\bibitem [{\citenamefont {Dumas}\ \emph {et~al.}(2024)\citenamefont {Dumas}, \citenamefont {Groleau-Paré}, \citenamefont {McDonald}, \citenamefont {Muñoz-Arias}, \citenamefont {Lledó}, \citenamefont {D’Anjou},\ and\ \citenamefont {Blais}}]{dumas_measurement-induced_2024}%
  \BibitemOpen
  \bibfield  {author} {\bibinfo {author} {\bibfnamefont {M.~F.}\ \bibnamefont {Dumas}}, \bibinfo {author} {\bibfnamefont {B.}~\bibnamefont {Groleau-Paré}}, \bibinfo {author} {\bibfnamefont {A.}~\bibnamefont {McDonald}}, \bibinfo {author} {\bibfnamefont {M.~H.}\ \bibnamefont {Muñoz-Arias}}, \bibinfo {author} {\bibfnamefont {C.}~\bibnamefont {Lledó}}, \bibinfo {author} {\bibfnamefont {B.}~\bibnamefont {D’Anjou}},\ and\ \bibinfo {author} {\bibfnamefont {A.}~\bibnamefont {Blais}},\ }\bibfield  {title} {\bibinfo {title} {Measurement-{Induced} {Transmon} {Ionization}},\ }\href {https://doi.org/10.1103/PhysRevX.14.041023} {\bibfield  {journal} {\bibinfo  {journal} {Physical Review X}\ }\textbf {\bibinfo {volume} {14}},\ \bibinfo {pages} {041023} (\bibinfo {year} {2024})}\BibitemShut {NoStop}%
\bibitem [{\citenamefont {Ofek}\ \emph {et~al.}(2016)\citenamefont {Ofek}, \citenamefont {Petrenko}, \citenamefont {Heeres}, \citenamefont {Reinhold}, \citenamefont {Leghtas}, \citenamefont {Vlastakis}, \citenamefont {Liu}, \citenamefont {Frunzio}, \citenamefont {Girvin}, \citenamefont {Jiang}, \citenamefont {Mirrahimi}, \citenamefont {Devoret},\ and\ \citenamefont {Schoelkopf}}]{ofek_extending_2016}%
  \BibitemOpen
  \bibfield  {author} {\bibinfo {author} {\bibfnamefont {N.}~\bibnamefont {Ofek}}, \bibinfo {author} {\bibfnamefont {A.}~\bibnamefont {Petrenko}}, \bibinfo {author} {\bibfnamefont {R.}~\bibnamefont {Heeres}}, \bibinfo {author} {\bibfnamefont {P.}~\bibnamefont {Reinhold}}, \bibinfo {author} {\bibfnamefont {Z.}~\bibnamefont {Leghtas}}, \bibinfo {author} {\bibfnamefont {B.}~\bibnamefont {Vlastakis}}, \bibinfo {author} {\bibfnamefont {Y.}~\bibnamefont {Liu}}, \bibinfo {author} {\bibfnamefont {L.}~\bibnamefont {Frunzio}}, \bibinfo {author} {\bibfnamefont {S.~M.}\ \bibnamefont {Girvin}}, \bibinfo {author} {\bibfnamefont {L.}~\bibnamefont {Jiang}}, \bibinfo {author} {\bibfnamefont {M.}~\bibnamefont {Mirrahimi}}, \bibinfo {author} {\bibfnamefont {M.~H.}\ \bibnamefont {Devoret}},\ and\ \bibinfo {author} {\bibfnamefont {R.~J.}\ \bibnamefont {Schoelkopf}},\ }\bibfield  {title} {\bibinfo {title} {Extending the lifetime of a quantum bit with error correction in superconducting circuits},\ }\href
  {https://doi.org/10.1038/nature18949} {\bibfield  {journal} {\bibinfo  {journal} {Nature}\ }\textbf {\bibinfo {volume} {536}},\ \bibinfo {pages} {441} (\bibinfo {year} {2016})}\BibitemShut {NoStop}%
\bibitem [{\citenamefont {{Google Quantum AI}}(2023)}]{google_quantum_ai_suppressing_2023}%
  \BibitemOpen
  \bibfield  {author} {\bibinfo {author} {\bibnamefont {{Google Quantum AI}}},\ }\bibfield  {title} {\bibinfo {title} {Suppressing quantum errors by scaling a surface code logical qubit},\ }\href {https://doi.org/10.1038/s41586-022-05434-1} {\bibfield  {journal} {\bibinfo  {journal} {Nature}\ }\textbf {\bibinfo {volume} {614}},\ \bibinfo {pages} {676} (\bibinfo {year} {2023})}\BibitemShut {NoStop}%
\bibitem [{\citenamefont {Sivak}\ \emph {et~al.}(2023)\citenamefont {Sivak}, \citenamefont {Eickbusch}, \citenamefont {Royer}, \citenamefont {Singh}, \citenamefont {Tsioutsios}, \citenamefont {Ganjam}, \citenamefont {Miano}, \citenamefont {Brock}, \citenamefont {Ding}, \citenamefont {Frunzio}, \citenamefont {Girvin}, \citenamefont {Schoelkopf},\ and\ \citenamefont {Devoret}}]{sivak_real-time_2023}%
  \BibitemOpen
  \bibfield  {author} {\bibinfo {author} {\bibfnamefont {V.~V.}\ \bibnamefont {Sivak}}, \bibinfo {author} {\bibfnamefont {A.}~\bibnamefont {Eickbusch}}, \bibinfo {author} {\bibfnamefont {B.}~\bibnamefont {Royer}}, \bibinfo {author} {\bibfnamefont {S.}~\bibnamefont {Singh}}, \bibinfo {author} {\bibfnamefont {I.}~\bibnamefont {Tsioutsios}}, \bibinfo {author} {\bibfnamefont {S.}~\bibnamefont {Ganjam}}, \bibinfo {author} {\bibfnamefont {A.}~\bibnamefont {Miano}}, \bibinfo {author} {\bibfnamefont {B.~L.}\ \bibnamefont {Brock}}, \bibinfo {author} {\bibfnamefont {A.~Z.}\ \bibnamefont {Ding}}, \bibinfo {author} {\bibfnamefont {L.}~\bibnamefont {Frunzio}}, \bibinfo {author} {\bibfnamefont {S.~M.}\ \bibnamefont {Girvin}}, \bibinfo {author} {\bibfnamefont {R.~J.}\ \bibnamefont {Schoelkopf}},\ and\ \bibinfo {author} {\bibfnamefont {M.~H.}\ \bibnamefont {Devoret}},\ }\bibfield  {title} {\bibinfo {title} {Real-time quantum error correction beyond break-even},\ }\href {https://doi.org/10.1038/s41586-023-05782-6} {\bibfield
  {journal} {\bibinfo  {journal} {Nature}\ }\textbf {\bibinfo {volume} {616}},\ \bibinfo {pages} {50} (\bibinfo {year} {2023})}\BibitemShut {NoStop}%
\bibitem [{\citenamefont {{Google Quantum AI}}(2024)}]{google_quantum_ai_quantum_2024}%
  \BibitemOpen
  \bibfield  {author} {\bibinfo {author} {\bibnamefont {{Google Quantum AI}}},\ }\href {http://arxiv.org/abs/2408.13687} {\bibinfo {title} {Quantum error correction below the surface code threshold}} (\bibinfo {year} {2024}),\ \bibinfo {note} {arXiv:2408.13687 [quant-ph]}\BibitemShut {NoStop}%
\bibitem [{\citenamefont {Clerk}\ \emph {et~al.}(2010)\citenamefont {Clerk}, \citenamefont {Devoret}, \citenamefont {Girvin}, \citenamefont {Marquardt},\ and\ \citenamefont {Schoelkopf}}]{clerk_introduction_2010}%
  \BibitemOpen
  \bibfield  {author} {\bibinfo {author} {\bibfnamefont {A.~A.}\ \bibnamefont {Clerk}}, \bibinfo {author} {\bibfnamefont {M.~H.}\ \bibnamefont {Devoret}}, \bibinfo {author} {\bibfnamefont {S.~M.}\ \bibnamefont {Girvin}}, \bibinfo {author} {\bibfnamefont {F.}~\bibnamefont {Marquardt}},\ and\ \bibinfo {author} {\bibfnamefont {R.~J.}\ \bibnamefont {Schoelkopf}},\ }\bibfield  {title} {\bibinfo {title} {Introduction to quantum noise, measurement, and amplification},\ }\href {https://doi.org/10.1103/RevModPhys.82.1155} {\bibfield  {journal} {\bibinfo  {journal} {Reviews of Modern Physics}\ }\textbf {\bibinfo {volume} {82}},\ \bibinfo {pages} {1155} (\bibinfo {year} {2010})}\BibitemShut {NoStop}%
\bibitem [{\citenamefont {Blais}\ \emph {et~al.}(2004)\citenamefont {Blais}, \citenamefont {Huang}, \citenamefont {Wallraff}, \citenamefont {Girvin},\ and\ \citenamefont {Schoelkopf}}]{blais_cavity_2004}%
  \BibitemOpen
  \bibfield  {author} {\bibinfo {author} {\bibfnamefont {A.}~\bibnamefont {Blais}}, \bibinfo {author} {\bibfnamefont {R.-S.}\ \bibnamefont {Huang}}, \bibinfo {author} {\bibfnamefont {A.}~\bibnamefont {Wallraff}}, \bibinfo {author} {\bibfnamefont {S.~M.}\ \bibnamefont {Girvin}},\ and\ \bibinfo {author} {\bibfnamefont {R.~J.}\ \bibnamefont {Schoelkopf}},\ }\bibfield  {title} {\bibinfo {title} {Cavity quantum electrodynamics for superconducting electrical circuits: {An} architecture for quantum computation},\ }\href {https://doi.org/10.1103/PhysRevA.69.062320} {\bibfield  {journal} {\bibinfo  {journal} {Physical Review A}\ }\textbf {\bibinfo {volume} {69}},\ \bibinfo {pages} {062320} (\bibinfo {year} {2004})}\BibitemShut {NoStop}%
\bibitem [{\citenamefont {Wallraff}\ \emph {et~al.}(2004)\citenamefont {Wallraff}, \citenamefont {Schuster}, \citenamefont {Blais}, \citenamefont {Frunzio}, \citenamefont {Huang}, \citenamefont {Majer}, \citenamefont {Kumar}, \citenamefont {Girvin},\ and\ \citenamefont {Schoelkopf}}]{wallraff_strong_2004}%
  \BibitemOpen
  \bibfield  {author} {\bibinfo {author} {\bibfnamefont {A.}~\bibnamefont {Wallraff}}, \bibinfo {author} {\bibfnamefont {D.~I.}\ \bibnamefont {Schuster}}, \bibinfo {author} {\bibfnamefont {A.}~\bibnamefont {Blais}}, \bibinfo {author} {\bibfnamefont {L.}~\bibnamefont {Frunzio}}, \bibinfo {author} {\bibfnamefont {R.-S.}\ \bibnamefont {Huang}}, \bibinfo {author} {\bibfnamefont {J.}~\bibnamefont {Majer}}, \bibinfo {author} {\bibfnamefont {S.}~\bibnamefont {Kumar}}, \bibinfo {author} {\bibfnamefont {S.~M.}\ \bibnamefont {Girvin}},\ and\ \bibinfo {author} {\bibfnamefont {R.~J.}\ \bibnamefont {Schoelkopf}},\ }\bibfield  {title} {\bibinfo {title} {Strong coupling of a single photon to a superconducting qubit using circuit quantum electrodynamics},\ }\href {https://doi.org/10.1038/nature02851} {\bibfield  {journal} {\bibinfo  {journal} {Nature}\ }\textbf {\bibinfo {volume} {431}},\ \bibinfo {pages} {162} (\bibinfo {year} {2004})}\BibitemShut {NoStop}%
\bibitem [{\citenamefont {Mallet}\ \emph {et~al.}(2009)\citenamefont {Mallet}, \citenamefont {Ong}, \citenamefont {Palacios-Laloy}, \citenamefont {Nguyen}, \citenamefont {Bertet}, \citenamefont {Vion},\ and\ \citenamefont {Esteve}}]{mallet_single-shot_2009}%
  \BibitemOpen
  \bibfield  {author} {\bibinfo {author} {\bibfnamefont {F.}~\bibnamefont {Mallet}}, \bibinfo {author} {\bibfnamefont {F.~R.}\ \bibnamefont {Ong}}, \bibinfo {author} {\bibfnamefont {A.}~\bibnamefont {Palacios-Laloy}}, \bibinfo {author} {\bibfnamefont {F.}~\bibnamefont {Nguyen}}, \bibinfo {author} {\bibfnamefont {P.}~\bibnamefont {Bertet}}, \bibinfo {author} {\bibfnamefont {D.}~\bibnamefont {Vion}},\ and\ \bibinfo {author} {\bibfnamefont {D.}~\bibnamefont {Esteve}},\ }\bibfield  {title} {\bibinfo {title} {Single-shot qubit readout in circuit quantum electrodynamics},\ }\href {https://doi.org/10.1038/nphys1400} {\bibfield  {journal} {\bibinfo  {journal} {Nature Physics}\ }\textbf {\bibinfo {volume} {5}},\ \bibinfo {pages} {791} (\bibinfo {year} {2009})}\BibitemShut {NoStop}%
\bibitem [{\citenamefont {Reed}\ \emph {et~al.}(2010)\citenamefont {Reed}, \citenamefont {DiCarlo}, \citenamefont {Johnson}, \citenamefont {Sun}, \citenamefont {Schuster}, \citenamefont {Frunzio},\ and\ \citenamefont {Schoelkopf}}]{reed_high-fidelity_2010}%
  \BibitemOpen
  \bibfield  {author} {\bibinfo {author} {\bibfnamefont {M.~D.}\ \bibnamefont {Reed}}, \bibinfo {author} {\bibfnamefont {L.}~\bibnamefont {DiCarlo}}, \bibinfo {author} {\bibfnamefont {B.~R.}\ \bibnamefont {Johnson}}, \bibinfo {author} {\bibfnamefont {L.}~\bibnamefont {Sun}}, \bibinfo {author} {\bibfnamefont {D.~I.}\ \bibnamefont {Schuster}}, \bibinfo {author} {\bibfnamefont {L.}~\bibnamefont {Frunzio}},\ and\ \bibinfo {author} {\bibfnamefont {R.~J.}\ \bibnamefont {Schoelkopf}},\ }\bibfield  {title} {\bibinfo {title} {High-{Fidelity} {Readout} in {Circuit} {Quantum} {Electrodynamics} {Using} the {Jaynes}-{Cummings} {Nonlinearity}},\ }\href {https://doi.org/10.1103/PhysRevLett.105.173601} {\bibfield  {journal} {\bibinfo  {journal} {Physical Review Letters}\ }\textbf {\bibinfo {volume} {105}},\ \bibinfo {pages} {173601} (\bibinfo {year} {2010})}\BibitemShut {NoStop}%
\bibitem [{\citenamefont {Jeffrey}\ \emph {et~al.}(2014)\citenamefont {Jeffrey}, \citenamefont {Sank}, \citenamefont {Mutus}, \citenamefont {White}, \citenamefont {Kelly}, \citenamefont {Barends}, \citenamefont {Chen}, \citenamefont {Chen}, \citenamefont {Chiaro}, \citenamefont {Dunsworth}, \citenamefont {Megrant}, \citenamefont {O’Malley}, \citenamefont {Neill}, \citenamefont {Roushan}, \citenamefont {Vainsencher}, \citenamefont {Wenner}, \citenamefont {Cleland},\ and\ \citenamefont {Martinis}}]{jeffrey_fast_2014}%
  \BibitemOpen
  \bibfield  {author} {\bibinfo {author} {\bibfnamefont {E.}~\bibnamefont {Jeffrey}}, \bibinfo {author} {\bibfnamefont {D.}~\bibnamefont {Sank}}, \bibinfo {author} {\bibfnamefont {J.}~\bibnamefont {Mutus}}, \bibinfo {author} {\bibfnamefont {T.}~\bibnamefont {White}}, \bibinfo {author} {\bibfnamefont {J.}~\bibnamefont {Kelly}}, \bibinfo {author} {\bibfnamefont {R.}~\bibnamefont {Barends}}, \bibinfo {author} {\bibfnamefont {Y.}~\bibnamefont {Chen}}, \bibinfo {author} {\bibfnamefont {Z.}~\bibnamefont {Chen}}, \bibinfo {author} {\bibfnamefont {B.}~\bibnamefont {Chiaro}}, \bibinfo {author} {\bibfnamefont {A.}~\bibnamefont {Dunsworth}}, \bibinfo {author} {\bibfnamefont {A.}~\bibnamefont {Megrant}}, \bibinfo {author} {\bibfnamefont {P.}~\bibnamefont {O’Malley}}, \bibinfo {author} {\bibfnamefont {C.}~\bibnamefont {Neill}}, \bibinfo {author} {\bibfnamefont {P.}~\bibnamefont {Roushan}}, \bibinfo {author} {\bibfnamefont {A.}~\bibnamefont {Vainsencher}}, \bibinfo {author} {\bibfnamefont {J.}~\bibnamefont {Wenner}},
  \bibinfo {author} {\bibfnamefont {A.}~\bibnamefont {Cleland}},\ and\ \bibinfo {author} {\bibfnamefont {J.~M.}\ \bibnamefont {Martinis}},\ }\bibfield  {title} {\bibinfo {title} {Fast {Accurate} {State} {Measurement} with {Superconducting} {Qubits}},\ }\href {https://doi.org/10.1103/PhysRevLett.112.190504} {\bibfield  {journal} {\bibinfo  {journal} {Physical Review Letters}\ }\textbf {\bibinfo {volume} {112}},\ \bibinfo {pages} {190504} (\bibinfo {year} {2014})}\BibitemShut {NoStop}%
\bibitem [{\citenamefont {Walter}\ \emph {et~al.}(2017)\citenamefont {Walter}, \citenamefont {Kurpiers}, \citenamefont {Gasparinetti}, \citenamefont {Magnard}, \citenamefont {Potočnik}, \citenamefont {Salathé}, \citenamefont {Pechal}, \citenamefont {Mondal}, \citenamefont {Oppliger}, \citenamefont {Eichler},\ and\ \citenamefont {Wallraff}}]{walter_rapid_2017}%
  \BibitemOpen
  \bibfield  {author} {\bibinfo {author} {\bibfnamefont {T.}~\bibnamefont {Walter}}, \bibinfo {author} {\bibfnamefont {P.}~\bibnamefont {Kurpiers}}, \bibinfo {author} {\bibfnamefont {S.}~\bibnamefont {Gasparinetti}}, \bibinfo {author} {\bibfnamefont {P.}~\bibnamefont {Magnard}}, \bibinfo {author} {\bibfnamefont {A.}~\bibnamefont {Potočnik}}, \bibinfo {author} {\bibfnamefont {Y.}~\bibnamefont {Salathé}}, \bibinfo {author} {\bibfnamefont {M.}~\bibnamefont {Pechal}}, \bibinfo {author} {\bibfnamefont {M.}~\bibnamefont {Mondal}}, \bibinfo {author} {\bibfnamefont {M.}~\bibnamefont {Oppliger}}, \bibinfo {author} {\bibfnamefont {C.}~\bibnamefont {Eichler}},\ and\ \bibinfo {author} {\bibfnamefont {A.}~\bibnamefont {Wallraff}},\ }\bibfield  {title} {\bibinfo {title} {Rapid {High}-{Fidelity} {Single}-{Shot} {Dispersive} {Readout} of {Superconducting} {Qubits}},\ }\href {https://doi.org/10.1103/PhysRevApplied.7.054020} {\bibfield  {journal} {\bibinfo  {journal} {Physical Review Applied}\ }\textbf {\bibinfo {volume}
  {7}},\ \bibinfo {pages} {054020} (\bibinfo {year} {2017})}\BibitemShut {NoStop}%
\bibitem [{\citenamefont {Dassonneville}\ \emph {et~al.}(2020)\citenamefont {Dassonneville}, \citenamefont {Ramos}, \citenamefont {Milchakov}, \citenamefont {Planat}, \citenamefont {Dumur}, \citenamefont {Foroughi}, \citenamefont {Puertas}, \citenamefont {Leger}, \citenamefont {Bharadwaj}, \citenamefont {Delaforce}, \citenamefont {Naud}, \citenamefont {Hasch-Guichard}, \citenamefont {Garcia-Ripoll}, \citenamefont {Roch},\ and\ \citenamefont {Buisson}}]{dassonneville_fast_2020}%
  \BibitemOpen
  \bibfield  {author} {\bibinfo {author} {\bibfnamefont {R.}~\bibnamefont {Dassonneville}}, \bibinfo {author} {\bibfnamefont {T.}~\bibnamefont {Ramos}}, \bibinfo {author} {\bibfnamefont {V.}~\bibnamefont {Milchakov}}, \bibinfo {author} {\bibfnamefont {L.}~\bibnamefont {Planat}}, \bibinfo {author} {\bibfnamefont {E.}~\bibnamefont {Dumur}}, \bibinfo {author} {\bibfnamefont {F.}~\bibnamefont {Foroughi}}, \bibinfo {author} {\bibfnamefont {J.}~\bibnamefont {Puertas}}, \bibinfo {author} {\bibfnamefont {S.}~\bibnamefont {Leger}}, \bibinfo {author} {\bibfnamefont {K.}~\bibnamefont {Bharadwaj}}, \bibinfo {author} {\bibfnamefont {J.}~\bibnamefont {Delaforce}}, \bibinfo {author} {\bibfnamefont {C.}~\bibnamefont {Naud}}, \bibinfo {author} {\bibfnamefont {W.}~\bibnamefont {Hasch-Guichard}}, \bibinfo {author} {\bibfnamefont {J.}~\bibnamefont {Garcia-Ripoll}}, \bibinfo {author} {\bibfnamefont {N.}~\bibnamefont {Roch}},\ and\ \bibinfo {author} {\bibfnamefont {O.}~\bibnamefont {Buisson}},\ }\bibfield  {title} {\bibinfo
  {title} {Fast {High}-{Fidelity} {Quantum} {Nondemolition} {Qubit} {Readout} via a {Nonperturbative} {Cross}-{Kerr} {Coupling}},\ }\href {https://doi.org/10.1103/PhysRevX.10.011045} {\bibfield  {journal} {\bibinfo  {journal} {Physical Review X}\ }\textbf {\bibinfo {volume} {10}},\ \bibinfo {pages} {011045} (\bibinfo {year} {2020})}\BibitemShut {NoStop}%
\bibitem [{\citenamefont {Swiadek}\ \emph {et~al.}(2024)\citenamefont {Swiadek}, \citenamefont {Shillito}, \citenamefont {Magnard}, \citenamefont {Remm}, \citenamefont {Hellings}, \citenamefont {Lacroix}, \citenamefont {Ficheux}, \citenamefont {Zanuz}, \citenamefont {Norris}, \citenamefont {Blais}, \citenamefont {Krinner},\ and\ \citenamefont {Wallraff}}]{swiadek_enhancing_2024}%
  \BibitemOpen
  \bibfield  {author} {\bibinfo {author} {\bibfnamefont {F.}~\bibnamefont {Swiadek}}, \bibinfo {author} {\bibfnamefont {R.}~\bibnamefont {Shillito}}, \bibinfo {author} {\bibfnamefont {P.}~\bibnamefont {Magnard}}, \bibinfo {author} {\bibfnamefont {A.}~\bibnamefont {Remm}}, \bibinfo {author} {\bibfnamefont {C.}~\bibnamefont {Hellings}}, \bibinfo {author} {\bibfnamefont {N.}~\bibnamefont {Lacroix}}, \bibinfo {author} {\bibfnamefont {Q.}~\bibnamefont {Ficheux}}, \bibinfo {author} {\bibfnamefont {D.~C.}\ \bibnamefont {Zanuz}}, \bibinfo {author} {\bibfnamefont {G.~J.}\ \bibnamefont {Norris}}, \bibinfo {author} {\bibfnamefont {A.}~\bibnamefont {Blais}}, \bibinfo {author} {\bibfnamefont {S.}~\bibnamefont {Krinner}},\ and\ \bibinfo {author} {\bibfnamefont {A.}~\bibnamefont {Wallraff}},\ }\bibfield  {title} {\bibinfo {title} {Enhancing {Dispersive} {Readout} of {Superconducting} {Qubits} through {Dynamic} {Control} of the {Dispersive} {Shift}: {Experiment} and {Theory}},\ }\href
  {https://doi.org/10.1103/PRXQuantum.5.040326} {\bibfield  {journal} {\bibinfo  {journal} {PRX Quantum}\ }\textbf {\bibinfo {volume} {5}},\ \bibinfo {pages} {040326} (\bibinfo {year} {2024})}\BibitemShut {NoStop}%
\bibitem [{\citenamefont {Spring}\ \emph {et~al.}(2024)\citenamefont {Spring}, \citenamefont {Milanovic}, \citenamefont {Sunada}, \citenamefont {Wang}, \citenamefont {van Loo}, \citenamefont {Tamate},\ and\ \citenamefont {Nakamura}}]{spring_fast_2024}%
  \BibitemOpen
  \bibfield  {author} {\bibinfo {author} {\bibfnamefont {P.~A.}\ \bibnamefont {Spring}}, \bibinfo {author} {\bibfnamefont {L.}~\bibnamefont {Milanovic}}, \bibinfo {author} {\bibfnamefont {Y.}~\bibnamefont {Sunada}}, \bibinfo {author} {\bibfnamefont {S.}~\bibnamefont {Wang}}, \bibinfo {author} {\bibfnamefont {A.~F.}\ \bibnamefont {van Loo}}, \bibinfo {author} {\bibfnamefont {S.}~\bibnamefont {Tamate}},\ and\ \bibinfo {author} {\bibfnamefont {Y.}~\bibnamefont {Nakamura}},\ }\href {http://arxiv.org/abs/2409.04967} {\bibinfo {title} {Fast multiplexed superconducting qubit readout with intrinsic {Purcell} filtering}} (\bibinfo {year} {2024}),\ \bibinfo {note} {arXiv:2409.04967 [quant-ph]}\BibitemShut {NoStop}%
\bibitem [{\citenamefont {Gambetta}\ \emph {et~al.}(2007)\citenamefont {Gambetta}, \citenamefont {Braff}, \citenamefont {Wallraff}, \citenamefont {Girvin},\ and\ \citenamefont {Schoelkopf}}]{gambetta_protocols_2007}%
  \BibitemOpen
  \bibfield  {author} {\bibinfo {author} {\bibfnamefont {J.}~\bibnamefont {Gambetta}}, \bibinfo {author} {\bibfnamefont {W.~A.}\ \bibnamefont {Braff}}, \bibinfo {author} {\bibfnamefont {A.}~\bibnamefont {Wallraff}}, \bibinfo {author} {\bibfnamefont {S.~M.}\ \bibnamefont {Girvin}},\ and\ \bibinfo {author} {\bibfnamefont {R.~J.}\ \bibnamefont {Schoelkopf}},\ }\bibfield  {title} {\bibinfo {title} {Protocols for optimal readout of qubits using a continuous quantum nondemolition measurement},\ }\href {https://doi.org/10.1103/PhysRevA.76.012325} {\bibfield  {journal} {\bibinfo  {journal} {Physical Review A}\ }\textbf {\bibinfo {volume} {76}},\ \bibinfo {pages} {012325} (\bibinfo {year} {2007})}\BibitemShut {NoStop}%
\bibitem [{\citenamefont {Zhang}\ \emph {et~al.}(2019)\citenamefont {Zhang}, \citenamefont {Lester}, \citenamefont {Gao}, \citenamefont {Jiang}, \citenamefont {Schoelkopf},\ and\ \citenamefont {Girvin}}]{zhang_engineering_2019}%
  \BibitemOpen
  \bibfield  {author} {\bibinfo {author} {\bibfnamefont {Y.}~\bibnamefont {Zhang}}, \bibinfo {author} {\bibfnamefont {B.~J.}\ \bibnamefont {Lester}}, \bibinfo {author} {\bibfnamefont {Y.~Y.}\ \bibnamefont {Gao}}, \bibinfo {author} {\bibfnamefont {L.}~\bibnamefont {Jiang}}, \bibinfo {author} {\bibfnamefont {R.~J.}\ \bibnamefont {Schoelkopf}},\ and\ \bibinfo {author} {\bibfnamefont {S.~M.}\ \bibnamefont {Girvin}},\ }\bibfield  {title} {\bibinfo {title} {Engineering bilinear mode coupling in circuit {QED}: {Theory} and experiment},\ }\href {https://doi.org/10.1103/PhysRevA.99.012314} {\bibfield  {journal} {\bibinfo  {journal} {Physical Review A}\ }\textbf {\bibinfo {volume} {99}},\ \bibinfo {pages} {012314} (\bibinfo {year} {2019})}\BibitemShut {NoStop}%
\bibitem [{\citenamefont {Petrescu}\ \emph {et~al.}(2020)\citenamefont {Petrescu}, \citenamefont {Malekakhlagh},\ and\ \citenamefont {Türeci}}]{petrescu_lifetime_2020}%
  \BibitemOpen
  \bibfield  {author} {\bibinfo {author} {\bibfnamefont {A.}~\bibnamefont {Petrescu}}, \bibinfo {author} {\bibfnamefont {M.}~\bibnamefont {Malekakhlagh}},\ and\ \bibinfo {author} {\bibfnamefont {H.~E.}\ \bibnamefont {Türeci}},\ }\bibfield  {title} {\bibinfo {title} {Lifetime renormalization of driven weakly anharmonic superconducting qubits. {II}. {The} readout problem},\ }\href {https://doi.org/10.1103/PhysRevB.101.134510} {\bibfield  {journal} {\bibinfo  {journal} {Physical Review B}\ }\textbf {\bibinfo {volume} {101}},\ \bibinfo {pages} {134510} (\bibinfo {year} {2020})}\BibitemShut {NoStop}%
\bibitem [{\citenamefont {Hanai}\ \emph {et~al.}(2021)\citenamefont {Hanai}, \citenamefont {McDonald},\ and\ \citenamefont {Clerk}}]{hanai_intrinsic_2021}%
  \BibitemOpen
  \bibfield  {author} {\bibinfo {author} {\bibfnamefont {R.}~\bibnamefont {Hanai}}, \bibinfo {author} {\bibfnamefont {A.}~\bibnamefont {McDonald}},\ and\ \bibinfo {author} {\bibfnamefont {A.}~\bibnamefont {Clerk}},\ }\bibfield  {title} {\bibinfo {title} {Intrinsic mechanisms for drive-dependent {Purcell} decay in superconducting quantum circuits},\ }\href {https://doi.org/10.1103/PhysRevResearch.3.043228} {\bibfield  {journal} {\bibinfo  {journal} {Physical Review Research}\ }\textbf {\bibinfo {volume} {3}},\ \bibinfo {pages} {043228} (\bibinfo {year} {2021})}\BibitemShut {NoStop}%
\bibitem [{\citenamefont {Bista}\ \emph {et~al.}(2025)\citenamefont {Bista}, \citenamefont {Thibodeau}, \citenamefont {Nie}, \citenamefont {Chow}, \citenamefont {Clark},\ and\ \citenamefont {Kou}}]{bista_readout-induced_2025}%
  \BibitemOpen
  \bibfield  {author} {\bibinfo {author} {\bibfnamefont {A.}~\bibnamefont {Bista}}, \bibinfo {author} {\bibfnamefont {M.}~\bibnamefont {Thibodeau}}, \bibinfo {author} {\bibfnamefont {K.}~\bibnamefont {Nie}}, \bibinfo {author} {\bibfnamefont {K.}~\bibnamefont {Chow}}, \bibinfo {author} {\bibfnamefont {B.~K.}\ \bibnamefont {Clark}},\ and\ \bibinfo {author} {\bibfnamefont {A.}~\bibnamefont {Kou}},\ }\href {https://doi.org/10.48550/arXiv.2501.17807} {\bibinfo {title} {Readout-induced leakage of the fluxonium qubit}} (\bibinfo {year} {2025}),\ \bibinfo {note} {arXiv:2501.17807 [quant-ph]}\BibitemShut {NoStop}%
\bibitem [{\citenamefont {Lescanne}\ \emph {et~al.}(2019)\citenamefont {Lescanne}, \citenamefont {Verney}, \citenamefont {Ficheux}, \citenamefont {Devoret}, \citenamefont {Huard}, \citenamefont {Mirrahimi},\ and\ \citenamefont {Leghtas}}]{lescanne_escape_2019}%
  \BibitemOpen
  \bibfield  {author} {\bibinfo {author} {\bibfnamefont {R.}~\bibnamefont {Lescanne}}, \bibinfo {author} {\bibfnamefont {L.}~\bibnamefont {Verney}}, \bibinfo {author} {\bibfnamefont {Q.}~\bibnamefont {Ficheux}}, \bibinfo {author} {\bibfnamefont {M.~H.}\ \bibnamefont {Devoret}}, \bibinfo {author} {\bibfnamefont {B.}~\bibnamefont {Huard}}, \bibinfo {author} {\bibfnamefont {M.}~\bibnamefont {Mirrahimi}},\ and\ \bibinfo {author} {\bibfnamefont {Z.}~\bibnamefont {Leghtas}},\ }\bibfield  {title} {\bibinfo {title} {Escape of a {Driven} {Quantum} {Josephson} {Circuit} into {Unconfined} {States}},\ }\href {https://doi.org/10.1103/PhysRevApplied.11.014030} {\bibfield  {journal} {\bibinfo  {journal} {Physical Review Applied}\ }\textbf {\bibinfo {volume} {11}},\ \bibinfo {pages} {014030} (\bibinfo {year} {2019})},\ \bibinfo {note} {publisher: American Physical Society}\BibitemShut {NoStop}%
\bibitem [{\citenamefont {Hazra}\ \emph {et~al.}(2024)\citenamefont {Hazra}, \citenamefont {Dai}, \citenamefont {Connolly}, \citenamefont {Kurilovich}, \citenamefont {Wang}, \citenamefont {Frunzio},\ and\ \citenamefont {Devoret}}]{hazra_benchmarking_2024}%
  \BibitemOpen
  \bibfield  {author} {\bibinfo {author} {\bibfnamefont {S.}~\bibnamefont {Hazra}}, \bibinfo {author} {\bibfnamefont {W.}~\bibnamefont {Dai}}, \bibinfo {author} {\bibfnamefont {T.}~\bibnamefont {Connolly}}, \bibinfo {author} {\bibfnamefont {P.~D.}\ \bibnamefont {Kurilovich}}, \bibinfo {author} {\bibfnamefont {Z.}~\bibnamefont {Wang}}, \bibinfo {author} {\bibfnamefont {L.}~\bibnamefont {Frunzio}},\ and\ \bibinfo {author} {\bibfnamefont {M.~H.}\ \bibnamefont {Devoret}},\ }\href {https://doi.org/10.48550/arXiv.2407.10934} {\bibinfo {title} {Benchmarking the readout of a superconducting qubit for repeated measurements}} (\bibinfo {year} {2024}),\ \bibinfo {note} {arXiv:2407.10934}\BibitemShut {NoStop}%
\bibitem [{\citenamefont {Aliferis}\ and\ \citenamefont {Terhal}(2007)}]{aliferis_fault-tolerant_2007}%
  \BibitemOpen
  \bibfield  {author} {\bibinfo {author} {\bibfnamefont {P.}~\bibnamefont {Aliferis}}\ and\ \bibinfo {author} {\bibfnamefont {B.~M.}\ \bibnamefont {Terhal}},\ }\bibfield  {title} {\bibinfo {title} {Fault-tolerant quantum computation for local leakage faults},\ }\href@noop {} {\bibfield  {journal} {\bibinfo  {journal} {Quantum Info. Comput.}\ }\textbf {\bibinfo {volume} {7}},\ \bibinfo {pages} {139} (\bibinfo {year} {2007})}\BibitemShut {NoStop}%
\bibitem [{\citenamefont {Fowler}(2013)}]{fowler_coping_2013}%
  \BibitemOpen
  \bibfield  {author} {\bibinfo {author} {\bibfnamefont {A.~G.}\ \bibnamefont {Fowler}},\ }\bibfield  {title} {\bibinfo {title} {Coping with qubit leakage in topological codes},\ }\href {https://doi.org/10.1103/PhysRevA.88.042308} {\bibfield  {journal} {\bibinfo  {journal} {Physical Review A}\ }\textbf {\bibinfo {volume} {88}},\ \bibinfo {pages} {042308} (\bibinfo {year} {2013})}\BibitemShut {NoStop}%
\bibitem [{\citenamefont {Ghosh}\ \emph {et~al.}(2013)\citenamefont {Ghosh}, \citenamefont {Fowler}, \citenamefont {Martinis},\ and\ \citenamefont {Geller}}]{ghosh_understanding_2013}%
  \BibitemOpen
  \bibfield  {author} {\bibinfo {author} {\bibfnamefont {J.}~\bibnamefont {Ghosh}}, \bibinfo {author} {\bibfnamefont {A.~G.}\ \bibnamefont {Fowler}}, \bibinfo {author} {\bibfnamefont {J.~M.}\ \bibnamefont {Martinis}},\ and\ \bibinfo {author} {\bibfnamefont {M.~R.}\ \bibnamefont {Geller}},\ }\bibfield  {title} {\bibinfo {title} {Understanding the effects of leakage in superconducting quantum-error-detection circuits},\ }\href {https://doi.org/10.1103/PhysRevA.88.062329} {\bibfield  {journal} {\bibinfo  {journal} {Physical Review A}\ }\textbf {\bibinfo {volume} {88}},\ \bibinfo {pages} {062329} (\bibinfo {year} {2013})}\BibitemShut {NoStop}%
\bibitem [{\citenamefont {Suchara}\ \emph {et~al.}(2015)\citenamefont {Suchara}, \citenamefont {Cross},\ and\ \citenamefont {Gambetta}}]{suchara_leakage_2015}%
  \BibitemOpen
  \bibfield  {author} {\bibinfo {author} {\bibfnamefont {M.}~\bibnamefont {Suchara}}, \bibinfo {author} {\bibfnamefont {A.~W.}\ \bibnamefont {Cross}},\ and\ \bibinfo {author} {\bibfnamefont {J.~M.}\ \bibnamefont {Gambetta}},\ }\bibfield  {title} {\bibinfo {title} {Leakage suppression in the {Toric} code},\ }\href@noop {} {\bibfield  {journal} {\bibinfo  {journal} {Quantum Info. Comput.}\ }\textbf {\bibinfo {volume} {15}},\ \bibinfo {pages} {997} (\bibinfo {year} {2015})}\BibitemShut {NoStop}%
\bibitem [{\citenamefont {Magnard}\ \emph {et~al.}(2018)\citenamefont {Magnard}, \citenamefont {Kurpiers}, \citenamefont {Royer}, \citenamefont {Walter}, \citenamefont {Besse}, \citenamefont {Gasparinetti}, \citenamefont {Pechal}, \citenamefont {Heinsoo}, \citenamefont {Storz}, \citenamefont {Blais},\ and\ \citenamefont {Wallraff}}]{magnard_fast_2018}%
  \BibitemOpen
  \bibfield  {author} {\bibinfo {author} {\bibfnamefont {P.}~\bibnamefont {Magnard}}, \bibinfo {author} {\bibfnamefont {P.}~\bibnamefont {Kurpiers}}, \bibinfo {author} {\bibfnamefont {B.}~\bibnamefont {Royer}}, \bibinfo {author} {\bibfnamefont {T.}~\bibnamefont {Walter}}, \bibinfo {author} {\bibfnamefont {J.-C.}\ \bibnamefont {Besse}}, \bibinfo {author} {\bibfnamefont {S.}~\bibnamefont {Gasparinetti}}, \bibinfo {author} {\bibfnamefont {M.}~\bibnamefont {Pechal}}, \bibinfo {author} {\bibfnamefont {J.}~\bibnamefont {Heinsoo}}, \bibinfo {author} {\bibfnamefont {S.}~\bibnamefont {Storz}}, \bibinfo {author} {\bibfnamefont {A.}~\bibnamefont {Blais}},\ and\ \bibinfo {author} {\bibfnamefont {A.}~\bibnamefont {Wallraff}},\ }\bibfield  {title} {\bibinfo {title} {Fast and {Unconditional} {All}-{Microwave} {Reset} of a {Superconducting} {Qubit}},\ }\href {https://doi.org/10.1103/PhysRevLett.121.060502} {\bibfield  {journal} {\bibinfo  {journal} {Physical Review Letters}\ }\textbf {\bibinfo {volume} {121}},\ \bibinfo
  {pages} {060502} (\bibinfo {year} {2018})}\BibitemShut {NoStop}%
\bibitem [{\citenamefont {Bultink}\ \emph {et~al.}(2020)\citenamefont {Bultink}, \citenamefont {O’Brien}, \citenamefont {Vollmer}, \citenamefont {Muthusubramanian}, \citenamefont {Beekman}, \citenamefont {Rol}, \citenamefont {Fu}, \citenamefont {Tarasinski}, \citenamefont {Ostroukh}, \citenamefont {Varbanov}, \citenamefont {Bruno},\ and\ \citenamefont {DiCarlo}}]{bultink_protecting_2020}%
  \BibitemOpen
  \bibfield  {author} {\bibinfo {author} {\bibfnamefont {C.~C.}\ \bibnamefont {Bultink}}, \bibinfo {author} {\bibfnamefont {T.~E.}\ \bibnamefont {O’Brien}}, \bibinfo {author} {\bibfnamefont {R.}~\bibnamefont {Vollmer}}, \bibinfo {author} {\bibfnamefont {N.}~\bibnamefont {Muthusubramanian}}, \bibinfo {author} {\bibfnamefont {M.~W.}\ \bibnamefont {Beekman}}, \bibinfo {author} {\bibfnamefont {M.~A.}\ \bibnamefont {Rol}}, \bibinfo {author} {\bibfnamefont {X.}~\bibnamefont {Fu}}, \bibinfo {author} {\bibfnamefont {B.}~\bibnamefont {Tarasinski}}, \bibinfo {author} {\bibfnamefont {V.}~\bibnamefont {Ostroukh}}, \bibinfo {author} {\bibfnamefont {B.}~\bibnamefont {Varbanov}}, \bibinfo {author} {\bibfnamefont {A.}~\bibnamefont {Bruno}},\ and\ \bibinfo {author} {\bibfnamefont {L.}~\bibnamefont {DiCarlo}},\ }\bibfield  {title} {\bibinfo {title} {Protecting quantum entanglement from leakage and qubit errors via repetitive parity measurements},\ }\href {https://doi.org/10.1126/sciadv.aay3050} {\bibfield  {journal}
  {\bibinfo  {journal} {Science Advances}\ }\textbf {\bibinfo {volume} {6}},\ \bibinfo {pages} {eaay3050} (\bibinfo {year} {2020})}\BibitemShut {NoStop}%
\bibitem [{\citenamefont {Varbanov}\ \emph {et~al.}(2020)\citenamefont {Varbanov}, \citenamefont {Battistel}, \citenamefont {Tarasinski}, \citenamefont {Ostroukh}, \citenamefont {O’Brien}, \citenamefont {DiCarlo},\ and\ \citenamefont {Terhal}}]{varbanov_leakage_2020}%
  \BibitemOpen
  \bibfield  {author} {\bibinfo {author} {\bibfnamefont {B.~M.}\ \bibnamefont {Varbanov}}, \bibinfo {author} {\bibfnamefont {F.}~\bibnamefont {Battistel}}, \bibinfo {author} {\bibfnamefont {B.~M.}\ \bibnamefont {Tarasinski}}, \bibinfo {author} {\bibfnamefont {V.~P.}\ \bibnamefont {Ostroukh}}, \bibinfo {author} {\bibfnamefont {T.~E.}\ \bibnamefont {O’Brien}}, \bibinfo {author} {\bibfnamefont {L.}~\bibnamefont {DiCarlo}},\ and\ \bibinfo {author} {\bibfnamefont {B.~M.}\ \bibnamefont {Terhal}},\ }\bibfield  {title} {\bibinfo {title} {Leakage detection for a transmon-based surface code},\ }\href {https://doi.org/10.1038/s41534-020-00330-w} {\bibfield  {journal} {\bibinfo  {journal} {npj Quantum Information}\ }\textbf {\bibinfo {volume} {6}},\ \bibinfo {pages} {1} (\bibinfo {year} {2020})}\BibitemShut {NoStop}%
\bibitem [{\citenamefont {McEwen~et al.}(2021)}]{mcewen_et_al_removing_2021}%
  \BibitemOpen
  \bibfield  {author} {\bibinfo {author} {\bibfnamefont {M.}~\bibnamefont {McEwen~et al.}},\ }\bibfield  {title} {\bibinfo {title} {Removing leakage-induced correlated errors in superconducting quantum error correction},\ }\href {https://doi.org/10.1038/s41467-021-21982-y} {\bibfield  {journal} {\bibinfo  {journal} {Nature Communications}\ }\textbf {\bibinfo {volume} {12}},\ \bibinfo {pages} {1761} (\bibinfo {year} {2021})}\BibitemShut {NoStop}%
\bibitem [{\citenamefont {Miao}\ and\ \citenamefont {McEwen~et al.}(2023)}]{miao_overcoming_2023}%
  \BibitemOpen
  \bibfield  {author} {\bibinfo {author} {\bibfnamefont {K.~C.}\ \bibnamefont {Miao}}\ and\ \bibinfo {author} {\bibfnamefont {M.}~\bibnamefont {McEwen~et al.}},\ }\bibfield  {title} {\bibinfo {title} {Overcoming leakage in quantum error correction},\ }\href {https://doi.org/10.1038/s41567-023-02226-w} {\bibfield  {journal} {\bibinfo  {journal} {Nature Physics}\ }\textbf {\bibinfo {volume} {19}},\ \bibinfo {pages} {1780} (\bibinfo {year} {2023})}\BibitemShut {NoStop}%
\bibitem [{\citenamefont {Shillito}\ \emph {et~al.}(2022)\citenamefont {Shillito}, \citenamefont {Petrescu}, \citenamefont {Cohen}, \citenamefont {Beall}, \citenamefont {Hauru}, \citenamefont {Ganahl}, \citenamefont {Lewis}, \citenamefont {Vidal},\ and\ \citenamefont {Blais}}]{shillito_dynamics_2022}%
  \BibitemOpen
  \bibfield  {author} {\bibinfo {author} {\bibfnamefont {R.}~\bibnamefont {Shillito}}, \bibinfo {author} {\bibfnamefont {A.}~\bibnamefont {Petrescu}}, \bibinfo {author} {\bibfnamefont {J.}~\bibnamefont {Cohen}}, \bibinfo {author} {\bibfnamefont {J.}~\bibnamefont {Beall}}, \bibinfo {author} {\bibfnamefont {M.}~\bibnamefont {Hauru}}, \bibinfo {author} {\bibfnamefont {M.}~\bibnamefont {Ganahl}}, \bibinfo {author} {\bibfnamefont {A.~G.}\ \bibnamefont {Lewis}}, \bibinfo {author} {\bibfnamefont {G.}~\bibnamefont {Vidal}},\ and\ \bibinfo {author} {\bibfnamefont {A.}~\bibnamefont {Blais}},\ }\bibfield  {title} {\bibinfo {title} {Dynamics of {Transmon} {Ionization}},\ }\href {https://doi.org/10.1103/PhysRevApplied.18.034031} {\bibfield  {journal} {\bibinfo  {journal} {Physical Review Applied}\ }\textbf {\bibinfo {volume} {18}},\ \bibinfo {pages} {034031} (\bibinfo {year} {2022})}\BibitemShut {NoStop}%
\bibitem [{\citenamefont {Xiao}\ \emph {et~al.}(2023)\citenamefont {Xiao}, \citenamefont {Venkatraman}, \citenamefont {Cortiñas}, \citenamefont {Chowdhury},\ and\ \citenamefont {Devoret}}]{xiao_diagrammatic_2023}%
  \BibitemOpen
  \bibfield  {author} {\bibinfo {author} {\bibfnamefont {X.}~\bibnamefont {Xiao}}, \bibinfo {author} {\bibfnamefont {J.}~\bibnamefont {Venkatraman}}, \bibinfo {author} {\bibfnamefont {R.~G.}\ \bibnamefont {Cortiñas}}, \bibinfo {author} {\bibfnamefont {S.}~\bibnamefont {Chowdhury}},\ and\ \bibinfo {author} {\bibfnamefont {M.~H.}\ \bibnamefont {Devoret}},\ }\href {http://arxiv.org/abs/2304.13656} {\bibinfo {title} {A diagrammatic method to compute the effective {Hamiltonian} of driven nonlinear oscillators}} (\bibinfo {year} {2023}),\ \bibinfo {note} {arXiv:2304.13656}\BibitemShut {NoStop}%
\bibitem [{\citenamefont {Cohen}\ \emph {et~al.}(2023)\citenamefont {Cohen}, \citenamefont {Petrescu}, \citenamefont {Shillito},\ and\ \citenamefont {Blais}}]{cohen_reminiscence_2023}%
  \BibitemOpen
  \bibfield  {author} {\bibinfo {author} {\bibfnamefont {J.}~\bibnamefont {Cohen}}, \bibinfo {author} {\bibfnamefont {A.}~\bibnamefont {Petrescu}}, \bibinfo {author} {\bibfnamefont {R.}~\bibnamefont {Shillito}},\ and\ \bibinfo {author} {\bibfnamefont {A.}~\bibnamefont {Blais}},\ }\bibfield  {title} {\bibinfo {title} {Reminiscence of {Classical} {Chaos} in {Driven} {Transmons}},\ }\href {https://doi.org/10.1103/PRXQuantum.4.020312} {\bibfield  {journal} {\bibinfo  {journal} {PRX Quantum}\ }\textbf {\bibinfo {volume} {4}},\ \bibinfo {pages} {020312} (\bibinfo {year} {2023})}\BibitemShut {NoStop}%
\bibitem [{\citenamefont {Nesterov}\ and\ \citenamefont {Pechenezhskiy}(2024)}]{nesterov_measurement-induced_2024}%
  \BibitemOpen
  \bibfield  {author} {\bibinfo {author} {\bibfnamefont {K.~N.}\ \bibnamefont {Nesterov}}\ and\ \bibinfo {author} {\bibfnamefont {I.~V.}\ \bibnamefont {Pechenezhskiy}},\ }\bibfield  {title} {\bibinfo {title} {Measurement-induced state transitions in dispersive qubit-readout schemes},\ }\href {https://doi.org/10.1103/PhysRevApplied.22.064038} {\bibfield  {journal} {\bibinfo  {journal} {Physical Review Applied}\ }\textbf {\bibinfo {volume} {22}},\ \bibinfo {pages} {064038} (\bibinfo {year} {2024})}\BibitemShut {NoStop}%
\bibitem [{\citenamefont {Gao}\ \emph {et~al.}(2018)\citenamefont {Gao}, \citenamefont {Lester}, \citenamefont {Zhang}, \citenamefont {Wang}, \citenamefont {Rosenblum}, \citenamefont {Frunzio}, \citenamefont {Jiang}, \citenamefont {Girvin},\ and\ \citenamefont {Schoelkopf}}]{gao_programmable_2018}%
  \BibitemOpen
  \bibfield  {author} {\bibinfo {author} {\bibfnamefont {Y.~Y.}\ \bibnamefont {Gao}}, \bibinfo {author} {\bibfnamefont {B.~J.}\ \bibnamefont {Lester}}, \bibinfo {author} {\bibfnamefont {Y.}~\bibnamefont {Zhang}}, \bibinfo {author} {\bibfnamefont {C.}~\bibnamefont {Wang}}, \bibinfo {author} {\bibfnamefont {S.}~\bibnamefont {Rosenblum}}, \bibinfo {author} {\bibfnamefont {L.}~\bibnamefont {Frunzio}}, \bibinfo {author} {\bibfnamefont {L.}~\bibnamefont {Jiang}}, \bibinfo {author} {\bibfnamefont {S.}~\bibnamefont {Girvin}},\ and\ \bibinfo {author} {\bibfnamefont {R.~J.}\ \bibnamefont {Schoelkopf}},\ }\bibfield  {title} {\bibinfo {title} {Programmable {Interference} between {Two} {Microwave} {Quantum} {Memories}},\ }\href {https://doi.org/10.1103/PhysRevX.8.021073} {\bibfield  {journal} {\bibinfo  {journal} {Physical Review X}\ }\textbf {\bibinfo {volume} {8}},\ \bibinfo {pages} {021073} (\bibinfo {year} {2018})},\ \bibinfo {note} {publisher: American Physical Society}\BibitemShut {NoStop}%
\bibitem [{\citenamefont {Chapman}\ \emph {et~al.}(2023)\citenamefont {Chapman}, \citenamefont {de~Graaf}, \citenamefont {Xue}, \citenamefont {Zhang}, \citenamefont {Teoh}, \citenamefont {Curtis}, \citenamefont {Tsunoda}, \citenamefont {Eickbusch}, \citenamefont {Read}, \citenamefont {Koottandavida}, \citenamefont {Mundhada}, \citenamefont {Frunzio}, \citenamefont {Devoret}, \citenamefont {Girvin},\ and\ \citenamefont {Schoelkopf}}]{chapman_high--off-ratio_2023}%
  \BibitemOpen
  \bibfield  {author} {\bibinfo {author} {\bibfnamefont {B.~J.}\ \bibnamefont {Chapman}}, \bibinfo {author} {\bibfnamefont {S.~J.}\ \bibnamefont {de~Graaf}}, \bibinfo {author} {\bibfnamefont {S.~H.}\ \bibnamefont {Xue}}, \bibinfo {author} {\bibfnamefont {Y.}~\bibnamefont {Zhang}}, \bibinfo {author} {\bibfnamefont {J.}~\bibnamefont {Teoh}}, \bibinfo {author} {\bibfnamefont {J.~C.}\ \bibnamefont {Curtis}}, \bibinfo {author} {\bibfnamefont {T.}~\bibnamefont {Tsunoda}}, \bibinfo {author} {\bibfnamefont {A.}~\bibnamefont {Eickbusch}}, \bibinfo {author} {\bibfnamefont {A.~P.}\ \bibnamefont {Read}}, \bibinfo {author} {\bibfnamefont {A.}~\bibnamefont {Koottandavida}}, \bibinfo {author} {\bibfnamefont {S.~O.}\ \bibnamefont {Mundhada}}, \bibinfo {author} {\bibfnamefont {L.}~\bibnamefont {Frunzio}}, \bibinfo {author} {\bibfnamefont {M.}~\bibnamefont {Devoret}}, \bibinfo {author} {\bibfnamefont {S.}~\bibnamefont {Girvin}},\ and\ \bibinfo {author} {\bibfnamefont {R.}~\bibnamefont {Schoelkopf}},\ }\bibfield  {title}
  {\bibinfo {title} {High-{On}-{Off}-{Ratio} {Beam}-{Splitter} {Interaction} for {Gates} on {Bosonically} {Encoded} {Qubits}},\ }\href {https://doi.org/10.1103/PRXQuantum.4.020355} {\bibfield  {journal} {\bibinfo  {journal} {PRX Quantum}\ }\textbf {\bibinfo {volume} {4}},\ \bibinfo {pages} {020355} (\bibinfo {year} {2023})}\BibitemShut {NoStop}%
\bibitem [{\citenamefont {Lu}\ \emph {et~al.}(2023)\citenamefont {Lu}, \citenamefont {Maiti}, \citenamefont {Garmon}, \citenamefont {Ganjam}, \citenamefont {Zhang}, \citenamefont {Claes}, \citenamefont {Frunzio}, \citenamefont {Girvin},\ and\ \citenamefont {Schoelkopf}}]{lu_high-fidelity_2023}%
  \BibitemOpen
  \bibfield  {author} {\bibinfo {author} {\bibfnamefont {Y.}~\bibnamefont {Lu}}, \bibinfo {author} {\bibfnamefont {A.}~\bibnamefont {Maiti}}, \bibinfo {author} {\bibfnamefont {J.~W.~O.}\ \bibnamefont {Garmon}}, \bibinfo {author} {\bibfnamefont {S.}~\bibnamefont {Ganjam}}, \bibinfo {author} {\bibfnamefont {Y.}~\bibnamefont {Zhang}}, \bibinfo {author} {\bibfnamefont {J.}~\bibnamefont {Claes}}, \bibinfo {author} {\bibfnamefont {L.}~\bibnamefont {Frunzio}}, \bibinfo {author} {\bibfnamefont {S.~M.}\ \bibnamefont {Girvin}},\ and\ \bibinfo {author} {\bibfnamefont {R.~J.}\ \bibnamefont {Schoelkopf}},\ }\bibfield  {title} {\bibinfo {title} {High-fidelity parametric beamsplitting with a parity-protected converter},\ }\href {https://doi.org/10.1038/s41467-023-41104-0} {\bibfield  {journal} {\bibinfo  {journal} {Nature Communications}\ }\textbf {\bibinfo {volume} {14}},\ \bibinfo {pages} {5767} (\bibinfo {year} {2023})}\BibitemShut {NoStop}%
\bibitem [{\citenamefont {Eickbusch}\ \emph {et~al.}(2022)\citenamefont {Eickbusch}, \citenamefont {Sivak}, \citenamefont {Ding}, \citenamefont {Elder}, \citenamefont {Jha}, \citenamefont {Venkatraman}, \citenamefont {Royer}, \citenamefont {Girvin}, \citenamefont {Schoelkopf},\ and\ \citenamefont {Devoret}}]{eickbusch_fast_2022}%
  \BibitemOpen
  \bibfield  {author} {\bibinfo {author} {\bibfnamefont {A.}~\bibnamefont {Eickbusch}}, \bibinfo {author} {\bibfnamefont {V.}~\bibnamefont {Sivak}}, \bibinfo {author} {\bibfnamefont {A.~Z.}\ \bibnamefont {Ding}}, \bibinfo {author} {\bibfnamefont {S.~S.}\ \bibnamefont {Elder}}, \bibinfo {author} {\bibfnamefont {S.~R.}\ \bibnamefont {Jha}}, \bibinfo {author} {\bibfnamefont {J.}~\bibnamefont {Venkatraman}}, \bibinfo {author} {\bibfnamefont {B.}~\bibnamefont {Royer}}, \bibinfo {author} {\bibfnamefont {S.~M.}\ \bibnamefont {Girvin}}, \bibinfo {author} {\bibfnamefont {R.~J.}\ \bibnamefont {Schoelkopf}},\ and\ \bibinfo {author} {\bibfnamefont {M.~H.}\ \bibnamefont {Devoret}},\ }\bibfield  {title} {\bibinfo {title} {Fast universal control of an oscillator with weak dispersive coupling to a qubit},\ }\href {https://doi.org/10.1038/s41567-022-01776-9} {\bibfield  {journal} {\bibinfo  {journal} {Nature Physics}\ }\textbf {\bibinfo {volume} {18}},\ \bibinfo {pages} {1464} (\bibinfo {year} {2022})}\BibitemShut {NoStop}%
\bibitem [{noa()}]{noauthor_see_nodate}%
  \BibitemOpen
  \href@noop {} {\bibinfo {title} {See supplementary materials.}}\BibitemShut {Stop}%
\bibitem [{\citenamefont {Nigg}\ \emph {et~al.}(2012)\citenamefont {Nigg}, \citenamefont {Paik}, \citenamefont {Vlastakis}, \citenamefont {Kirchmair}, \citenamefont {Shankar}, \citenamefont {Frunzio}, \citenamefont {Devoret}, \citenamefont {Schoelkopf},\ and\ \citenamefont {Girvin}}]{nigg_black-box_2012}%
  \BibitemOpen
  \bibfield  {author} {\bibinfo {author} {\bibfnamefont {S.~E.}\ \bibnamefont {Nigg}}, \bibinfo {author} {\bibfnamefont {H.}~\bibnamefont {Paik}}, \bibinfo {author} {\bibfnamefont {B.}~\bibnamefont {Vlastakis}}, \bibinfo {author} {\bibfnamefont {G.}~\bibnamefont {Kirchmair}}, \bibinfo {author} {\bibfnamefont {S.}~\bibnamefont {Shankar}}, \bibinfo {author} {\bibfnamefont {L.}~\bibnamefont {Frunzio}}, \bibinfo {author} {\bibfnamefont {M.~H.}\ \bibnamefont {Devoret}}, \bibinfo {author} {\bibfnamefont {R.~J.}\ \bibnamefont {Schoelkopf}},\ and\ \bibinfo {author} {\bibfnamefont {S.~M.}\ \bibnamefont {Girvin}},\ }\bibfield  {title} {\bibinfo {title} {Black-{Box} {Superconducting} {Circuit} {Quantization}},\ }\href {https://doi.org/10.1103/PhysRevLett.108.240502} {\bibfield  {journal} {\bibinfo  {journal} {Physical Review Letters}\ }\textbf {\bibinfo {volume} {108}},\ \bibinfo {pages} {240502} (\bibinfo {year} {2012})}\BibitemShut {NoStop}%
\bibitem [{\citenamefont {Singh}\ \emph {et~al.}(2025)\citenamefont {Singh}, \citenamefont {Refael}, \citenamefont {Clerk},\ and\ \citenamefont {Rosenfeld}}]{singh_impact_2025}%
  \BibitemOpen
  \bibfield  {author} {\bibinfo {author} {\bibfnamefont {S.}~\bibnamefont {Singh}}, \bibinfo {author} {\bibfnamefont {G.}~\bibnamefont {Refael}}, \bibinfo {author} {\bibfnamefont {A.}~\bibnamefont {Clerk}},\ and\ \bibinfo {author} {\bibfnamefont {E.}~\bibnamefont {Rosenfeld}},\ }\href {https://doi.org/10.48550/arXiv.2412.14788} {\bibinfo {title} {Impact of {Josephson} junction array modes on fluxonium readout}} (\bibinfo {year} {2025}),\ \bibinfo {note} {arXiv:2412.14788 [cond-mat]}\BibitemShut {NoStop}%
\bibitem [{\citenamefont {Kishmar}\ \emph {et~al.}(2025)\citenamefont {Kishmar}, \citenamefont {Kurilovich}, \citenamefont {Klots}, \citenamefont {Connolly}, \citenamefont {Aleiner},\ and\ \citenamefont {Kurilovich}}]{kishmar_quasiparticle-induced_2025}%
  \BibitemOpen
  \bibfield  {author} {\bibinfo {author} {\bibfnamefont {M.}~\bibnamefont {Kishmar}}, \bibinfo {author} {\bibfnamefont {P.~D.}\ \bibnamefont {Kurilovich}}, \bibinfo {author} {\bibfnamefont {A.}~\bibnamefont {Klots}}, \bibinfo {author} {\bibfnamefont {T.}~\bibnamefont {Connolly}}, \bibinfo {author} {\bibfnamefont {I.~L.}\ \bibnamefont {Aleiner}},\ and\ \bibinfo {author} {\bibfnamefont {V.~D.}\ \bibnamefont {Kurilovich}},\ }\href {https://doi.org/10.48550/arXiv.2505.00769} {\bibinfo {title} {Quasiparticle-induced decoherence of a driven superconducting qubit}} (\bibinfo {year} {2025}),\ \bibinfo {note} {arXiv:2505.00769 [quant-ph]}\BibitemShut {NoStop}%
\bibitem [{\citenamefont {Chowdhury}\ \emph {et~al.}(2025)\citenamefont {Chowdhury}, \citenamefont {Hays}, \citenamefont {Jha}, \citenamefont {Serniak}, \citenamefont {Orlando}, \citenamefont {Grover},\ and\ \citenamefont {Oliver}}]{chowdhury_theory_2025}%
  \BibitemOpen
  \bibfield  {author} {\bibinfo {author} {\bibfnamefont {S.}~\bibnamefont {Chowdhury}}, \bibinfo {author} {\bibfnamefont {M.}~\bibnamefont {Hays}}, \bibinfo {author} {\bibfnamefont {S.~R.}\ \bibnamefont {Jha}}, \bibinfo {author} {\bibfnamefont {K.}~\bibnamefont {Serniak}}, \bibinfo {author} {\bibfnamefont {T.~P.}\ \bibnamefont {Orlando}}, \bibinfo {author} {\bibfnamefont {J.~A.}\ \bibnamefont {Grover}},\ and\ \bibinfo {author} {\bibfnamefont {W.~D.}\ \bibnamefont {Oliver}},\ }\href {https://doi.org/10.48550/arXiv.2505.00773} {\bibinfo {title} {Theory of {Quasiparticle} {Generation} by {Microwave} {Drives} in {Superconducting} {Qubits}}} (\bibinfo {year} {2025}),\ \bibinfo {note} {arXiv:2505.00773 [quant-ph]}\BibitemShut {NoStop}%
\bibitem [{\citenamefont {Féchant}\ \emph {et~al.}(2025)\citenamefont {Féchant}, \citenamefont {Dumas}, \citenamefont {Bénâtre}, \citenamefont {Gosling}, \citenamefont {Lenhard}, \citenamefont {Spiecker}, \citenamefont {Wernsdorfer}, \citenamefont {D'Anjou}, \citenamefont {Blais},\ and\ \citenamefont {Pop}}]{fechant_offset_2025}%
  \BibitemOpen
  \bibfield  {author} {\bibinfo {author} {\bibfnamefont {M.}~\bibnamefont {Féchant}}, \bibinfo {author} {\bibfnamefont {M.~F.}\ \bibnamefont {Dumas}}, \bibinfo {author} {\bibfnamefont {D.}~\bibnamefont {Bénâtre}}, \bibinfo {author} {\bibfnamefont {N.}~\bibnamefont {Gosling}}, \bibinfo {author} {\bibfnamefont {P.}~\bibnamefont {Lenhard}}, \bibinfo {author} {\bibfnamefont {M.}~\bibnamefont {Spiecker}}, \bibinfo {author} {\bibfnamefont {W.}~\bibnamefont {Wernsdorfer}}, \bibinfo {author} {\bibfnamefont {B.}~\bibnamefont {D'Anjou}}, \bibinfo {author} {\bibfnamefont {A.}~\bibnamefont {Blais}},\ and\ \bibinfo {author} {\bibfnamefont {I.~M.}\ \bibnamefont {Pop}},\ }\href {https://doi.org/10.48550/arXiv.2505.00674} {\bibinfo {title} {Offset {Charge} {Dependence} of {Measurement}-{Induced} {Transitions} in {Transmons}}} (\bibinfo {year} {2025}),\ \bibinfo {note} {arXiv:2505.00674 [quant-ph]}\BibitemShut {NoStop}%
\bibitem [{\citenamefont {Wang}\ \emph {et~al.}(2025)\citenamefont {Wang}, \citenamefont {D'Anjou}, \citenamefont {Gigon}, \citenamefont {Blais},\ and\ \citenamefont {Blok}}]{wang_probing_2025}%
  \BibitemOpen
  \bibfield  {author} {\bibinfo {author} {\bibfnamefont {Z.}~\bibnamefont {Wang}}, \bibinfo {author} {\bibfnamefont {B.}~\bibnamefont {D'Anjou}}, \bibinfo {author} {\bibfnamefont {P.}~\bibnamefont {Gigon}}, \bibinfo {author} {\bibfnamefont {A.}~\bibnamefont {Blais}},\ and\ \bibinfo {author} {\bibfnamefont {M.~S.}\ \bibnamefont {Blok}},\ }\href {https://doi.org/10.48550/arXiv.2505.00639} {\bibinfo {title} {Probing excited-state dynamics of transmon ionization}} (\bibinfo {year} {2025}),\ \bibinfo {note} {arXiv:2505.00639 [quant-ph]}\BibitemShut {NoStop}%
\end{thebibliography}%


\begin{thebibliography}{10}%
\makeatletter
\providecommand \@ifxundefined [1]{%
 \@ifx{#1\undefined}
}%
\providecommand \@ifnum [1]{%
 \ifnum #1\expandafter \@firstoftwo
 \else \expandafter \@secondoftwo
 \fi
}%
\providecommand \@ifx [1]{%
 \ifx #1\expandafter \@firstoftwo
 \else \expandafter \@secondoftwo
 \fi
}%
\providecommand \natexlab [1]{#1}%
\providecommand \enquote  [1]{``#1''}%
\providecommand \bibnamefont  [1]{#1}%
\providecommand \bibfnamefont [1]{#1}%
\providecommand \citenamefont [1]{#1}%
\providecommand \href@noop [0]{\@secondoftwo}%
\providecommand \href [0]{\begingroup \@sanitize@url \@href}%
\providecommand \@href[1]{\@@startlink{#1}\@@href}%
\providecommand \@@href[1]{\endgroup#1\@@endlink}%
\providecommand \@sanitize@url [0]{\catcode `\\12\catcode `\$12\catcode `\&12\catcode `\#12\catcode `\^12\catcode `\_12\catcode `\%12\relax}%
\providecommand \@@startlink[1]{}%
\providecommand \@@endlink[0]{}%
\providecommand \url  [0]{\begingroup\@sanitize@url \@url }%
\providecommand \@url [1]{\endgroup\@href {#1}{\urlprefix }}%
\providecommand \urlprefix  [0]{URL }%
\providecommand \Eprint [0]{\href }%
\providecommand \doibase [0]{https://doi.org/}%
\providecommand \selectlanguage [0]{\@gobble}%
\providecommand \bibinfo  [0]{\@secondoftwo}%
\providecommand \bibfield  [0]{\@secondoftwo}%
\providecommand \translation [1]{[#1]}%
\providecommand \BibitemOpen [0]{}%
\providecommand \bibitemStop [0]{}%
\providecommand \bibitemNoStop [0]{.\EOS\space}%
\providecommand \EOS [0]{\spacefactor3000\relax}%
\providecommand \BibitemShut  [1]{\csname bibitem#1\endcsname}%
\let\auto@bib@innerbib\@empty
%</preamble>
\bibitem [{\citenamefont {Kurilovich}\ \emph {et~al.}(2025)\citenamefont {Kurilovich}, \citenamefont {Connolly}, \citenamefont {Bøttcher}, \citenamefont {Weiss}, \citenamefont {Hazra}, \citenamefont {Joshi}, \citenamefont {Ding}, \citenamefont {Nho}, \citenamefont {Diamond}, \citenamefont {Kurilovich}, \citenamefont {Dai}, \citenamefont {Fatemi}, \citenamefont {Frunzio}, \citenamefont {Glazman},\ and\ \citenamefont {Devoret}}]{kurilovich_high-frequency_2025}%
  \BibitemOpen
  \bibfield  {author} {\bibinfo {author} {\bibfnamefont {P.~D.}\ \bibnamefont {Kurilovich}}, \bibinfo {author} {\bibfnamefont {T.}~\bibnamefont {Connolly}}, \bibinfo {author} {\bibfnamefont {C.~G.~L.}\ \bibnamefont {Bøttcher}}, \bibinfo {author} {\bibfnamefont {D.~K.}\ \bibnamefont {Weiss}}, \bibinfo {author} {\bibfnamefont {S.}~\bibnamefont {Hazra}}, \bibinfo {author} {\bibfnamefont {V.~R.}\ \bibnamefont {Joshi}}, \bibinfo {author} {\bibfnamefont {A.~Z.}\ \bibnamefont {Ding}}, \bibinfo {author} {\bibfnamefont {H.}~\bibnamefont {Nho}}, \bibinfo {author} {\bibfnamefont {S.}~\bibnamefont {Diamond}}, \bibinfo {author} {\bibfnamefont {V.~D.}\ \bibnamefont {Kurilovich}}, \bibinfo {author} {\bibfnamefont {W.}~\bibnamefont {Dai}}, \bibinfo {author} {\bibfnamefont {V.}~\bibnamefont {Fatemi}}, \bibinfo {author} {\bibfnamefont {L.}~\bibnamefont {Frunzio}}, \bibinfo {author} {\bibfnamefont {L.~I.}\ \bibnamefont {Glazman}},\ and\ \bibinfo {author} {\bibfnamefont {M.~H.}\ \bibnamefont {Devoret}},\ }\href
  {https://doi.org/10.48550/arXiv.2501.09161} {\bibinfo {title} {High-frequency readout free from transmon multi-excitation resonances}} (\bibinfo {year} {2025}),\ \bibinfo {note} {arXiv:2501.09161 [quant-ph]}\BibitemShut {NoStop}%
\bibitem [{\citenamefont {Lecocq}\ \emph {et~al.}(2011)\citenamefont {Lecocq}, \citenamefont {Pop}, \citenamefont {Peng}, \citenamefont {Matei}, \citenamefont {Crozes}, \citenamefont {Fournier}, \citenamefont {Naud}, \citenamefont {Guichard},\ and\ \citenamefont {Buisson}}]{lecocq_junction_2011}%
  \BibitemOpen
  \bibfield  {author} {\bibinfo {author} {\bibfnamefont {F.}~\bibnamefont {Lecocq}}, \bibinfo {author} {\bibfnamefont {I.~M.}\ \bibnamefont {Pop}}, \bibinfo {author} {\bibfnamefont {Z.}~\bibnamefont {Peng}}, \bibinfo {author} {\bibfnamefont {I.}~\bibnamefont {Matei}}, \bibinfo {author} {\bibfnamefont {T.}~\bibnamefont {Crozes}}, \bibinfo {author} {\bibfnamefont {T.}~\bibnamefont {Fournier}}, \bibinfo {author} {\bibfnamefont {C.}~\bibnamefont {Naud}}, \bibinfo {author} {\bibfnamefont {W.}~\bibnamefont {Guichard}},\ and\ \bibinfo {author} {\bibfnamefont {O.}~\bibnamefont {Buisson}},\ }\bibfield  {title} {\bibinfo {title} {Junction fabrication by shadow evaporation without a suspended bridge},\ }\href {https://doi.org/10.1088/0957-4484/22/31/315302} {\bibfield  {journal} {\bibinfo  {journal} {Nanotechnology}\ }\textbf {\bibinfo {volume} {22}},\ \bibinfo {pages} {315302} (\bibinfo {year} {2011})}\BibitemShut {NoStop}%
\bibitem [{\citenamefont {Place}\ \emph {et~al.}(2021)\citenamefont {Place}, \citenamefont {Rodgers}, \citenamefont {Mundada}, \citenamefont {Smitham}, \citenamefont {Fitzpatrick}, \citenamefont {Leng}, \citenamefont {Premkumar}, \citenamefont {Bryon}, \citenamefont {Vrajitoarea}, \citenamefont {Sussman}, \citenamefont {Cheng}, \citenamefont {Madhavan}, \citenamefont {Babla}, \citenamefont {Le}, \citenamefont {Gang}, \citenamefont {Jäck}, \citenamefont {Gyenis}, \citenamefont {Yao}, \citenamefont {Cava}, \citenamefont {de~Leon},\ and\ \citenamefont {Houck}}]{place_new_2021}%
  \BibitemOpen
  \bibfield  {author} {\bibinfo {author} {\bibfnamefont {A.~P.~M.}\ \bibnamefont {Place}}, \bibinfo {author} {\bibfnamefont {L.~V.~H.}\ \bibnamefont {Rodgers}}, \bibinfo {author} {\bibfnamefont {P.}~\bibnamefont {Mundada}}, \bibinfo {author} {\bibfnamefont {B.~M.}\ \bibnamefont {Smitham}}, \bibinfo {author} {\bibfnamefont {M.}~\bibnamefont {Fitzpatrick}}, \bibinfo {author} {\bibfnamefont {Z.}~\bibnamefont {Leng}}, \bibinfo {author} {\bibfnamefont {A.}~\bibnamefont {Premkumar}}, \bibinfo {author} {\bibfnamefont {J.}~\bibnamefont {Bryon}}, \bibinfo {author} {\bibfnamefont {A.}~\bibnamefont {Vrajitoarea}}, \bibinfo {author} {\bibfnamefont {S.}~\bibnamefont {Sussman}}, \bibinfo {author} {\bibfnamefont {G.}~\bibnamefont {Cheng}}, \bibinfo {author} {\bibfnamefont {T.}~\bibnamefont {Madhavan}}, \bibinfo {author} {\bibfnamefont {H.~K.}\ \bibnamefont {Babla}}, \bibinfo {author} {\bibfnamefont {X.~H.}\ \bibnamefont {Le}}, \bibinfo {author} {\bibfnamefont {Y.}~\bibnamefont {Gang}}, \bibinfo {author} {\bibfnamefont
  {B.}~\bibnamefont {Jäck}}, \bibinfo {author} {\bibfnamefont {A.}~\bibnamefont {Gyenis}}, \bibinfo {author} {\bibfnamefont {N.}~\bibnamefont {Yao}}, \bibinfo {author} {\bibfnamefont {R.~J.}\ \bibnamefont {Cava}}, \bibinfo {author} {\bibfnamefont {N.~P.}\ \bibnamefont {de~Leon}},\ and\ \bibinfo {author} {\bibfnamefont {A.~A.}\ \bibnamefont {Houck}},\ }\bibfield  {title} {\bibinfo {title} {New material platform for superconducting transmon qubits with coherence times exceeding 0.3 milliseconds},\ }\href {https://doi.org/10.1038/s41467-021-22030-5} {\bibfield  {journal} {\bibinfo  {journal} {Nature Communications}\ }\textbf {\bibinfo {volume} {12}},\ \bibinfo {pages} {1779} (\bibinfo {year} {2021})}\BibitemShut {NoStop}%
\bibitem [{\citenamefont {Ganjam}\ \emph {et~al.}(2024)\citenamefont {Ganjam}, \citenamefont {Wang}, \citenamefont {Lu}, \citenamefont {Banerjee}, \citenamefont {Lei}, \citenamefont {Krayzman}, \citenamefont {Kisslinger}, \citenamefont {Zhou}, \citenamefont {Li}, \citenamefont {Jia}, \citenamefont {Liu}, \citenamefont {Frunzio},\ and\ \citenamefont {Schoelkopf}}]{ganjam_surpassing_2024}%
  \BibitemOpen
  \bibfield  {author} {\bibinfo {author} {\bibfnamefont {S.}~\bibnamefont {Ganjam}}, \bibinfo {author} {\bibfnamefont {Y.}~\bibnamefont {Wang}}, \bibinfo {author} {\bibfnamefont {Y.}~\bibnamefont {Lu}}, \bibinfo {author} {\bibfnamefont {A.}~\bibnamefont {Banerjee}}, \bibinfo {author} {\bibfnamefont {C.~U.}\ \bibnamefont {Lei}}, \bibinfo {author} {\bibfnamefont {L.}~\bibnamefont {Krayzman}}, \bibinfo {author} {\bibfnamefont {K.}~\bibnamefont {Kisslinger}}, \bibinfo {author} {\bibfnamefont {C.}~\bibnamefont {Zhou}}, \bibinfo {author} {\bibfnamefont {R.}~\bibnamefont {Li}}, \bibinfo {author} {\bibfnamefont {Y.}~\bibnamefont {Jia}}, \bibinfo {author} {\bibfnamefont {M.}~\bibnamefont {Liu}}, \bibinfo {author} {\bibfnamefont {L.}~\bibnamefont {Frunzio}},\ and\ \bibinfo {author} {\bibfnamefont {R.~J.}\ \bibnamefont {Schoelkopf}},\ }\bibfield  {title} {\bibinfo {title} {Surpassing millisecond coherence in on chip superconducting quantum memories by optimizing materials and circuit design},\ }\href
  {https://doi.org/10.1038/s41467-024-47857-6} {\bibfield  {journal} {\bibinfo  {journal} {Nature Communications}\ }\textbf {\bibinfo {volume} {15}},\ \bibinfo {pages} {3687} (\bibinfo {year} {2024})}\BibitemShut {NoStop}%
\bibitem [{\citenamefont {Serniak}\ \emph {et~al.}(2019)\citenamefont {Serniak}, \citenamefont {Diamond}, \citenamefont {Hays}, \citenamefont {Fatemi}, \citenamefont {Shankar}, \citenamefont {Frunzio}, \citenamefont {Schoelkopf},\ and\ \citenamefont {Devoret}}]{serniak_direct_2019}%
  \BibitemOpen
  \bibfield  {author} {\bibinfo {author} {\bibfnamefont {K.}~\bibnamefont {Serniak}}, \bibinfo {author} {\bibfnamefont {S.}~\bibnamefont {Diamond}}, \bibinfo {author} {\bibfnamefont {M.}~\bibnamefont {Hays}}, \bibinfo {author} {\bibfnamefont {V.}~\bibnamefont {Fatemi}}, \bibinfo {author} {\bibfnamefont {S.}~\bibnamefont {Shankar}}, \bibinfo {author} {\bibfnamefont {L.}~\bibnamefont {Frunzio}}, \bibinfo {author} {\bibfnamefont {R.}~\bibnamefont {Schoelkopf}},\ and\ \bibinfo {author} {\bibfnamefont {M.}~\bibnamefont {Devoret}},\ }\bibfield  {title} {\bibinfo {title} {Direct dispersive monitoring of charge parity in offset-charge-sensitive transmons},\ }\href {https://doi.org/10.1103/PhysRevApplied.12.014052} {\bibfield  {journal} {\bibinfo  {journal} {Physical Review Applied}\ }\textbf {\bibinfo {volume} {12}},\ \bibinfo {pages} {014052} (\bibinfo {year} {2019})}\BibitemShut {NoStop}%
\bibitem [{\citenamefont {Connolly}\ \emph {et~al.}(2024)\citenamefont {Connolly}, \citenamefont {Kurilovich}, \citenamefont {Diamond}, \citenamefont {Nho}, \citenamefont {Bøttcher}, \citenamefont {Glazman}, \citenamefont {Fatemi},\ and\ \citenamefont {Devoret}}]{connolly_coexistence_2024}%
  \BibitemOpen
  \bibfield  {author} {\bibinfo {author} {\bibfnamefont {T.}~\bibnamefont {Connolly}}, \bibinfo {author} {\bibfnamefont {P.~D.}\ \bibnamefont {Kurilovich}}, \bibinfo {author} {\bibfnamefont {S.}~\bibnamefont {Diamond}}, \bibinfo {author} {\bibfnamefont {H.}~\bibnamefont {Nho}}, \bibinfo {author} {\bibfnamefont {C.~G.}\ \bibnamefont {Bøttcher}}, \bibinfo {author} {\bibfnamefont {L.~I.}\ \bibnamefont {Glazman}}, \bibinfo {author} {\bibfnamefont {V.}~\bibnamefont {Fatemi}},\ and\ \bibinfo {author} {\bibfnamefont {M.~H.}\ \bibnamefont {Devoret}},\ }\bibfield  {title} {\bibinfo {title} {Coexistence of {Nonequilibrium} {Density} and {Equilibrium} {Energy} {Distribution} of {Quasiparticles} in a {Superconducting} {Qubit}},\ }\href {https://doi.org/10.1103/PhysRevLett.132.217001} {\bibfield  {journal} {\bibinfo  {journal} {Physical Review Letters}\ }\textbf {\bibinfo {volume} {132}},\ \bibinfo {pages} {217001} (\bibinfo {year} {2024})}\BibitemShut {NoStop}%
\bibitem [{\citenamefont {Rehammar}\ and\ \citenamefont {Gasparinetti}(2023)}]{rehammar_low-pass_2023}%
  \BibitemOpen
  \bibfield  {author} {\bibinfo {author} {\bibfnamefont {R.}~\bibnamefont {Rehammar}}\ and\ \bibinfo {author} {\bibfnamefont {S.}~\bibnamefont {Gasparinetti}},\ }\bibfield  {title} {\bibinfo {title} {Low-{Pass} {Filter} {With} {Ultrawide} {Stopband} for {Quantum} {Computing} {Applications}},\ }\href {https://doi.org/10.1109/TMTT.2023.3238543} {\bibfield  {journal} {\bibinfo  {journal} {IEEE Transactions on Microwave Theory and Techniques}\ }\textbf {\bibinfo {volume} {71}},\ \bibinfo {pages} {3075} (\bibinfo {year} {2023})}\BibitemShut {NoStop}%
\bibitem [{\citenamefont {Nigg}\ \emph {et~al.}(2012)\citenamefont {Nigg}, \citenamefont {Paik}, \citenamefont {Vlastakis}, \citenamefont {Kirchmair}, \citenamefont {Shankar}, \citenamefont {Frunzio}, \citenamefont {Devoret}, \citenamefont {Schoelkopf},\ and\ \citenamefont {Girvin}}]{nigg_black-box_2012}%
  \BibitemOpen
  \bibfield  {author} {\bibinfo {author} {\bibfnamefont {S.~E.}\ \bibnamefont {Nigg}}, \bibinfo {author} {\bibfnamefont {H.}~\bibnamefont {Paik}}, \bibinfo {author} {\bibfnamefont {B.}~\bibnamefont {Vlastakis}}, \bibinfo {author} {\bibfnamefont {G.}~\bibnamefont {Kirchmair}}, \bibinfo {author} {\bibfnamefont {S.}~\bibnamefont {Shankar}}, \bibinfo {author} {\bibfnamefont {L.}~\bibnamefont {Frunzio}}, \bibinfo {author} {\bibfnamefont {M.~H.}\ \bibnamefont {Devoret}}, \bibinfo {author} {\bibfnamefont {R.~J.}\ \bibnamefont {Schoelkopf}},\ and\ \bibinfo {author} {\bibfnamefont {S.~M.}\ \bibnamefont {Girvin}},\ }\bibfield  {title} {\bibinfo {title} {Black-{Box} {Superconducting} {Circuit} {Quantization}},\ }\href {https://doi.org/10.1103/PhysRevLett.108.240502} {\bibfield  {journal} {\bibinfo  {journal} {Physical Review Letters}\ }\textbf {\bibinfo {volume} {108}},\ \bibinfo {pages} {240502} (\bibinfo {year} {2012})}\BibitemShut {NoStop}%
\bibitem [{\citenamefont {Koch}\ \emph {et~al.}(2007)\citenamefont {Koch}, \citenamefont {Yu}, \citenamefont {Gambetta}, \citenamefont {Houck}, \citenamefont {Schuster}, \citenamefont {Majer}, \citenamefont {Blais}, \citenamefont {Devoret}, \citenamefont {Girvin},\ and\ \citenamefont {Schoelkopf}}]{koch_charge-insensitive_2007}%
  \BibitemOpen
  \bibfield  {author} {\bibinfo {author} {\bibfnamefont {J.}~\bibnamefont {Koch}}, \bibinfo {author} {\bibfnamefont {T.~M.}\ \bibnamefont {Yu}}, \bibinfo {author} {\bibfnamefont {J.}~\bibnamefont {Gambetta}}, \bibinfo {author} {\bibfnamefont {A.~A.}\ \bibnamefont {Houck}}, \bibinfo {author} {\bibfnamefont {D.~I.}\ \bibnamefont {Schuster}}, \bibinfo {author} {\bibfnamefont {J.}~\bibnamefont {Majer}}, \bibinfo {author} {\bibfnamefont {A.}~\bibnamefont {Blais}}, \bibinfo {author} {\bibfnamefont {M.~H.}\ \bibnamefont {Devoret}}, \bibinfo {author} {\bibfnamefont {S.~M.}\ \bibnamefont {Girvin}},\ and\ \bibinfo {author} {\bibfnamefont {R.~J.}\ \bibnamefont {Schoelkopf}},\ }\bibfield  {title} {\bibinfo {title} {Charge-insensitive qubit design derived from the {Cooper} pair box},\ }\href {https://doi.org/10.1103/PhysRevA.76.042319} {\bibfield  {journal} {\bibinfo  {journal} {Physical Review A}\ }\textbf {\bibinfo {volume} {76}},\ \bibinfo {pages} {042319} (\bibinfo {year} {2007})}\BibitemShut {NoStop}%
\bibitem [{\citenamefont {Willsch}\ \emph {et~al.}(2024)\citenamefont {Willsch}, \citenamefont {Rieger}, \citenamefont {Winkel}, \citenamefont {Willsch}, \citenamefont {Dickel}, \citenamefont {Krause}, \citenamefont {Ando}, \citenamefont {Lescanne}, \citenamefont {Leghtas}, \citenamefont {Bronn}, \citenamefont {Deb}, \citenamefont {Lanes}, \citenamefont {Minev}, \citenamefont {Dennig}, \citenamefont {Geisert}, \citenamefont {Günzler}, \citenamefont {Ihssen}, \citenamefont {Paluch}, \citenamefont {Reisinger}, \citenamefont {Hanna}, \citenamefont {Bae}, \citenamefont {Schüffelgen}, \citenamefont {Grützmacher}, \citenamefont {Buimaga-Iarinca}, \citenamefont {Morari}, \citenamefont {Wernsdorfer}, \citenamefont {DiVincenzo}, \citenamefont {Michielsen}, \citenamefont {Catelani},\ and\ \citenamefont {Pop}}]{willsch_observation_2024}%
  \BibitemOpen
  \bibfield  {author} {\bibinfo {author} {\bibfnamefont {D.}~\bibnamefont {Willsch}}, \bibinfo {author} {\bibfnamefont {D.}~\bibnamefont {Rieger}}, \bibinfo {author} {\bibfnamefont {P.}~\bibnamefont {Winkel}}, \bibinfo {author} {\bibfnamefont {M.}~\bibnamefont {Willsch}}, \bibinfo {author} {\bibfnamefont {C.}~\bibnamefont {Dickel}}, \bibinfo {author} {\bibfnamefont {J.}~\bibnamefont {Krause}}, \bibinfo {author} {\bibfnamefont {Y.}~\bibnamefont {Ando}}, \bibinfo {author} {\bibfnamefont {R.}~\bibnamefont {Lescanne}}, \bibinfo {author} {\bibfnamefont {Z.}~\bibnamefont {Leghtas}}, \bibinfo {author} {\bibfnamefont {N.~T.}\ \bibnamefont {Bronn}}, \bibinfo {author} {\bibfnamefont {P.}~\bibnamefont {Deb}}, \bibinfo {author} {\bibfnamefont {O.}~\bibnamefont {Lanes}}, \bibinfo {author} {\bibfnamefont {Z.~K.}\ \bibnamefont {Minev}}, \bibinfo {author} {\bibfnamefont {B.}~\bibnamefont {Dennig}}, \bibinfo {author} {\bibfnamefont {S.}~\bibnamefont {Geisert}}, \bibinfo {author} {\bibfnamefont {S.}~\bibnamefont {Günzler}},
  \bibinfo {author} {\bibfnamefont {S.}~\bibnamefont {Ihssen}}, \bibinfo {author} {\bibfnamefont {P.}~\bibnamefont {Paluch}}, \bibinfo {author} {\bibfnamefont {T.}~\bibnamefont {Reisinger}}, \bibinfo {author} {\bibfnamefont {R.}~\bibnamefont {Hanna}}, \bibinfo {author} {\bibfnamefont {J.~H.}\ \bibnamefont {Bae}}, \bibinfo {author} {\bibfnamefont {P.}~\bibnamefont {Schüffelgen}}, \bibinfo {author} {\bibfnamefont {D.}~\bibnamefont {Grützmacher}}, \bibinfo {author} {\bibfnamefont {L.}~\bibnamefont {Buimaga-Iarinca}}, \bibinfo {author} {\bibfnamefont {C.}~\bibnamefont {Morari}}, \bibinfo {author} {\bibfnamefont {W.}~\bibnamefont {Wernsdorfer}}, \bibinfo {author} {\bibfnamefont {D.~P.}\ \bibnamefont {DiVincenzo}}, \bibinfo {author} {\bibfnamefont {K.}~\bibnamefont {Michielsen}}, \bibinfo {author} {\bibfnamefont {G.}~\bibnamefont {Catelani}},\ and\ \bibinfo {author} {\bibfnamefont {I.~M.}\ \bibnamefont {Pop}},\ }\bibfield  {title} {\bibinfo {title} {Observation of {Josephson} harmonics in tunnel junctions},\
  }\href {https://doi.org/10.1038/s41567-024-02400-8} {\bibfield  {journal} {\bibinfo  {journal} {Nature Physics}\ }\textbf {\bibinfo {volume} {20}},\ \bibinfo {pages} {815} (\bibinfo {year} {2024})}\BibitemShut {NoStop}%
\end{thebibliography}%
\end{document}

\myexternaldocument{ms}
\beginsupplement
\title{Supplementary information for ``Full characterization of measurement-induced transitions of a superconducting qubit''}
\author{Thomas~Connolly}
\thanks{These two authors contributed equally.\\
tom.connolly@yale.edu, pavel.kurilovich@yale.edu}
\affiliation{Departments of Applied Physics and Physics, Yale University, New Haven, Connecticut 06520, USA}
\author{Pavel~D.~Kurilovich}
\thanks{These two authors contributed equally.\\
tom.connolly@yale.edu, pavel.kurilovich@yale.edu}
\affiliation{Departments of Applied Physics and Physics, Yale University, New Haven, Connecticut 06520, USA}
\author{Vladislav~D.~Kurilovich}\thanks{{Present address: Google Quantum AI, 301 Mentor Dr, Goleta, CA93111, USA}}
\affiliation{Departments of Applied Physics and Physics, Yale University, New Haven, Connecticut 06520, USA}

\author{Charlotte~G.~L.~B\o ttcher}
\thanks{{Present address: Department of Applied Physics, Stanford University, Stanford, California 94305, USA}}
\affiliation{Departments of Applied Physics and Physics, Yale University, New Haven, Connecticut 06520, USA}

\author{Sumeru~Hazra}
\affiliation{Departments of Applied Physics and Physics, Yale University, New Haven, Connecticut 06520, USA}
\author{Wei~Dai}
\affiliation{Departments of Applied Physics and Physics, Yale University, New Haven, Connecticut 06520, USA}

\author{Andy~Z.~Ding}
\affiliation{Departments of Applied Physics and Physics, Yale University, New Haven, Connecticut 06520, USA}

\author{Vidul~R.~Joshi}
\thanks{{Present address: Microsoft Quantum}}
\affiliation{Departments of Applied Physics and Physics, Yale University, New Haven, Connecticut 06520, USA}

\author{Heekun~Nho}
\affiliation{Departments of Applied Physics and Physics, Yale University, New Haven, Connecticut 06520, USA}

\author{Spencer~Diamond}
\affiliation{Departments of Applied Physics and Physics, Yale University, New Haven, Connecticut 06520, USA}

\author{Daniel~K.~Weiss}
\thanks{{Present address: Quantum Circuits, Inc., New Haven, CT, USA}}
\affiliation{Departments of Applied Physics and Physics, Yale University, New Haven, Connecticut 06520, USA}
\affiliation{Yale Quantum Institute, Yale University, New Haven, Connecticut 06511, USA}

\author{Valla~Fatemi}
\affiliation{Departments of Applied Physics and Physics, Yale University, New Haven, Connecticut 06520, USA}
\affiliation{School of Applied and Engineering Physics, Cornell University, Ithaca, New York 14853, USA}
\author{Luigi Frunzio}
\affiliation{Departments of Applied Physics and Physics, Yale University, New Haven, Connecticut 06520, USA}
\author{Leonid~I.~Glazman}
\affiliation{Departments of Applied Physics and Physics, Yale University, New Haven, Connecticut 06520, USA}
\affiliation{Yale Quantum Institute, Yale University, New Haven, Connecticut 06511, USA}
\author{Michel~H.~Devoret}\thanks{michel.devoret@yale.edu\\
{Present address: Physics Dept., U.C. Santa Barbara, Santa Barbara, California 93106, USA and Google Quantum AI, 301 Mentor Dr, Goleta, California 93111, USA}}
\affiliation{Departments of Applied Physics and Physics, Yale University, New Haven, Connecticut 06520, USA}

\date{\today}

\maketitle

\tableofcontents
\newpage
\section{Experimental setup}
\subsection{Wiring diagram}
\begin{figure}[h!]
  \begin{center}
    \includegraphics[scale = 1.0]{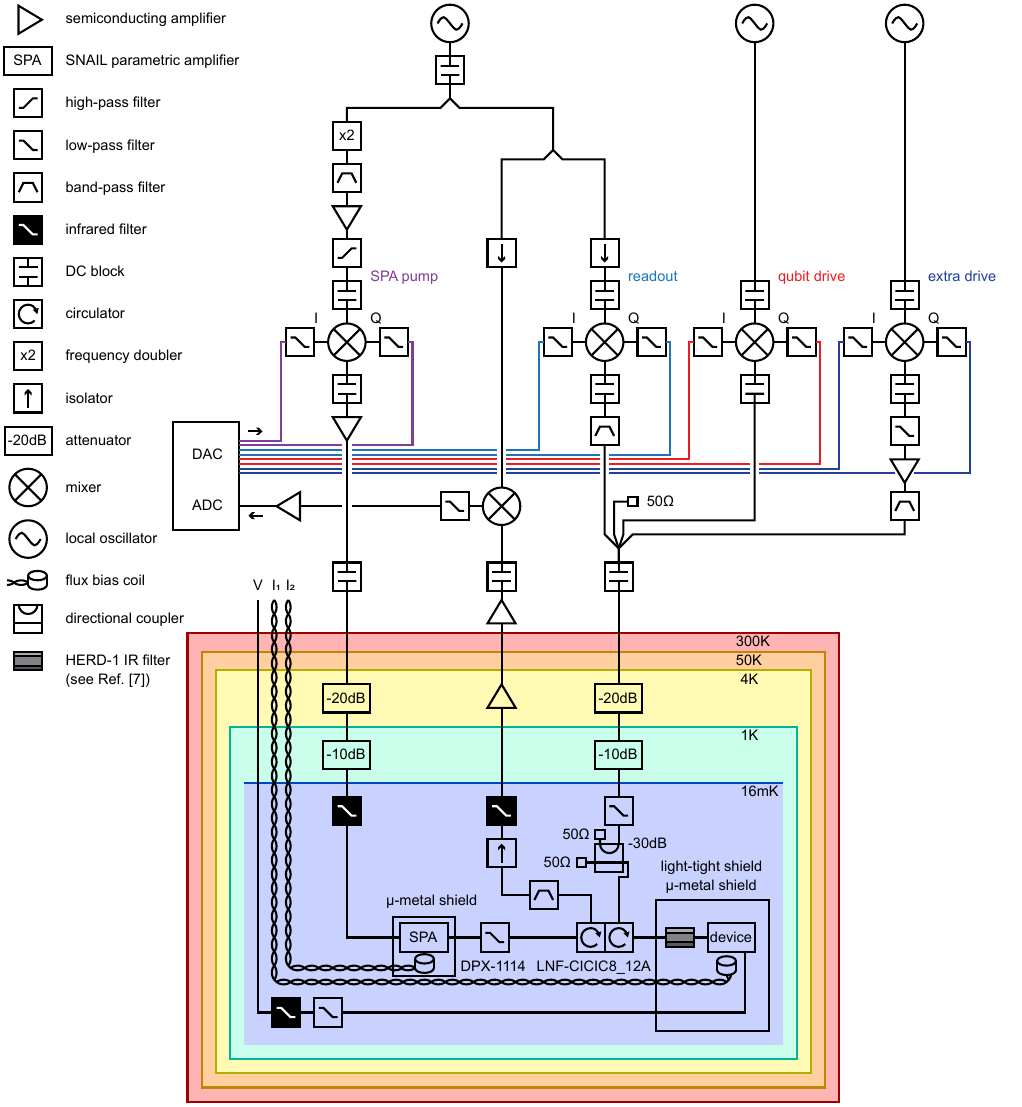}
    \caption{Wiring diagram of the setup. The setup is the same as in Ref.~\cite{kurilovich_high-frequency_2025}. The ``extra drive'' generator supplies the drive used for the rate measurements.}
    \label{fig:wiring}
  \end{center}
\end{figure}

\subsection{Fabrication and packaging details}
%The fabrication procedure is described in detail in Ref.~\cite{kurilovich_high-frequency_2025}. Here, we briefly recap the procedure.

\begin{figure}[h]
  \begin{center}
    \includegraphics[scale = 1]{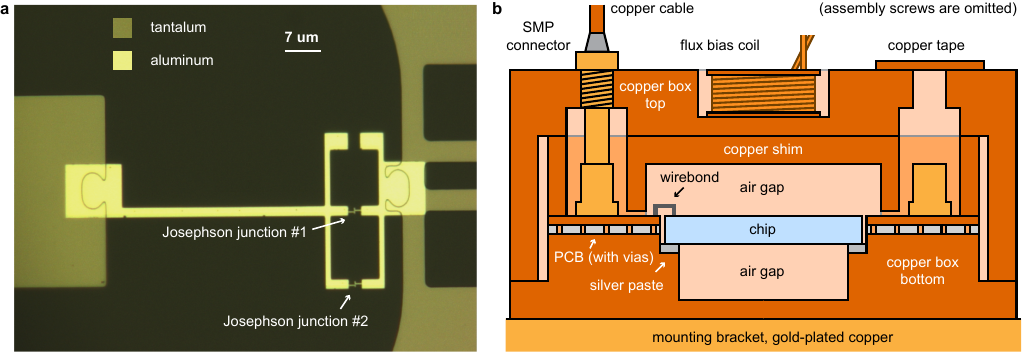}
    \caption{Device and packaging used in our work. Image taken from Ref.~\cite{kurilovich_high-frequency_2025}. (a) The Josephson junctions of our device. The Al/AlOx/Al junctions are arranged in a loop to form a superconducting quantum interference device (SQUID). SQUID is interfaced with the tantalum pads and the ground plane. The flux lines on the right from the SQUID and the additional SQUID arm are not used. The flux lines are shorted to the ground plane with wirebonds and the extra SQUID arm is left open. (b) Schematic of the package hosting the chip (scale not preserved).}
    \label{fig:fab}
  \end{center}
\end{figure}

The device described in the main text is deposited on an annealed EFG Sapphire substrate. The Josephson junctions are made out of aluminum via the bridge-free technique \cite{lecocq_junction_2011}. Large features of the device are made of tantalum \cite{place_new_2021, ganjam_surpassing_2024}. A zoom in on the region of the device containing the Josephson junctions is demonstrated in Figure~\ref{fig:fab}(a). The tantalum base layer is fabricated following the procedue from Ref.~\cite{ganjam_surpassing_2024}. The fabrication of the Josephson junctions follows Ref.~\cite{serniak_direct_2019} with two differences. First, to form the junctions, we oxidized the aluminum film at a higher pressure of 50 Torr for a longer duration of 30 minutes. Second, we used a different ion milling process compared to Ref.~\cite{serniak_direct_2019}. The process was similar to that of Ref.~\cite{ganjam_surpassing_2024}. 

The sapphire chip with our transmon is seated in a copper package, see Figure~\ref{fig:fab}(b) for a schematic. The chip is thermalized to the package with silver paste deposited in the corners. The package is placed in a shield protecting it from stray infrared radiation, see Ref.~\cite{connolly_coexistence_2024}. HERD-1 filter is used \cite{rehammar_low-pass_2023} to suppress infrared radiation incident from the readout channel.

\section{Derivation of $|m\rangle\rightarrow |m+2\rangle$ transition rate due to inelastic scattering}

As explained in the main text of the manuscript, the inelastic scattering of drive photons can cause the transmon to transition from one of the computational states $|m\rangle$ (where $m=\{0,1\}$) to a non-computational state $|m+2\rangle$. In this process, the frequency of the drive photon $\win$ is changed to $\wout = \win - \omega_{m+2,m}$, where $\hbar\omega_{m+2,m}=\epsilon_{m+2} - \epsilon_m$ and $\epsilon_m$ is the energy of state $|m\rangle$. The photon at frequency $\wout$ leaves into the environment of the transmon.  This process was the limiting factor for high-frequency dispersive readout that we recently put forward as an improvement over more standard dispersive readout, see Ref.~\cite{kurilovich_high-frequency_2025}.

Here, we present the detailed derivation of the rate of the $|m\rangle\rightarrow |m+2\rangle$ process due to the inelastic scattering. The core idea of the derivation is to treat the transmon non-linearity as a perturbation. It is conceptually similar to black-box quantization approach of Ref.~\cite{nigg_black-box_2012} extended to the case where modes have a continuous spectrum. The derivation consists of four parts. First, we decompose the operator of the phase $\varphi$ at the Josephson junction in terms of the normal modes of the \textit{linearized} Hamiltonian of the system. In this decomposition, it is crucial to account for the hybridization between the transmon and the environment modes. The normal mode decomposition is described in Section~\ref{sec:normal_mode}. Second, we use the obtained decomposition to compute the matrix element of the $T$-matrix between the scattering states. The $T$ matrix is computed in Born's approximation, \textit{i.e.}, to the lowest order in transmon non-linearity. Third, we employ the resulting matrix element in Fermi's golden rule for the transition rate. This yields the transition rate in terms of parameters characterizing the drive power and coupling between the transmon and the environment. Second and third steps are explained in Section~\ref{sec:T_matrix}. Finally, we relate parameters entering the expression for the rate to measurable quantities, such as AC Stark shift experienced by the qubit and the real part of the impedance $Z[\omega]$ of the transmon island. This step is explained in Section~\ref{sec:measurable_params}.
Throughout our derivations we assume zero temperature of the environment. This is justified since frequency $\wout/2\pi$ typical for our experiments is on the order of several GHz (the frequency scale associated to the fridge temperature is $\lesssim 0.4\:\mathrm{GHz}$). The results can be straightforwardly generalized to the case of a non-zero temperature.

\subsection{Decomposition of phase operator into normal modes of transmon and environment}
\label{sec:normal_mode}
As described above, the first step in our derivation is the decomposition of the operator $\varphi$ of the phase at the Josephson junction through normal modes of the linearized Hamiltonian. Let us start with the full version of the Hamiltonian. It reads
\begin{equation}
\label{eq:full_hamiltonian}
    H = H_{t} + H_{b} + V,\quad H_{t} = 4E_C^{(0)} (N - n_g)^2 - E_J \cos \varphi, \quad H_{b} = \sum_{k} \hbar\omega_k (b^\dagger_k b_k + 1/2), \quad V = \sum_{k} N(\lambda_k b^\dagger_k + \lambda_k^\star b_k).
\end{equation}
Here, $N$ is the operator of the number of Cooper pairs on the transmon island; this operator is conjugate to the phase operator, $\left[N, \varphi\right]=-i$. Parameter $E_C^{(0)}$ is the ``bare'' charging energy of the transmon. As will be described momentarily, this charging energy is renormalized by coupling between the transmon and the environment. In the charging energy term of the Hamiltonian, parameter $n_g$ is the transmon offset charge~\cite{koch_charge-insensitive_2007}. Parameter $E_J$ is the Josephson energy of the junction. Bosonic creation and annihilation operators $b_k^\dagger$ and $b_k$ describe the modes in the environment of the transmon. They satisfy the standard commutation relation $[b_k, b_k^\dagger] = 1$. Parameter $\omega_k$ is the frequency of the $k$-th environment mode. Term $V$ in the Hamiltonian describes the coupling between the transmon and the environment modes. Parameters $\lambda_k$ characterize the strength of this coupling for different modes. The presence of the readout resonator manifests as a peak in $|\lambda_k|^2$ centered around the resonator frequency. 

%we use Hamiltonian $H = H_{t} + H_{b} + V$ that models the transmon coupled to a transmission line through a readout resonator. Term $H_{t} = 4E_C (N - n_g)^2 - E_J \cos \varphi$ describes the transmon; $N$ is the Cooper pair number on the island and $\varphi$ is the superconducting phase. Term $H_{b} = \sum_{k} \hbar\omega_k (a^\dagger_k a_k + 1/2)$ describes the transmission line modes labeled by index $k$; $\omega_k$ and $a_k$ are the corresponding frequencies and annihilation operators, respectively. Term $V$ describes the coupling between the transmon and the transmission line mediated by the readout resonator, $V=\sum_{k}N(\lambda_k a^\dagger_k + \lambda_k^\star a_k)$. In what follows, we link the coupling constants $\lambda_k$ to the impedance $Z[\omega]$ introduced above.

Having introduced the notations, we linearize the transmon Hamiltonian given by equation \eqref{eq:full_hamiltonian}. This yields
\begin{equation}
\label{eq:lin_hamiltonian}
    H^{\rm lin}=\frac{8E_{C}^{(0)} N^2}{2} + \frac{E_J\varphi^2}{2}  + \sum_{k} \hbar\omega_k (a^\dagger_k a_k + 1/2) + \sum_k N(\lambda_k a^\dagger_k + \lambda_k^\star a_k).
\end{equation}
Note that we omitted the offset charge $n_g$ in the transmon Hamiltonian \cite{koch_charge-insensitive_2007}; its effects are not important for the rates of inelastic scattering in the limit $E_J/E_C \gg 1$ where the phase slips can be neglected. Next, we diagonalize the linearized Hamiltonian given by Eq.~\eqref{eq:lin_hamiltonian}.
To this end, we introduce the canonically conjugate ``phase'' and ``charge'' operators for the environment modes,
\begin{equation}
    N_k = \frac{a_k e^{i\phi_k} + a_k^\dagger e^{-i\phi_k}}{\sqrt{2}},\quad \varphi_k = \frac{i a_k e^{i\phi_k} - i a_k^\dagger e^{-i\phi_k}}{\sqrt{2}},\quad e^{i\phi_k} = \frac{\lambda_k}{|\lambda_k|}.
\end{equation}
In terms of these operators, we rewrite the Hamiltonian as
\begin{equation}
    H^{\rm lin}=\frac{8E_C^{(0)} N^2}{2} + \frac{E_J\varphi^2}{2} + \sum_k \sqrt{2}|\lambda_k| N_k N + \sum_k\hbar\omega_k\left(\frac{N_k^2}{2} + \frac{\varphi_k^2}{2}\right).
\end{equation}
The normal modes can then be found in a standard way by solving the equations of motion in the frequency domain. As a result, we can expand the phase operator in terms of the normal modes,
%We proceed by applying a canonical transformation
%\begin{equation}
%    \varphi = \tilde{\varphi} / E_J^{1/2}, \quad N = E_J^{1/2} \tilde{N},\quad \varphi_k = \tpk / (\hbar\omega_k)^{1/2}, \quad N_k = (\hbar \omega_k)^{1/2} \tilde{N}_k.
%\end{equation}
%This yields
%\begin{equation}
%    H=\frac{8E_JE_C^{(0)} \tn^2}{2} + \frac{\tp^2}{2} + \sum_k \sqrt{2 E_J \hbar\omega_k}|\lambda_k| \tnk \tn + \sum_k\left(\frac{(\hbar\omega_k)^2\tnk^2}{2} + \frac{\tpk^2}{2}\right).
%\end{equation}
%We diagonalize this Hamiltonian perturbatively, to the lowest order in $\lambda_k$. With this accuracy, we can neglect the influence of environment modes on each other. The problem, therefore, reduces to that of \textit{two} coupled oscillators. As a result, we find
%\begin{equation}
%    \tp = \sqrt{\frac{\hbar\wq}{2}} i(a - a^\dagger) + \sum_k \frac{\sqrt{2E_J \hbar\omega_k}|\lambda_k|}{(\hbar\omega_k)^2 - 8 E_J E_C^{(0)}}\tpk.
%\end{equation}
%Here, the qubit frequency $\wq$ includes the loading due to the coupling to environment,
\begin{equation}
    \label{eq:phase_decomposition}
    \ph = \sqrt{\frac{\hbar\wq}{2E_J}} i(a - a^\dagger) + \sum_k (ia_k \mu_k^\star - i a_k^\dagger \mu_k), \quad \mu_k = \frac{\hbar\omega_k \lambda_k}{(\hbar\omega_k)^2 - (\hbar\wq)^2},
\end{equation}
where
\begin{equation}
    \label{eq:ec_renorm}
    \hbar\wq = \sqrt{8 E_J E_C},\quad E_C = E_C^{(0)} - \sum_k \frac{1}{4}\frac{\hbar\omega_k |\lambda_k|^2}{(\hbar\omega_k)^2 - (\hbar \wq)^2}.
\end{equation}
Here, operator $a$ describes the qubit mode dressed by its coupling to the environment; operators $a_k$ describe the dressed environment modes. Eq.~\eqref{eq:ec_renorm} describes the renormalization of the charging energy of the transmon due to its coupling to the environment.

We note that because of the Purcell effect the qubit mode slowly decays with time. In Eq.~\eqref{eq:phase_decomposition}, we batched quasi-elastic levels in the vicinity of the qubit frequency into a single qubit mode $a$. Correspondingly, the sum in Eq.~\eqref{eq:phase_decomposition} excludes modes with frequencies within the decay linewidth of the qubit frequency. The separation between the qubit and the environment in Eq.~\eqref{eq:phase_decomposition} allows us to consider initial states where the qubit is excited.

%Here, the qubit frequency $\wq$ includes the loading due to the coupling to environment,
%where we changed $8 E_J E_C^{(0)}$ in the denominator to $(\hbar\wq)^2$, which is allowed within the considered accuracy. Equation~\eqref{eq:phase_decomposition} allows us to compute the matrix elements of the $T$-matrix as we describe in the next section.

%Qubit decays slowly. Batch-quasi-elastic levels into a qubit mode. Sum excludes states within qubit on the scale gamma. Allows us to consider the problem where the initial state describes the excited qubit.

\subsection{Matrix element of the $T$-matrix and the transition rate}
\label{sec:T_matrix}
Next, as described in the main text, we need to evaluate the matrix element of the $T$-matrix between the scattering states. The initial state of the system is $|\Psi_{i}\rangle \propto (a_{\rm in}^\dagger)^{n_{\rm in}} |\Omega\rangle |m\rangle$, where $|\Omega\rangle$ is the vacuum state of the transmission line and $|m\rangle$ is the state of the transmon; the photon occupation number $n_{\rm in}$ and the mode with frequency $\omega_{\rm in}$ characterize the applied drive (index $k=\mathrm{in}$ corresponds to the mode with frequency $\win$). The final state, where one of the drive photons is inelastically scattered, is $|\Psi_{f}\rangle\propto (a_{\rm in}^\dagger)^{n_{\rm in}-1}a_{\rm out}^\dagger|\Omega\rangle |m+2\rangle$. Here, a new mode at frequency of the outgoing photon, $\wout = \win - \omega_{m+2,m}$, is populated (we denoted the corresponding state index as $k=\mathrm{out}$).
To the lowest order in non-linearity, the $T$-matrix is given by $-E_J \varphi^4/4!$. Then, using the normal mode decomposition given by Eq.~\eqref{eq:phase_decomposition} we find the matrix element
\begin{gather}
    |\mathcal{M}_{fi}| 
 =|\langle\Psi_f|E_J \varphi^4 / 4!|\Psi_i\rangle| = \frac{2!}{4!}
    \begin{pmatrix}
        4\\2
    \end{pmatrix}
    E_J \sqrt{\frac{\hbar\wq}{2E_J}}\sqrt{\frac{\hbar\wq}{2E_J}} |\mu_{\rm in}| |\mu_{\rm out}| |\langle\Psi_i|a^\dagger a^\dagger a^\dagger_{\rm out} a_{\rm in}|\Psi_{f}\rangle| = \notag \\ = \sqrt{m+1}\sqrt{m+2}\frac{1}{2}\frac{1}{2} \hbar\wq \sqrt{n}_{\rm in} |\mu_{\rm in}||\mu_{\rm out}|.\label{eq:matrix_element_final}
\end{gather}
Then, using Fermi's golden rule, we can evaluate the rate of inelastic scattering and thus the qubit $|m\rangle\rightarrow |m+2\rangle$ transition rate. We obtain
\begin{equation}
    \label{eq:rate_weird_params}
    \Gamma_{m\rightarrow m+2} = (m+1)(m+2) \frac{2\pi}{\hbar} \nu_{\rm out} \frac{\hbar^2\wq^2}{16} |\mu_{\rm in}|^2 |\mu_{\rm out}|^2 n_{\rm in}
\end{equation}
where $\nu_{\rm out}$ is the density of states in the environment at the frequency of the outgoing photons.

In the described calculation, we disregarded the potential degeneracy of the mode with $k=\mathrm{out}$. In the presence of such a degeneracy, $|\mu_{\rm out}|^2$ in Eq.~\eqref{eq:rate_weird_params} should be averaged over the iso-energetic surface.
 %%%%%%%%%%%%%%%%%%%%%%%%%%%%%%%%%%%%%%%%%%%%%%%%%%%%%%%%%%%%%%%%%%%%%%%%%%%%%%%%%%%%%%%%%%%%%%%%%%%%%%%%%%%%%%%%%%%%%%%%%%%%
\subsection{Expressing the final result for the $|m\rangle \rightarrow |m+2\rangle$ rate in terms of measurable quantities}
\label{sec:measurable_params}
Equation~\eqref{eq:rate_weird_params} expresses the qubit state transition rate in terms of parameters of the drive, of the qubit itself, and of its environment. For example, the number of photons in the incoming mode, $n_{in}$, is directly proportional to the drive power. However, neither of the parameters $n_{\rm in}$, $\mu_{\rm in}$, $\mu_{\rm out}$, and $\nu_{\rm out}$ is directly measurable in our experiment. In the present section, we relate these model parameters to physical observables that can be directly measured. We first show that $|\mu_{\rm in}|^2 n_{\rm in}$ is directly proportional to the AC Stark shift experienced by the qubit due to the presence of the drive photons. Stark shift can easily be measured experimentally, see, for example, Section~\ref{sec:ac_stark}. The combination $\nu_{\rm out}|\mu_{\rm out}|^2$ can be linked to the impedance of the transmon at the frequency of the outgoing photons $\wout$ and certain combination of frequencies of the drive and of the qubit.
\subsubsection{Relation between the parameters of the drive and the transmon AC Stark shift}
To evaluate the AC Stark shift $\delta\omega$ of the $|0\rangle \rightarrow |1\rangle$ transition of our transmon, we treat its non-linearity perturbatively. The term in the Hamiltonian describing the lowest order correction to the spectrum due to the nonlinearity is given by $H^{\rm non-lin} = - E_J \varphi^4/4!$. Substituting the decomposition of the phase operator given by Eq.~\eqref{eq:phase_decomposition} into $H^{\rm non-lin}$ and leaving only resonant terms, we obtain
\begin{equation}
    H^{\rm non-lin} = -\frac{E_J}{4!} 4! \sqrt{\frac{\hbar\wq}{2E_J}} \sqrt{\frac{\hbar\wq}{2E_J}} |\mu_{\rm in}|^2 a_{\rm in}^\dagger a_{\rm in}a^\dagger a + ...
\end{equation}
Exchanging $a_{\rm in}^\dagger a_{\rm in}$ for the number of photons in the drive mode $n_{\rm in}$, we find the desired Stark shift is \textit{negative}, its absolute value is
\begin{equation}
    \label{eq:stark}
    \delta\omega = \frac{\wq}{2} |\mu_{\rm in}|^2 n_{\rm in}.
\end{equation}
This combination of parameters explicitly enters the expression for the rate of inelastic scattering, Eq.~\eqref{eq:rate_weird_params}. 
\subsubsection{Relation between the parameters of the environment and the impedance of the transmon island}
Next, we relate parameters $\nu_{\rm out}$ and $\mu_{\rm out}$ in Eq.~\eqref{eq:rate_weird_params} to the real part of the impedance of the transmon island. The impedance is measured with respect to ground, see Figure 1 of the main text.
According to Kubo formula, the impedance can be expressed through the correlation function of the flux operator $\Phi$ as
\begin{equation}
    \label{eq:Z_and_chi}
    Z[\omega] = \frac{\omega}{\hbar}\int_0^\infty dt \langle[\Phi(t),\Phi(0)]\rangle e^{i(\omega + i\epsilon) t}.
\end{equation}
Next, we substitute $\varphi = 2\pi \Phi / \Phi_0$ (where $\Phi_0 = h/2e$ is the flux quantum) in Eq.~\eqref{eq:Z_and_chi}. Then, we use the expansion of the phase operator through the normal modes, Eq.~\eqref{eq:phase_decomposition}. This yields
%\begin{equation}
%        \chi[\omega] = -\frac{i}{\hbar}\int_0^\infty dt \left(\pzpf^2[e^{i\wq t} a_q^\dagger + e^{-i\wq t} a_q, a_q^\dagger + a_q] + \sum_k \mu_k^2 [e^{i\omega_k t} a_k^\dagger + e^{-i\omega_k t} a_k, a^\dagger_k + a_k]\right) e^{i(\omega + i\epsilon) t}
%\end{equation}
\begin{equation}
    \label{eq:chi_integral}
    Z[\omega] = \frac{\omega}{\hbar} \left(\frac{\Phi_0}{2\pi}\right)^2 \int_0^\infty dt \left(\pzpf^2 (e^{-i\wq t} - e^{i\wq t}) + \sum_k |\mu_k|^2 (e^{-i\omega_k t} - e^{i\omega_k t} ) \right) e^{i(\omega + i\epsilon) t}.
\end{equation}
For our purposes, only the \textit{real} (dissipative) part of impedance is important. By performing the integration in Eq.~\eqref{eq:chi_integral}, we obtain
\begin{equation}
        \mathrm{Re}Z[\omega] =\frac{\omega}{\hbar} \left(\frac{\Phi_0}{2\pi}\right)^2 \sum_k \pi |\mu_k|^2 \delta(\omega - \omega_k).
\end{equation}
We can then compute the dissipative part of the impedance of the circuit as
\begin{equation}
\label{eq:ReZ_vs_stuff}
        \mathrm{Re}Z[\omega_{\rm out}] = \frac{\hbar\omega_{\rm out}}{2}\left(\frac{\Phi_0}{2\pi}\right)^2  \frac{2\pi}{\hbar}  |\mu_{\rm out}|^2 \nu_{\rm out}.
\end{equation}
Substituting Eq.~\eqref{eq:stark} and Eq.~\eqref{eq:ReZ_vs_stuff} into Eq.~\eqref{eq:rate_weird_params}, we obtain the final result for the rate, Eq.~1 of the main text,
\begin{equation}
\label{eq:raman_final}
    \Gamma_{m\rightarrow m+2} = (m+1)(m+2)  \frac{\wq}{\wout}\frac{2\pi\mathrm{Re}Z[\omega_{\rm out}]}{R_Q} \delta\omega.
\end{equation}

\begin{figure}[h]
  \begin{center}
    \includegraphics[scale = 1]{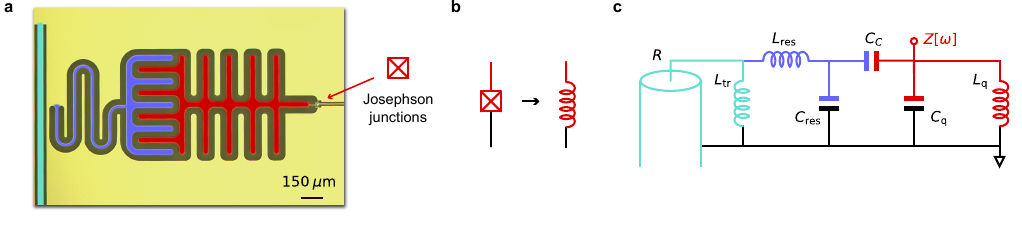}
    \caption{Lumped-element model of our device used to evaluate the impedance $Z[\omega]$ of the transmon island. The real part of this impedance determines the rate of inelastic scattering as described by Eq.~\eqref{eq:raman_final}. (a) False-color microscope image of the device. Quarter-wavelength readout resonator (blue) is capacitively coupled to the transmon island (red). The resonator is inductively coupled to the transmission line (cyan). (b) To evaluate the impedance, we replace the Josephson junctions of our device with linear inductors. (b) The resulting lumped-element model. }
    \label{fig:lumped}
  \end{center}
\end{figure}

\subsection{Calculation of the impedance $Z[\omega]$ for our device}
The rate of qubit transitions caused by inelastic scattering of drive photons, Eq.~\eqref{eq:raman_final}, depends on the real (dissipative) part of impedance of the transmon island $Z[\omega]$. Here, the dissipation comes from coupling to the transmission line. In this section, we determine $\mathrm{Re}Z[\omega]$ of our device using classical circuit analysis. To this end, we employ a lumped element model of a transmon capacitively coupled to a readout resonator, see Fig.~\ref{fig:lumped}. In this model, we replace the Josephson junction of our transmon with a linear inductor. This replacement is well justified in the regime of small anharmonicity, $E_J\gg E_C$. We model the transmission line as a resistor with resistance $R = 50\:\Omega$. This resistor is \textit{inductively} coupled to the readout resonator.

Within our lumped-element model the qubit and the resonator frequencies are (see Figure~\ref{fig:lumped} for the definition of circuit parameters)
\begin{equation}
    \wres^2 = \frac{\cq + C_C}{\lres C_\Sigma^2},\quad \wq^2 = \frac{1}{\lqb(\cq + C_C)},\quad C_\Sigma^2 = \cres \cq + \cres C_C + \cq C_C.
\end{equation}
Here, we neglected corrections of order $\eta^2(\wq/\wres)^2$, where the coupling efficiency $\eta$ is defined as
\begin{equation}
\label{eq:eta}
    \eta = \frac{C_C}{\sqrt{(\cres + C_C)(\cq+C_C)}},\quad 0 < \eta < 1.    
 \end{equation}
Neglecting corrections of the same order, we can determine the real part of the impedance $Z[\omega]$ in the vicinity of the resonator frequency. We obtain\footnote{The real part of the impedance has an extra peak at the transmon frequency (in addition to resonator frequency). Within our model, the width of this peak is determined by Purcell effect \cite{kurilovich_high-frequency_2025}. We assume that the frequency is sufficiently detuned from this peak.}
\begin{equation}
\label{eq:rez_lump}
    \mathrm{Re}Z[\omega]=\frac{C_{C}^{2}}{(\cq+C_{C})^{2}}\frac{\left(\omega L_{\rm res}\right)\left({\omega L_{\mathrm{tr}}^{2}}/{L_{\rm res}R}\right)}{\left(1-\omega^{2}/{\wres^{2}}\right)^{2}+\omega^{2}L_{\mathrm{tr}}^{4}/L_{\rm res}^{2}R^{2}},
\end{equation}
where $L_\mathrm{tr}$ is the coupling inductor connecting the readout resonator to the transmission line (see Fig.~\ref{fig:lumped}). In deriving equation Eq.~\eqref{eq:rez_lump}, we additionally assumed $L_\mathrm{tr} \ll \lres, R/\omega$. Equation~\eqref{eq:rez_lump} can be alternatively rewritten as 
\begin{equation}
\label{eq:rez_massaged}
    \frac{\pi\mathrm{Re}Z[\omega]}{R_Q} = \frac{\eta^2}{1-\eta^2}\frac{E_C}{\hbar}\frac{\omega^2 \kappa}{(\omega^2 - \omega_{\rm res}^2)^2 + \omega^2 \kappa^2}
\end{equation}
where $\kappa$ is the linewidth of the readout resonator. Substituting $\mathrm{Re}\:Z[\omega]$ given by equation~\eqref{eq:rez_massaged} into Eq.~\ref{eq:raman_final} yields the rate of undesired transitions.

\subsection{Simplified expression for inelastic scattering rate in the case of a high-frequency readout}
Notably, in the regime $\wres \gg \wq$ and $\eta^2\ll1$ -- which is relevant for our device -- Eq.~\eqref{eq:raman_final} for the rate of qubit transitions can be substantially simplified. To this end, let us assume that the drive is applied at the frequency of the resonator $\wres$. Using the fact that the anharmonicity of the transmon is weak, we can approximate $\wout \approx \wres - 2\wq$. This allows us to approximate the impedance in Eq.~\eqref{eq:rez_massaged} as 
\begin{equation}
\label{eq:simpleZ}
        \frac{\pi\mathrm{Re}Z[\wout]}{R_Q} = \eta^2\frac{E_C}{\hbar}\frac{\wout^2 \kappa}{16 \wq^2 \wout^2} = \eta^2\frac{E_C}{\hbar}\frac{\kappa}{16 \wq^2}
\end{equation}
Here, $\kappa$ is defined in Eq.~\eqref{eq:rez_massaged} and $\eta$ is defined in Eq.~\eqref{eq:eta}. Then, substituting Eq.~\eqref{eq:simpleZ} into Eq.~\eqref{eq:raman_final}, we obtain 
\begin{equation}
\label{eq:raman_interm}
    \Gamma_{m\rightarrow m+2} = (m+1)(m+2)  \frac{\wq}{\wout}2\eta^2\frac{E_C}{\hbar}\frac{\kappa}{16 \wq^2} \delta\omega.
\end{equation}
Next, we use the relation connecting the coupling efficiency $\eta$ to the dispersive shift $\chi$ (see supplement of Ref.~\cite{kurilovich_high-frequency_2025} for details),
\begin{equation}
\chi=2\eta^2\frac{E_C}{\hbar} \frac{\wq}{\wres}.
\end{equation}
Substitution of this relation into equation \eqref{eq:raman_interm} yields the simplified expression for the transition rate in terms of measurable quantities:
\begin{equation}
    \label{eq:simple_raman}
    \Gamma_{m\rightarrow m+2} = (m+1)(m+2)  \frac{\kappa \chi}{16\wq^2} \delta\omega
\end{equation}
According to Eq.~\eqref{eq:simple_raman} the attempt to improve the qubit readout by increasing $\kappa$ and $\chi$ is heavily penalized by the increase of the inelastic scattering rate.
%\subsubsection{Inelastic scattering as a limitation for fast readout}
%Now, let us assume the optimal condition for qubit readout, $\chi = \kappa$ \cite{clerk_introduction_2010}. Under this condition, in the steady-state, the rate of acquiring information about the qubit state is simply quantified by the AC Stark shift \cite{kurilovich_high-frequency_2025}. Indeed, both the Stark shift and the measurement rate scale as $\propto \chi \bar{n}$ where $\bar{n}$ is the number of photons in the resonator. Thus, let us fix the steady state value of the AC Stark shift. The steady state is achieved in $t\sim 1/\kappa$. It is thus beneficial to increase $\kappa$. However, as can be seen 

%Then, for a fixed value of AC Stark shift $\delta \omega$ (which yields a fixed rate of acquiring information about the qubit state) and for a fixed qubit frequency $\wq$ the rate of transitions linearly increases with both $\kappa$ and $\chi$. For an optimal readout $\chi = \kappa$. Therefore, an attempt to improve the readout by increasing the dispersive shift will be heavily penalized by $\Gamma_{m\rightarrow m+2}\propto \chi^2$.

\section{Rates of other lowest-order inelastic scattering processes}

As mentioned in the main text, the Raman scattering described above is not the only inelastic process that can cause qubit state transitions. Other inelastic processes are possible. The goal of the present section is to investigate them and build their systematic theory.

Throughout this section, we consider inelastic processes of the same order in nonlinearity as the Raman scattering. There are two possible processes of this kind, each of them is of a four-wave mixing type.  In the first process, two drive photons with frequency $\win$ convert into a qubit $|m\rangle$ to $|m-1\rangle$ transition (or $|m\rangle$ to $|m+1\rangle$ transition) and a photon emitted at a frequency $\wout = 2\win + \omega_{m,m-1}$ ($\wout = 2\win - \omega_{m+1,m}$). Experimentally, this process is especially concerning because it can couple the qubit to its electromagnetic environment at frequency much higher than both $\wq$ and $\wres$. Such a high-frequency environment is often not well controlled. As we demonstrate, a spurious mode in the electromagnetic environment can be detrimental for the readout performance. This type of inelastic scattering is explored in Section~\ref{sec:two_photon} and the influence of spurious environment modes in discussed in Section~\ref{sec:two_photons_plus_mode}.

\begin{figure}[h]
  \begin{center}
    \includegraphics[scale = 1]{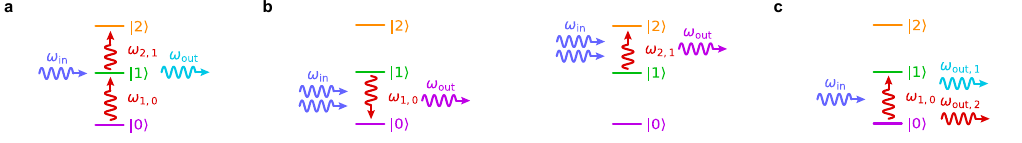}
    \caption{Different possible inelastic scattering processes allowed by the transmon non-linearity. To the lowest-order in nonlinearity, only four wave mixing processes are allowed. (a) Raman scattering described in the main text. An incoming photon gives two quanta of energy to the qubit and reflects at a lower frequency. Such a Raman scattering is the dominant leakage mechanism limiting the readout performance in our device. The rate of the Raman scattering is given by Eq.~\eqref{eq:raman_final}. (b) In a different process, two drive quanta combine with a qubit excitation and reflect as a single photon of a higher frequency. The rate of this process is given by Eq.~\eqref{eq:two_photon_relax}. A similar ``two-photon'' process can also excite the qubit. The rate is also given by Eq.~\eqref{eq:two_photon_relax} except a different outgoing frequency $\wout$ is needed. (c) A single photon can excite the qubit and convert to two photons leaving into the environment. The rate of this process is given by Eq.~\eqref{eq:ReZReZ}.}
    \label{fig:inelastic_processes}
  \end{center}
\end{figure}

In the second process, a single drive photon converts into a qubit transition $|m\rangle$ to $|m+1\rangle$ (or $|m\rangle$ to $|m-1\rangle$) and \textit{two} photons emitted into the transmission line. As we explain, this type of inelastic scattering is strongly suppressed, at least in the context of readout. This allowed us to ignore it in the main text of the manuscript. The described process is explored in Section~\ref{sec:one_photon_to_two}. The three possible types of inelastic processes that we consider -- the Raman process and the two processes introduced in this section -- are summarized in Figure~\ref{fig:inelastic_processes}.

\subsection{Two-photon inelastic scattering process leading to $|m\rangle\rightarrow |m-1\rangle$ or $|m\rangle\rightarrow |m+1\rangle$}
\label{sec:two_photon}
We set out by computing the rate of an inelastic scattering process in which two drive photons and a qubit excitation are converted to a photon going out into the environment at frequency $\wout = 2\win + \omega_{m, m-1}$. The derivation is conceptually similar to that of Eq.~\eqref{eq:raman_final}.

To describe this process, similarly to how it was done for the Raman scattering, we compute the $T$-matrix in Born's approximation, $T=-E_J \varphi^4 / 4!$. In the notations of Section~\ref{sec:T_matrix}, we take the initial state to be $|\Psi_{i}\rangle \propto (a_{\rm in}^\dagger)^{n_{\rm in}} |\Omega\rangle |m\rangle$. The final state is $|\Psi_{f}\rangle\propto (a_{\rm in}^\dagger)^{n_{\rm in}-2}a_{\rm out}^\dagger|\Omega\rangle |m-1\rangle$. The matrix element of the $T$-matrix can then be computed in the same way as in Eq.~\eqref{eq:matrix_element_final}. The result of the calculation is
\begin{equation}
\label{eq:M_two_photon}
|\mathcal{M}_{if}| = \frac{1}{2}E_J \pzpf |\mu_{\rm out}| |\mu_{\rm in}|^2 n_{\rm in}.
\end{equation}
We can employ Fermi Golden rule to evaluate the rate,
\begin{equation}
    \label{eq:rate_via_M}
    \Gamma_{m\rightarrow m-1} = \frac{2\pi}{\hbar} \nu_{\rm out} |\mathcal{M}_{if}|^2.
\end{equation}
We then substitute equation \eqref{eq:M_two_photon} into equation \eqref{eq:rate_via_M} to obtain the rate. Further, we relate the result for the rate to measureable quantities using Eq.~\eqref{eq:stark} and Eq.~\eqref{eq:ReZ_vs_stuff}. In this way, obtain
\begin{equation}
    \label{eq:two_photon_relax}
    \Gamma_{m\rightarrow m-1} = m\frac{\wq}{\omega_{\rm out}}\frac{\pi\mathrm{Re}Z[\omega_{\rm out}]}{R_Q}\frac{\delta\omega^2}{E_C/\hbar}.
\end{equation}
Notably, the rate scales \textit{quadratically} with the drive power. This is in contrast to the previously discussed Raman processes. The rate of the Raman transition scales \textit{linearly} with power.

\subsubsection{Simplified expression for the rate of the two-photon process in the case of high-frequency readout}

Here, we derive a simplified expression for the rate of the discussed two-photon process under a set of assumptions. First, we focus on the case of a high-frequency readout, $\wres\gg\wq$. Second, we assume that the coupling between the readout resonator and the qubit is weak, $\eta^2\ll1$. Finally, we assume that the dissipation into the environment is solely determined by coupling to the transmission line through the readout resonator. Therefore, we employ Eq.~\eqref{eq:rez_massaged} to model the environment of the transmon [this assumption is lifted in Section~\ref{sec:two_photons_plus_mode}]. Under these assumptions Eq.~\eqref{eq:two_photon_relax} simplifies to
\begin{equation}
    \label{eq:simple_rate}
    \Gamma_{m\rightarrow m-1} = \frac{2m}{9}\frac{\kappa\chi}{\wres^2}\frac{\delta\omega^2}{E_C/\hbar}.
\end{equation}
We note that for our typical readout, see Ref.~\cite{kurilovich_high-frequency_2025}, the drive power is such that $\delta\omega \sim E_C/\hbar$. At such powers, the rate of the two-photon process is strongly suppressed compared to the rate of the Raman process [compare Eq.~\eqref{eq:simple_rate} with Eq.~\eqref{eq:simple_raman}].

\subsubsection{Two-photon process in the presence of spurious modes in the device}
\label{sec:two_photons_plus_mode}

The process captured by Eq.~\eqref{eq:two_photon_relax} couples the transmon to its electromagnetic environment at frequency $\wout = 2\win + \wq$. In case of dispersive readout, $\win = \wres$, frequency scale $\wout$ significantly exceeds both the qubit frequency and the frequency of the readout resonator. This coupling is problematic since the high-frequency environment can contain uncontrolled spurious modes. Here, we simplify Eq.~\eqref{eq:two_photon_relax} in the presence of a mode in the transmon environment. We show that if the frequency $\wout$ matches that of a mode, the rate of qubit relaxation is enhanced; we relate this enhancement to the parameters of the mode.

Let us assume for simplicity that frequency $\wout$ is close to that of an spurious mode $\omega_s$. Let $\kappa_s$ be the linewidth of the mode and $g_s$ be the coupling between the transmon and the mode. The contribution of the spurious mode to the impedance ``seen'' by the qubit can be expressed as
%Practically this leads to the presence of the modes that manifest as sharp peaks in $\mathrm{Re}\:Z[\omega]$ at certain frequencies $\omega_s$. In the frequency of the outgoing radiation $\wout$ accidentally matches that of one of the modes, the rate of the undesired transitions is vastly enhanced. To describe this effect we assume that the outgoing radiation is in the vicinity of one of the mode. We assume that the linewidth of the mode is $\Gamma$. We also assume that the strength of the coupling between the qubit and the mode is $g_s$. In that case we obtain
\begin{equation}
    \label{eq:impedance_mode}
    \frac{\wq}{\omega}\frac{2\pi\mathrm{Re}Z[\omega]}{R_Q} = \frac{4E_C}{\hbar} \frac{{\kappa_s}/{2}}{\left({\kappa_s}/{2}\right)^2 + (\omega-\omega_s)^2}\frac{g^2_s\omega^2}{(\omega^2-\wq^2)^2}.
\end{equation}
Combining Eq.~\eqref{eq:impedance_mode} with Eq.~\eqref{eq:two_photon_relax} we obtain
\begin{equation}
    \Gamma_{m\rightarrow m-1} = m \frac{\kappa_s}{\left({\kappa_s}/{2}\right)^2 + (\wout-\omega_s)^2}\frac{g^2_s\wout^2}{(\wout^2-\wq^2)^2}\delta\omega^2.
\end{equation}
Exactly on the resonance, i.e., when $\wout = \omega_s$
\begin{equation}
    \Gamma_{m\rightarrow m-1} = m \frac{4}{\kappa_s}\frac{g^2_s\wout^2}{(\wout^2-\wq^2)^2}\delta\omega^2.
\end{equation}
%Let us take $m=1$, $\delta\omega/2\pi \sim 100\:\mathrm{MHz}$, $\wout/2\pi\sim 10\:\mathrm{GHz}$. This corresponds to the parameters of the readout experiment of Ref.~[]. For the sake of argument, let us take a weakly coupled mode, $g_m/2\pi \sim 1\:\mathrm{MHz}$. Assuming that the mode has linewidth $\Gamma/2\pi \sim 1\mhz$. In that case, we obtain $\Gamma_{m\rightarrow m-1} \sim 1\:\mathrm{ms}^{-1}$, comparable to the bare decoherence of our transmon. Importantly, the linewidth and the coupling strength used in the estimate above are rather typical for strongly-coupled material defects []. 

\subsection{Inelastic scattering process with two photons leaving into the environment}
\label{sec:one_photon_to_two}
Another possible inelastic scattering process takes a single photon from the drive, excites (or relaxes) the qubit to a neighboring level and produces \textit{two photons} leaving into the environment, see Figure~\ref{fig:inelastic_processes}(d). In this section, we compute the rate of the described process and show that this rate scales linearly with power. This qualitative behavior is similar to that of the Raman process, Figure~\ref{fig:inelastic_processes}(a). In the context of readout, however, the prefactor in the linear dependence is much smaller for the discussed process of compared to the Raman process.
%In fact, for the range of powers considered in the main text, this process is negligible compared to excitation and relaxation rates of our qubit in the absence of the drive.
For this reason, we omitted this process from the discussion of the main text.

%We demonstrate that this process is irrelevant for readout performance for the parameters of our circuit. The reason for why the rate of this process is small compared to the rate of Raman process is that is that the discussed process involves coupling to the transmission line twice. As such, compared to other of the discussed processes this process is suppressed by an extra factor of $\kappa/\wq$.
\subsubsection{Drive-induced $|m\rangle\rightarrow|m+1\rangle$ process with two photons leaving into the environment}

We set out by considering the excitation process where the transmon transitions from $|m\rangle$ to $|m+1\rangle$. The matrix element for this process is given by
\begin{equation}
    \label{eq:matr_elem_two_photon}
    |\mathcal{M}_{fi}| = |\langle\Psi_f|E_J \varphi^4 / 4!|\Psi_i\rangle|,\quad |\Psi_i\rangle \propto (a^\dagger_{\rm in})^{n_{\rm in}}|m\rangle |\Omega\rangle,\quad |\Psi_f \rangle \propto (a^\dagger_{\rm in})^{n_{\rm in}-1} a^\dagger_{\rm out_1}a^\dagger_{\rm out_2}|m+1\rangle |\Omega\rangle.
\end{equation}
We proceed by substituting the decomposition of the phase operator given by Eq.~\eqref{eq:phase_decomposition} into Eq.~\eqref{eq:matr_elem_two_photon}. This allows us to find the transition rate via the Fermi's golden rule,
\begin{equation}
    \Gamma_{m\rightarrow m + 1} = (m+1)\frac{2\pi}{\hbar} \frac{1}{2}\sum_{{\rm out_1}, {\rm out_2}} E_J^2 \frac{\hbar\wq}{2E_J}|\mu_{\rm out_1}|^2|\mu_{\rm out_2}|^2|\mu_{\rm in}|^2 n_{\rm in}\delta( \hbar\win - \hbar \omega_{m+1,m} - \hbar\omega_{\rm out_1} - \hbar\omega_{\rm out_2}).
\end{equation}
Next, we relate parameters entering this expression to measurable quantities using Eqs.~\eqref{eq:stark} and~\eqref{eq:ReZ_vs_stuff}. Changing the sums into integrals, we obtain
\begin{equation}
    \label{eq:ReZReZ}
    \Gamma_{m\rightarrow m + 1} = (m+1) 64\pi \frac{E_J}{\hbar} \delta\omega \int d\omega_{\rm out_1} d\omega_{\rm out_2} \frac{\mathrm{Re}Z[\omega_{\rm out_2}]}{\omega_{\rm out_2} R_Q} \frac{\mathrm{Re}Z[\omega_{\rm out_1}]}{\omega_{\rm out_1} R_Q}\delta(\hbar\win -\hbar \omega_{m+1,m} - \hbar\omega_{\rm out_1} - \hbar\omega_{\rm out_2}).
\end{equation}
Resolving the delta-function, we arrive at
\begin{equation}
    \label{eq:ReZReZ}
    \Gamma_{m\rightarrow m +1 } = (m+1) 64\pi \frac{E_J}{\hbar} \delta\omega \int_0^{\win - \omega_{m+1,m}} d\omega_{\rm out_1} \frac{\mathrm{Re}Z[\hbar\win -\hbar \omega_{m+1,m} - \hbar\omega_{\rm out_1}]}{(\hbar\win -\hbar \omega_{m+1,m} - \hbar\omega_{\rm out_1}) R_Q} \frac{\mathrm{Re}Z[\omega_{\rm out_1}]}{\omega_{\rm out_1} R_Q}.
\end{equation}
When dealing with Eqs.~\eqref{eq:ReZReZ}, we encounter a subtlety. Namely, the real part of the impedance contains a sharp peak at the frequency of the transmon. This peak corresponds to the Purcell-broadened qubit mode. The width of that peak is determined by the inverse of the Purcell-limited qubit lifetime $T_1$. This additional peak should be \textit{disregarded} when computing the rates of inelastic scattering. Accounting for the peak corresponds to a process where in addition to a qubit transition $|m\rangle\rightarrow |m+1\rangle$, another photon is emitted into the Purcell-broadened qubit mode. Such a transition corresponds to the process $|m\rangle \rightarrow |m+2\rangle$ described by Eq.~\eqref{eq:raman_final}. The peak should be disregarded to avoid double-counting.

Next, we estimate the integral in Eq.~\eqref{eq:ReZReZ} in the context of readout performance of our device. To this end, we assume that the incoming signal is applied at the resonator frequency, $\win = \wres$. Then, we substitute the impedance given by Eq.~\eqref{eq:rez_massaged} in Eq.~\eqref{eq:ReZReZ}. The integrand varies little across the integration range. This allows us to estimate (assuming additionally $\eta^2\ll 1$)
\begin{equation}
    \Gamma_{m\rightarrow m+1}\sim \frac{1}{\hbar}E_J \delta\omega \wres \left(\eta^2 \frac{E_C}{\hbar} \frac{\kappa}{\wres^3}\right)^2 \sim \delta\omega\frac{\chi^2}{E_C/\hbar} \frac{\kappa^2}{\wres^3}. 
\end{equation}
For our typical readout parameters $\delta\omega\sim E_C/\hbar$. In that case we can estimate the rate as $\Gamma_{m\rightarrow m + 1}\sim 10^{-5}\:\mathrm{s}^{-1}$.

\subsubsection{Drive-induced $|m\rangle\rightarrow|m-1\rangle$  process with two photons leaving into the environment}
Now we consider a similar process where the transmon transitions $|m\rangle \rightarrow |m-1\rangle$ and two photons are emitted into the environment. Similarly to Eq.~\eqref{eq:ReZReZ}, we obtain
\begin{equation}
    \label{eq:ReZReZ_relax}
    \Gamma_{m\rightarrow m - 1 } = m 64\pi \frac{E_J}{\hbar} \delta\omega \int_0^{\win + \omega_{m,m-1}} d\omega_{\rm out_1} \frac{\mathrm{Re}Z[\hbar\win +\hbar \omega_{m,m-1} - \hbar\omega_{\rm out_1}]}{(\hbar\win +\hbar \omega_{m,m-1} - \hbar\omega_{\rm out_1}) R_Q} \frac{\mathrm{Re}Z[\omega_{\rm out_1}]}{\omega_{\rm out_1} R_Q}
\end{equation}
Again, the Purcell-broadened peak in the impedance at the qubit frequency should be disregarded in Eq.~\eqref{eq:ReZReZ_relax}. Accounting for this peak corresponds to quasi-elastic scattering of the incoming photon of frequency $\omega_{\rm in}$. The contribution of the peak to the total photon scattering cross-section is proportional to the square of reflection phase shift (that may be used as the readout signal). Therefore, when computing the scattering rate, it is appropriate to assume that $Z[\omega]$ is given by Eq.~\eqref{eq:rez_massaged} at all frequencies. Under this assumption we can estimate the integral as 
%Now we estimate this integral in the context of readout for our circuit depicted in Fig.~\ref{fig:lumped}.
%In this estimate, we encounter a subtlety. Namely, the real part of the impedance contains a sharp peak at the frequency of the transmon. This peak is in addition to the impedance given by Eq.~\eqref{eq:rez_massaged}. As we now explain, the additional peak should be \textit{disregarded} when computing the rate of inelastic scattering. 
%Indeed, accounting for that peak would correspond to quasi-elastic scattering of the incoming photon of frequency $\omega_{\rm in}$. The width of that peak is determined by the inverse of the Purcell-limited qubit lifetime $T_1$. The contribution of the peak to the total photon scattering cross-section is proportional to the square of reflection phase shift used as the readout signal.
%Indeed, the peak at $\wq$ corresponds to the qubit mode broadened by the Purcell effect. Therefore, if one of the outgoing photons leaves through this peak, no transmon transition happens, transmon remains in the excited state. Therefore, when computing the scattering rate, it is appropriate to assume that $Z[\omega]$ is given by Eq.~\eqref{eq:rez_massaged} at all frequencies. Under this assumption we can estimate the integral as 
\begin{equation}
    \Gamma_{m\rightarrow m-1} \sim \delta\omega \frac{\chi^2}{E_C / \hbar} \frac{\wq \kappa}{\wres^3}.
\end{equation}
For our typical readout parameters $\Gamma_{m\rightarrow m - 1} \sim 10^{-2}\:\mathrm{s}^{-1}$.

%The impedance actually has a peak not only at the resonator frequency but also a sharp peak at the qubit frequency. See figure 1 of the main text. This peak corresponds to the qubit mode itself broadened by its Purcell decay through the transmission line. This peak has to be subtracted from the impedance since the final state here also contains an excitation in the qubit mode. Cutting away the peak, we can safely use equation blah when evaluating the impedance. In this case the integral is rather straightforward to estimate. We then obtain
%\begin{equation}
%\label{eq:two_photon_decay}
%\Gamma_{m\rightarrow m - 1} \sim \delta \omega \frac{\kappa \chi}{\wq^2} \left(\frac{\wq}{\win}\right)^3 \frac{\chi}{E_C}
%\end{equation}
%Comparing this with equation (simple-raman) we find that the transition rate in Eq.~\eqref{eq:two_photon_decay} is suppressed by many orders of magnitude compared to the Raman process discussed in the main text of the article. For this reason we can safely neglect this process. By the same token we can neglect the process in which the qubit excites from $0$ to $1$. In fact, the rate of this process can be estimated as
%\begin{equation}
%    \Gamma_{m\rightarrow m+1} \propto ...
%\end{equation}

\section{Additional data}
In this section, we provide the additional data that corroborates the discussion of the main text. To begin with, in Section~\ref{sec:0_to_other}, we demonstrate the state transition rates for qubit initialized in state $|0\rangle$ as a function of drive power and qubit frequency. These data complement similar results for qubit initialized in state $|1\rangle$, Figure~4 of the main text of the manuscript. We explain the relationship between various features observed in the two datasets.
Then, in Section~\ref{sec:multi_excitation} we provide additional data for the strongest feature observed in the transition rate $\Gamma_{1\rightarrow 4+}$, Figure~4(d) of the main text. We provide extra evidence that this feature stems from a multi-excitation resonance that takes the qubit from state $|1\rangle$ to state $|8\rangle$. We also explicitly demonstrate that the splitting of this feature in Figure~4(d) can be explained by the activation of readout resonator sub-bands. To this end, we compare the spacing of the split feature to a theoretical prediction.

\subsection{Rates of transmon state transitions for qubit initialized in $|0\rangle$}

\label{sec:0_to_other}
In Figure~4 of the main text, we explore qubit state transitions caused by a high-frequency drive applied close to the frequency of the readout resonator. To this end, in this figure, we plotted the dependence of transition rates between the transmon state $|1\rangle$ and other transmon states on the power of the drive (quantified by AC Stark shift experienced by the qubit) and qubit frequency. In this section, we complement these data with similar data for the transmon initialized in state $|0\rangle$. We compare the two datasets to each other and explain similarities and differences between them. 

\begin{figure*}[t]
  \begin{center}
    \includegraphics[scale = 1]{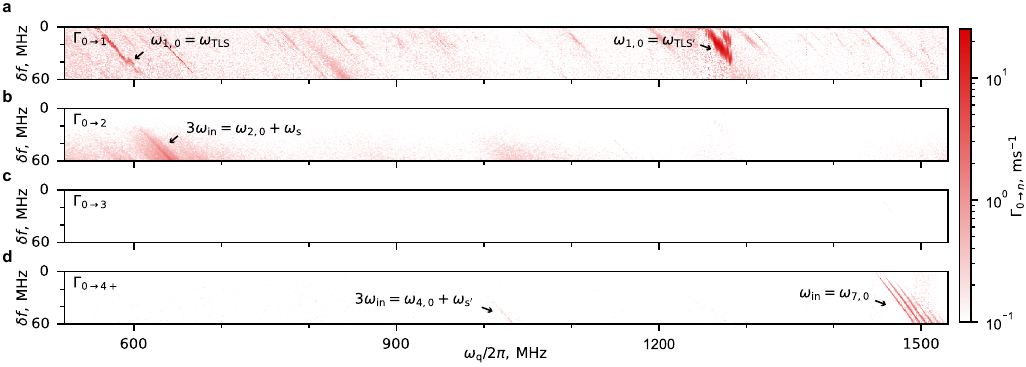}
    \caption{Rates of transmon state transitions plotted against drive power (quantified by the absolute value of the AC Stark shift $\delta f= \delta\omega/2\pi$) and qubit frequency (controlled with a flux bias). The transmon is initialized in its ground state $|0\rangle$. (a) Rate $\Gamma_{0\rightarrow 1}$. The smooth background  corresponds to thermal heating of the transmon. Stripey pattern results from coupling to material defects. The pattern is shifted compared to that in $\Gamma_{1\rightarrow2}$ roughly by qubit anharmonicity of 40 MHz (see Figure~4(b) of the main text). (b) Rate $\Gamma_{0\rightarrow 2}$. The smooth background in $\Gamma_{0\rightarrow 2}$ stems from inelastic scattering. Sharp feature stems from the coupling to a mode at $\omega_s/2\pi = 26.72\:\mathrm{GHz}$. The same mode is responsible for the peak in $\Gamma_{1\rightarrow 3}$ described in the main text. (c) Rate $\Gamma_{0\rightarrow 3}$ remains small in the shown range of qubit frequencies and drive powers. (d) Rate $\Gamma_{0\rightarrow 4+}$ of transitions to $|4\rangle$ and higher excited states. The strongest feature in $\Gamma_{0\rightarrow 4+}$ stems from a multi-photon resonance $|0\rangle\rightarrow |7\rangle$ where the drive frequency satisfies $\win = \omega_{7,0}$. Another sharp feature stems from coupling to the mode at $\omega_s^\prime/2\pi = 24.15\:\mathrm{GHz}$}
    \label{fig:big_map_sup}
  \end{center}
\end{figure*}

%However, not only qubit state $|1\rangle$ can experience the unwanted drive induced transitions. By the same token state $|0\rangle$ can also excite to states $|1\rangle$, $|2\rangle$, $|3\rangle$ and so on. In Figure~\ref{fig:big_map_sup} we explore these rates. 
The result of the rate measurement for qubit initialized in state $|0\rangle$ is shown in Figure~\ref{fig:big_map_sup}. The protocol for collecting and analyzing the data shown in the figure is identical to that used for Figure~4, see description in the main text. In fact, the data was collected simultaneously for the both figures. 

Figure~\ref{fig:big_map_sup} has a set of similarities to Figure~4 of the main text. Similar to the behavior of $\Gamma_{1\rightarrow 2}$ presented in the main text, rate $\Gamma_{0\rightarrow 1}$ is also dominated by thermal bath. The rate of heating is mostly independent of the drive power. An exception to that is the stripey pattern caused by coupling to material defects. In fact, each ``stripe'' corresponds to a condition $\omega_{1,0}[\delta\omega] = \omega_{\rm TLS}$ being met, where $\omega_{\rm TLS}$ is the frequency of a material defect coupled to the qubit, and $\omega_{1,0}[\delta\omega]$ is the AC Stark-shifted frequency of the qubit. Notably, the stripes in $\Gamma_{0\rightarrow 1}$ and $\Gamma_{1\rightarrow 2}$ are shifted relative to each other by about 40 MHz. This is because of the transmon anharmonicity and consequent difference $\omega_{1,0} - \omega_{2,1}\approx 40\:\mathrm{MHz}$. Notably, the position of the stipes in the heating rate $\Gamma_{0\rightarrow 1}$, Figure~\ref{fig:big_map_sup}(a), is the same as in the relaxation rate $\Gamma_{1\rightarrow 0}$, Figure~4(a) of the main text. This happens because in both cases the position of the stripes is determined by the condition $\omega_{1,0}[\delta\omega] = \omega_{\rm TLS}$.

The behavior of $\Gamma_{0\rightarrow 2}$ in Figure~\ref{fig:big_map_sup}(b) is similar to that of $\Gamma_{1\rightarrow 3}$ in Figure~4(b) of the main text. In both cases, the transition rate stems from inelastic scattering processes. Smooth background stems from the Raman scattering, see Section~III of the main text. Sharp feature stems from interaction with a spurious mode, as described in Section~V of the main text. There are, however, two important distinctions between Figure 4(b) and Figure~\ref{fig:big_map_sup}(b). First, the features in the two plots are shifted in qubit frequency by about 40 MHz. This occurs due to the difference of the transition frequencies $\omega_{0,2}$ and $\omega_{1,3}$ caused by qubit anharmonicity. Second, all features in $\Gamma_{0\rightarrow 2}$ have a rate by a factor of 3 smaller than the corresponding features in $\Gamma_{1\rightarrow 3}$. This is consistent with our theoretical expectation, compare for example Eq.~\eqref{eq:raman_final} for $m = 0$ and for $m = 1$.

Next, we note that the transition rate $\Gamma_{0\rightarrow 3}$ remains small in the range of powers and qubit frequencies shown in Figure~\ref{fig:big_map_sup}(c). By the same token, we expect the rate of transitions from $1$ to $4$ to be negligibly small. Unfortunately, in Figure 4 of the main text, we cannot separately resolve the rates of transitions to $4$ and to higher states ($5,6,...$).

Finally, we discuss the features in $\Gamma_{0\rightarrow 4+}$, see Figure~\ref{fig:big_map_sup}(d). All of these features also have a counterpart in Figure~4(d). In Figure~\ref{fig:big_map_sup}(d), we label the features according to their origin and also show the final state to which transmon becomes excited.

\subsection{Multi-excitation resonances in the transmon spectrum activated by the drive}

\begin{figure*}[t]
  \begin{center}
    \includegraphics[scale = 1]{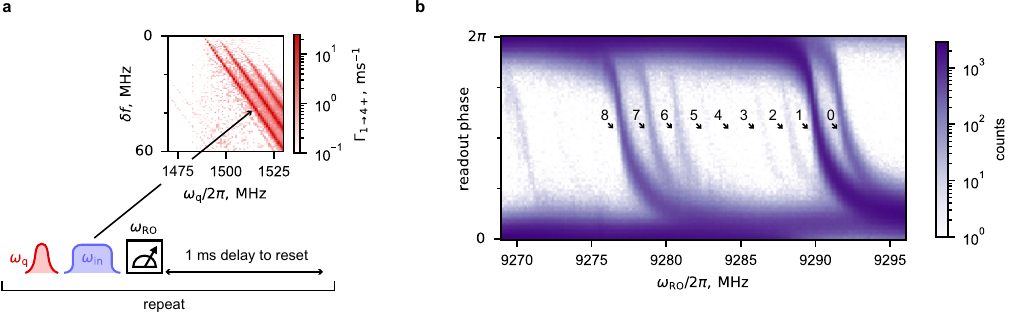}
    \caption{Investigation of the strongest feature in Figure~4(d) of the main text. We demonstrate the the transmon indeed excites from state $|1\rangle$ directly to state $|8\rangle$. (a) Pulse sequence used in the experiment. We prepare qubit in $|1\rangle$ (at $\wq/2\pi = 1530\:\mathrm{MHz}$). Then we apply a drive tone at $\win/2\pi = 9280\:\mathrm{MHz}$ activating the sharp feature in $\Gamma_{1\rightarrow 4+}$. Afterwards, we readout the resonator with a variable frequency readout tone. Finally, we let the system relax back to the ground state. (b) The result of the experiment. Phase-rolls correspond to progressive transmon states. It is clear from the figure that the transmon indeed excites to state $|8\rangle$ by our drive tone.}
    \label{fig:1_to_8_indeed}
  \end{center}
\end{figure*}

\label{sec:multi_excitation}
As is shown in Figure~4(d) of the main text, close to the highest achievable qubit frequency, the drive can resonantly excite a transition from $|1\rangle$ to one of the higher-excited states (which this measurement cannot directly resolve). Specifically, close to frequency 1500 MHz, an enhancement of the excitation rate $\Gamma_{1\rightarrow 4+}$ is observed. As mentioned in the main text, we believe that this enhancement happens when the frequency of the drive matches that of the $|1\rangle$ to $|8\rangle$ transition of our transmon, $\win = \omega_{8,1}$.

In Section~\ref{sec:wtf_is_the_feature}, we provide additional evidence that the discussed feature stems from a multi-excitation resonance $|1\rangle \rightarrow |8\rangle$. First, we demonstrate that the feature indeed occurs at a qubit frequency at which we theoretically expect the drive frequency to match $\omega_{8,1}$. Second, we explicitly show that the qubit excites to state $|8\rangle$ by performing the resonator measurement using pulses with a lowered frequency.

As is also apparent from Figure~4(d) of the main text, the transition line corresponding to the discussed resonance is split into many parallel lines. In our interpretation, the parallel lines stem from the activation of the readout resonator sidebands. Namely, they occur each time the condition $n\win = (n-1)\wres + \omega_{8,1}$ is fulfilled. In Section~\ref{sec:wtf_are_split_lines} we provide evidence in favor of this interpretation. Specifically, we show that the spacing of the split lines is well-explained by our interpretation.

\begin{figure*}
  \begin{center}
    \includegraphics[scale = 1]{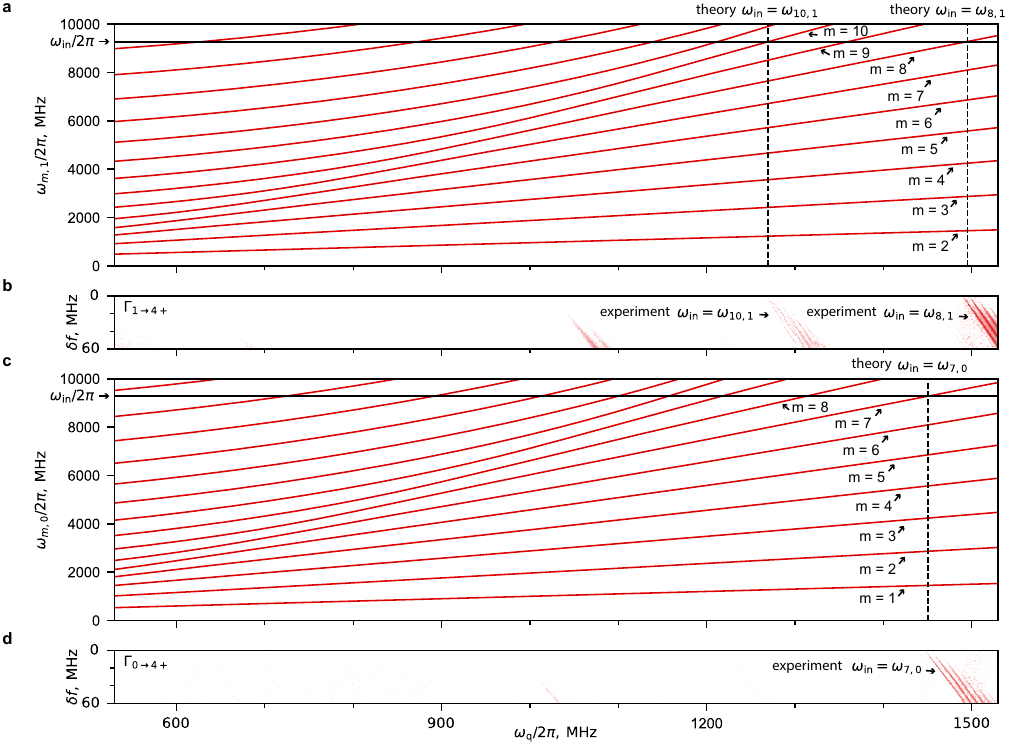}
    \caption{Theory and experiment for multi-excitation resonances activated by the drive. (a) Theoretically calculated spectrum of the transmon initialized in state $|1\rangle$ plotted for different qubit frequencies (tuned by changing $E_J$ while holding $E_C$ fixed). Subsequent lines show transitions to states $|2\rangle$, $|3\rangle$, and so on. Solid black line corresponds to the drive frequency of $\win/2\pi = 9280\:\mathrm{MHz}$ employed in our measurements. Close to $\wq/2\pi=1500\:\mathrm{MHz}$, the drive frequency matches that of the $|1\rangle$ to $|8\rangle$ transition. Close to $\wq/2\pi=1270\:\mathrm{MHz}$, the drive frequency matches that of the $|1\rangle$ to $|10\rangle$ transition. (b) Data from Figure~4(d) of the main text. The features close to $\wq/2\pi=1500\:\mathrm{MHz}$ and $\wq/2\pi=1270\:\mathrm{MHz}$ line up with theoretical expectation for $\win = \omega_{8,1}$ and $\omega_{10,1}$, respectively. We note that there is no feature at $\win=\omega_{9,1}$. The corresponding transition is forbidden by symmetry. (c) Theoretically calculated spectrum of the transmon initialized in state $|0\rangle$ plotted for different qubit frequencies. Close to $\wq/2\pi=1460\:\mathrm{MHz}$ the drive frequency matches that of the $|0\rangle$ to $|7\rangle$ transition. (d) Data from Figure~\ref{fig:big_map_sup}(d) of the main text. The feature close to $\wq/2\pi=1460\:\mathrm{MHz}$ lines up with theoretical expectation for $\win = \omega_{7,0}$.}
    \label{fig:mist}
  \end{center}
\end{figure*}

\subsubsection{Evidence that feature in $\Gamma_{1\rightarrow 4+}$ stems from $|1\rangle \rightarrow |8\rangle$ multi-excitation resonance}
\label{sec:wtf_is_the_feature}
The measurement presented in Figure 4(d) of the main text is incapable of resolving the final state to which the transmon gets excited to after the drive pulse. This measurement is performed at a readout frequency that is optimized to distinguish the low-lying transmon states. As a result, all states higher than $|4\rangle$ are lumped together and cannot be resolved.

To resolve the state to which our transmon is excited by activating the feature close to $\wq/2\pi = 1500\:{\rm MHz}$ in Figure~4(d), we perform the following measurement. First, prepare the system in state $|1\rangle$. Then, we apply an excitation tone which hits one of the sharp transition lines. After that, we perform a single-shot measurement of the readout resonator with a variable-frequency readout tone. After a millisecond delay -- needed to reset the qubit -- we repeat the measurement. After collecting sufficient statistics, we change the readout frequency and repeat the sequence. The pulse sequence is summarized in Figure~\ref{fig:1_to_8_indeed}(a). The results of this measurement are shown in Figure~\ref{fig:1_to_8_indeed}(b).

The figure shows the phase of the readout signal measured as a function of the readout frequency. Subsequent ``phase-rolls'' -- where the reflection phase winds by $2\pi$ -- correspond to a sequence of transmon states (see Section~\ref{sec:qubit_cavity} for details). 
From the figure it is clear that after activating the resonance in $\Gamma_{1\rightarrow 4+}$, the system does indeed transition to state $|8\rangle$.

In Figure~\ref{fig:mist}, we show that for the parameters of our device and for the driving applied at $\win/2\pi = 9280\:{\rm MHz}$, we indeed theoretically expect a resonance between the drive and the $|1\rangle\rightarrow |8\rangle$ transition that occurs close to $\wq/2\pi = 1500\:{\rm MHz}$. From this figure it is also clear that a weaker feature present in $\Gamma_{1\rightarrow 4+}$ around $\wq/2\pi = 1300\:{\rm MHz}$ is a resonance between the drive and the $|1\rangle\rightarrow |10\rangle$ transition. Similarly, the feature in $\Gamma_{0\rightarrow 4+}$ observed close to 1500 MHz is a resonance between the drive and the $|0\rangle\rightarrow |7\rangle$ transition. Its position is also well captured by the theory. The discrepancy $\sim5$ MHz can be attributed to either imperfect knowledge of parameter $E_C$ or to higher harmonics in the current phase relation of the Josephson junction \cite{willsch_observation_2024} (which can stem from stray inductnace of the leads).

\subsubsection{Splitting of the strongest feature in $\Gamma_{1\rightarrow 4+}$ explained by readout resonator sidebands}

\label{sec:wtf_are_split_lines}
\begin{figure*}
  \begin{center}
    \includegraphics[scale = 1]{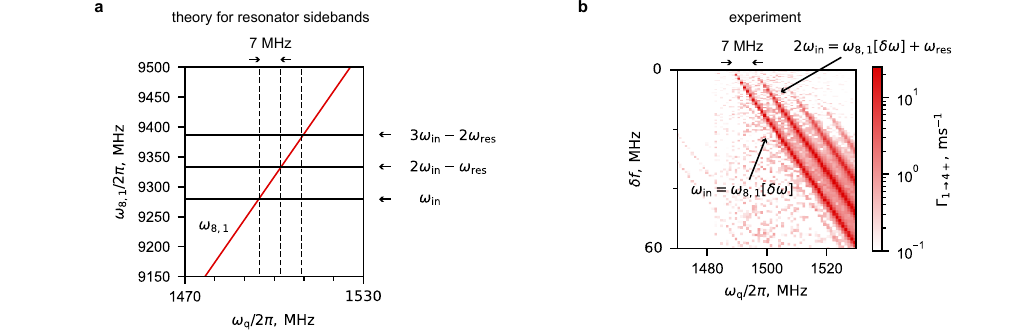}
    \caption{Resonator sidebands that stem from proximity of the drive frequency to that of the readout resonator. (a) Theoretically calculated frequency of $|1\rangle \rightarrow |8\rangle$ transition of the transmon (red solid line) as a function of its frequency (controlled by changing $E_J$). Black horizontal solid lines show the resonator sidebands of the drive frequency, $n\omega_{\mathrm{in}}-(n-1)\omega_{\mathrm{res}}$. Dashed vertical lines show at what qubit frequencies the drive sidebands cross the frequency of $|1\rangle$ to $|8\rangle$ transition. (b) The experimentally observed resonator sidebands [data from Figure~4(d)]. The spacing between the sharp features is consistent with the theoretical expectation of panel (a).}
    \label{fig:mist_2}
  \end{center}
\end{figure*}

As explained above, the feature in $\Gamma_{1\rightarrow 4+}$ around $\wq/2\pi=1500\:\mathrm{MHz}$ stems from the resonance between the drive and the $|1\rangle$ to $|8\rangle$ transition of the transmon. It occurs when the frequency matching condition is fulfilled, $\win = \omega_{8,1}$.
However, as is clear from Figure 4(d) of the main text, the feature is split into multiple parallel lines (see Figure~\ref{fig:mist_2} for a close up). In our interpretation, the splitting stems from the activation of the readout resonator sidebands. These sidebands correspond to a process in which several drive quanta are simultaneously absorbed to excite both the transmon and the readout resonator. The corresponding frequency matching condition reads $n \win = \omega_{8,1} + (n-1)\wres$. Since $\win - \wres > 0$, the sidebands occur at a qubit frequency higher than that corresponding to a direct process, $\win = \omega_{8,1}$. Because the drive frequency is close to that of the resonator (the detuning is only about 50 MHz), the frequencies of the sideband transitions are close to the frequency of the direct transition. In fact, theoretically, we expect the spacing in qubit frequency to be about 7 MHz, see Figure~\ref{fig:mist}(a). As we show in Figure~\ref{fig:mist}(b), this is indeed consistent with the observed spacing between the split lines in $\Gamma_{1\rightarrow 4+}$.
%To corroborate our understanding of the split lines we look at the power dependence of the rate. The first subband converts two drive photons into a qubit and cavity excitation. As such we expect the power dependence of the rate to be $\Gamma_{1\rightarrow 8} \propto \delta\omega^2$. in Figure ... we show that this is indeed the case.
%In Figure~\ref{fig:mist} we show that the resonator sideband interpretation correctly predicts the spacing between the split lines in $\Gamma_{1\rightarrow 4+}$.

\section{Details of the calibration procedure}
In this section, we describe the calibration procedures that we employ to measure the transmon transition rates in the presence of the drive. In the first calibration, we determine the frequency of the qubit and of the readout resonator as functions of the phase bias applied to the SQUID loop of our device. The knowledge of the readout resonator frequency allows us to perform transmon readout at different values of the phase bias. The calibrated qubit frequency is directly used as a variable (as opposed to phase bias), see for example Figure~3 and Figure~4 of the main text. The calibration of the resonator and of the qubit frequency is detailed in Section~\ref{sec:qubit_cavity}. The next step is to calibrate how the raw single-shot readout data is used to assign the transmon state. This has to be done across the full range of accessible qubit frequencies. This calibration should also take into account the change of the readout frequency with qubit frequency. The details of state assignment are explained in Section~\ref{sec:state_assignment}. Next, we calibrate the power of the off-resonant drive applied to the readout resonator. To this end, we measure the AC Stark shift of the qubit frequency in the presence of the drive. We do that for different qubit frequencies while holding the drive frequency fixed (close to the frequency of the readout resonator). For a set of qubit frequencies, we also calibrate the Stark shift across a range of \textit{drive} frequencies, see panels (e) and (g) of Figure~4 of the main text. The calibration of the drive power is described in Section~\ref{sec:ac_stark}. Finally, we calibrate the amplitude of the qubit $\pi/2$-pulse at different qubit frequencies. We highlight that certain features present in this calibration happen at the same qubit frequency as dips in the qubit relaxation time measurement (see Figure~\ref{fig:rabi_t1_vs_flux}). We relate this to Purcell effect. The calibration of pulse amplitude is discussed in Section~\ref{sec:rabi_and_t1_dip}.

%state assignment at different qubit frequencies. 

%To perform this calibration we measure the histogram of the measurement outcomes at different values of the qubit flux using the prior knowledge of the flux-dependence of the resonator frequency.

%se to scramble the qubit state. This is needed to efficiently sample transitions from both states $|0\rangle$ and $|1\rangle$ of the qubit. We explain how this calibration is performed in section blah. The next step in our calibration procedure is to calibrate the state assignment at different qubit frequencies. To perform this calibration we measure the histogram of the measurement outcomes at different values of the qubit flux using the prior knowledge of the flux-dependence of the resonator frequency. Next, we need to calibrate the power delivered by our drive at different values of the magnetic flux. We calibrate the power via the AC Stark shift experienced by the frequency of the $|0\rangle \rightarrow |1\rangle$ transition of our qubit. We thus measure the qubit Stark shift as a function of drive power at different values of the qubit frequency. For certain qubit frequencies explored in the main text, we also need to calibrate the Stark shift at a different value of the drive frequency (see for example panel (x) of Figure 4 of the main text). In the following text, we go through these steps sequentially and demonstrate how the calibrations are performed.

\subsection{Measurement of qubit and readout resonator frequencies as functions of phase bias}
\label{sec:qubit_cavity}
In this section, we describe the measurement of the resonator frequency and of the qubit frequency at different values of the phase bias applied to the SQUID loop of our transmon.

\begin{figure}[h!]
  \begin{center}
    \includegraphics[scale = 1]{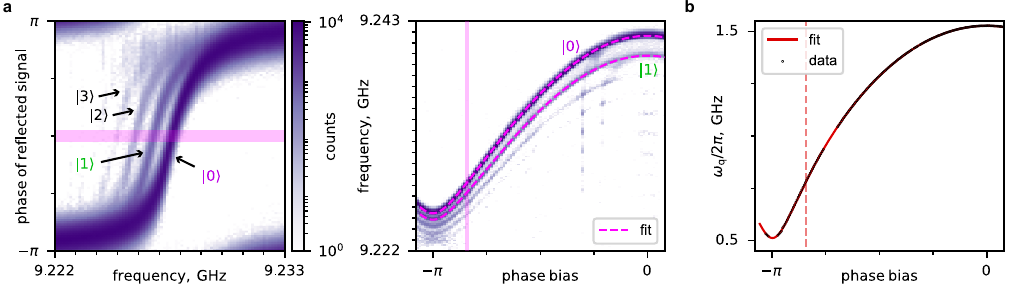}
    \caption{Resonator and qubit spectroscopy at different values of the SQUID phase bias (figure from Ref.~\cite{kurilovich_high-frequency_2025}. (a) Spectroscopy of the readout resonator. Left panel: histogram of the phase of the signal reflected from the resonator as a function of signal frequency. Different phase-rolls correspond to different transmon states (measurement is performed in a single-shot readout regime). The middle of each phase-roll indicates the resonator frequency $\omega_n$ for the corresponding transmon state $|n\rangle$. Right panel: resulting dependence of frequencies $\omega_n$ on the phase bias. Frequencies $\omega_n$ for states $|n\rangle$ with $n=0,1,2$ are visible in the full range of phase biases. (b) Qubit spectroscopy. The measurement shows the frequency of the $|0\rangle \rightarrow |1\rangle$ transition of our transmon as a function of the SQUID phase bias. The red vertical line corresponds to the operation point considered in the main text of Ref.~\cite{kurilovich_high-frequency_2025}.}
    \label{fig:qubit_vs_flux}
  \end{center}
\end{figure}

To measure the resonator frequency, we send a pulse with a variable carrier frequency through the readout channel and record the reflected signal. Having obtained enough statistics at a given carrier frequency, we change it and repeat the measurement. When the carrier frequency is changed, the phase of the reflected signal winds by $2\pi$ around the frequency $\omega_{n} = \omega_{\mathrm{r,bare}} + \chi_n$ for each transmon state $|n\rangle$. Here, $\omega_{\mathrm{r,bare}}$ is the frequency of the resonator not dressed by its coupling to the transmon and $\chi_n$ is resonator frequency pull due to its coupling to a transmon. Locating the center of such a ``phase-roll'', we determine $\omega_n$. In this measurement, thermal fluctuations populate the higher transmon states. This allows us to simultaneously observe phase-rolls corresponding to different $|n\rangle$ and extract the resulting frequencies $\omega_n$. The results of this measurement for different values of the phase bias are shown in Figure~\ref{fig:qubit_vs_flux}(a). The details of the theoretical fit are presented in Ref.~\cite{kurilovich_high-frequency_2025} (see Section~III of the supplementary information in this paper).

The calibration of the resonator frequency allows us to perform qubit readout at different values of the phase bias. With that, we can measure the phase-bias dependence of the qubit frequency. The results of this measurement are demonstrated in Figure~\ref{fig:qubit_vs_flux}(b). The theoretical fit is obtained using a model of a capacitively shunted SQUID. To avoid fit errors on the scale of a few MHz it is crucial to include the stray inductance of the leads in our model~\cite{willsch_observation_2024}. 

\subsection{State assignment at different values of the flux bias}
\label{sec:state_assignment}
In our measurement of rates, we need to assign the raw resonator single-shot measurements to transmon states. This assignment requires a separate calibration which we describe in this section.

For the measurement of the rates described in the main text, we use 4-microsecond readout pulses. This duration is substantially longer than that of our optimal pulses described in Ref.~\cite{kurilovich_high-frequency_2025}.  Such a long measurement duration is required to resolve multiple higher-excited states. Namely, we need to resolve states $|2\rangle$, $|3\rangle$, and $|4\rangle$ in addition to the computational states $|0\rangle$ and $|1\rangle$.

To calibrate the state assignment, we perform the following measurement. First, we scramble the state of the transmon using the calibrated qubit $\pi/2$-pulses (see Section~\ref{sec:rabi_and_t1_dip} for the calibration). Then, we apply a resonator readout pulse. For the frequency of this pulse, we aim between frequencies $\omega_1$ and $\omega_2$ (i.e., resonator frequency when transmon is in $|1\rangle$ and $|2\rangle$, respectively). We repeat this measurement many times back-to-back to collect enough statistics. Notably, the readout pulses cause transitions from the computational states to non-computational states (see main text for the discussion of why that happens). This allows us to sample the distributions for these states too.

\begin{figure}[h!]
  \begin{center}
    \includegraphics[scale = 1]{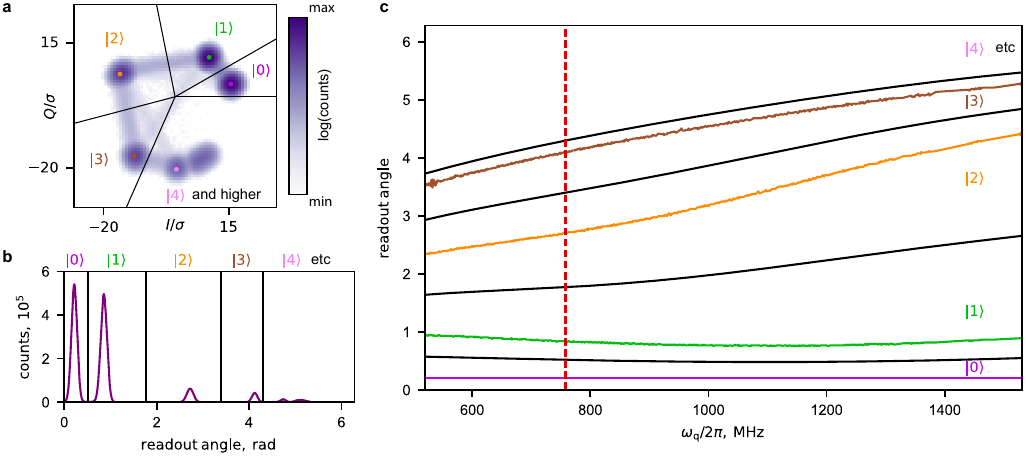}
    \caption{Calibration of the state state assignment across different qubit frequencies. (a) Readout histogram obtained after scrambling the qubit state. The qubit frequency is tuned to $\wq/2\pi = 760\mhz$. The brightest states are qubit states $|0\rangle$ and $|1\rangle$. Distributions corresponding to higher qubit states $|2\rangle$ and $|3\rangle$ are located sequentially in the clockwise direction. Black solid lines show the thresholds used for the state assignment. (b) The histogram of the argument of the complex measurement outcomes. Black vertical lines are the thresholds used for state assignment. (c) Measured angles of distributions for different transmon state at different transmon frequencies. Different transmons states are depicted with different colors. Black lines show the thresholds used for state assignment. Red vertical dashed line shows the qubit frequency of $\wq/2\pi = 760\mhz$ employed in panels (a) and (b).}
    \label{fig:blobs_vs_flux}
  \end{center}
\end{figure}

An example of the measured I-Q distribution is shown in Figure~\ref{fig:blobs_vs_flux}(a). To perform the state assignment, we project the data on a circle by computing the argument of the measured complex number. The histogram of the angle measurements is shown in Figure~\ref{fig:blobs_vs_flux}(b). To perform the state assignment, we draw thresholds that separate different distributions in Figure~\ref{fig:blobs_vs_flux}(b) from each other. In Figure~\ref{fig:blobs_vs_flux}(b) the thresholds are located (i) several variances away from the peak corresponding to $|0\rangle$ (the data is rotated so that this threshold is at zero angle), (ii) in the middle between peaks corresponding to $|0\rangle$ and $|1\rangle$, (iii) in the middle between $|1\rangle$ and $|2\rangle$, (iv) in the middle between $|2\rangle$ and $|3\rangle$, and (v) several variances away from the peak corresponding to $|3\rangle$. 
%We draw thresholds in the middle between the peaks corresponding to states $|0\rangle$ and $|1\rangle$ (as well as in the middle between $|1\rangle$ and $|2\rangle$, and in the middle between $|2\rangle$ and $|3\rangle$). We also draw threshold s

%The state of the transmon can be inferred by comparing the measured outcome  

%The thresholds are chosen to separate the distributions from each other.

We then repeat this measurement at different values of the qubit frequency (changed by threading flux through the SQUID loop). For convenience we always rotate the data such that the peak of the distribution corresponding to the state $|0\rangle$ has a fixed angle regardless of the qubit frequency. The resulting flux dependence is depicted in Figure~\ref{fig:blobs_vs_flux}(c).

\subsection{Calibration of drive power via the AC Stark shift}
\label{sec:ac_stark}
Next, we measure the AC Stark shift of our transmon qubit in the presence of the drive applied through the readout channel close to the frequency of the readout resonator. The Stark shift is proportional to the power of the drive. This allows us to use the Stark shift as a proxy for the power throughout the main text. We perform the measurement of the AC Stark shift at different values of the phase bias applied to the SQUID loop and at different values of the drive frequency.

To measure the AC Stark shift, we perform qubit spectroscopy measurement in the presence of a drive tone.
To this end, we first measure the initial state of the qubit. Then we apply a drive tone and, in the middle of this drive pulse, we apply an extra pulse of a variable frequency in the vicinity of the qubit transition. Finally, we perform another measurement at the end of this sequence to see where the qubit ends up. We rely on our measurement of the calibrated Rabi amplitude to choose an appropriate value of the qubit excitation pulse.

\begin{figure}[h!]
  \begin{center}
    \includegraphics[scale = 1.0]{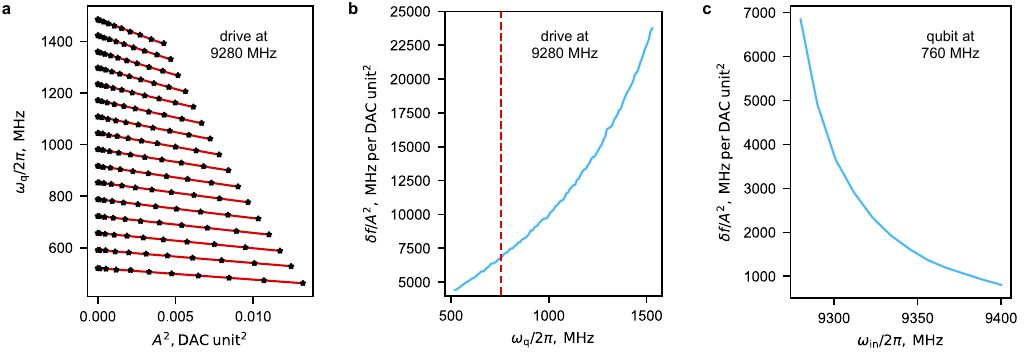}
    \caption{Calibration of the AC Stark shift of the qubit $|0\rangle\rightarrow |1\rangle$ transition caused by the driving applied through the readout channel. (a) Qubit frequency versus the power of the drive applied at the frequency $\win / 2\pi = 9.280\ghz$. The power of the drive is quantified by the square of the amplitude at the digital-to-analog converter of our setup. The dependence of the Stark shift on the power is linear. (b) The slope of the AC Stark shift versus power plotted as a function of the qubit frequency. (c) Slope of AC Stark shift versus power as a function of the drive frequency for a fixed qubit frequency of $\wq/2\pi = 760\mhz$.}
    \label{fig:spec_cal}
  \end{center}
\end{figure}

The results of this measurement for the drive frequency of 9280 MHz are shown in Figure~\ref{fig:spec_cal}(a,b). Panel (a) of this figure shows the dependence of the qubit frequency on the applied drive power. Here, we quantify power by $A^2$, where $A$ is the amplitude of the pulse produced by DAC the converter. This pulse modulates the amplitude of the drive provided by the generator, see Fig.~\ref{fig:wiring}. Different curves in Figure~\ref{fig:spec_cal}(a) correspond to different values of the phase bias applied to the SQUID loop of our transmon. As is clear from the figure, at finite drive power, the frequency of the qubit is Stark shifted down to lower values. As expected, the change of frequency with power is approximately linear. Panel (b) of Figure~\ref{fig:spec_cal} shows the absolute value of the slope of the AC Stark shift with power at different values of the qubit frequency. The drive needs to be much stronger close to half-quantum flux (minimal qubit frequency). This is because the dispersive shift is smaller at low qubit frequencies and because the readout resonator frequency is most detuned from the drive.

Finally, for some data presented in the main text of the paper, it is important to calibrate the AC Stark shift as a function of the drive frequency. We present an example of such a calibration in Figure~\ref{fig:spec_cal}(c). From the figure it is apparent that more power is needed to obtain a fixed value of the AC Stark shift at higher drive frequencies.

\subsection{Calibration of the Rabi amplitude}
\label{sec:rabi_and_t1_dip}
In the main text of the manuscript, we demonstrate the measurements of the qubit transition rates for qubit initialized in state $|0\rangle$ and in state $|1\rangle$. To prepare the qubit in state $|0\rangle$, it is sufficient to let the qubit relax for a prolonged time. In contrast, to prepare the qubit in state $|1\rangle$, the qubit has to be actively excited by a microwave pulse. In this section, we describe the calibration of the Rabi amplitude for such pulses. To measure how the transmon state transition rates depend on the qubit frequency (Figures~3 and 4 of the main text), we have to calibrate the Rabi amplitude in a range of qubit frequencies.

%In this section, we describe a measurement of Rabi amplitude at various qubit frequencies, where the qubit frequency is controlled using magnetic flux. This calibration allows us to selectively excite the qubit at different values of the flux bias.

Notably, in our setup, the Rabi drive is delivered through the readout line. The Rabi amplitude thus carries information on how strongly the readout line is coupled to the qubit. Indeed, we find that certain features in the dependence of the Rabi amplitude on the qubit frequency are correlated with features in the qubit relaxation time $T_1$. We relate this to Purcell effect.

\subsubsection{Measurement of Rabi amplitude}
Figure~\ref{fig:rabi_t1_vs_flux}(a) shows measured DAC amplitude required to drive the $\pi/2$ pulse at different frequencies of our qubit. We deliver this pulse through the readout channel. The measurement shows that our transmission line is poorly matched at the frequency of about a few gigahertz. This is to be expected since some of our microwave components (such as the circulators) are out of band in this frequency range.
%A particularly notable feature is the presence of the dip at the frequency $\sim 1230\:\mathrm{MHz}$. This dip implies that it the coupling between the transmission line and the qubit is especially strong at this frequency. This happens due to the presence of a mode in the transmission line which arises due to the impedance mismatch.

\subsubsection{Measurement of the qubit lifetime}
\begin{figure}[h!]
  \begin{center}
    \includegraphics[scale = 1]{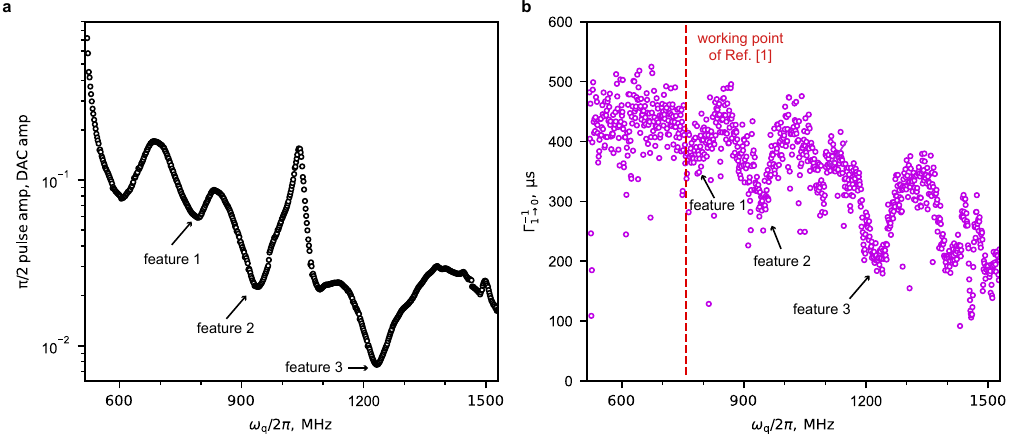}
    \caption{(a) Calibration of the Rabi amplitude for different qubit fluxes. Strong variation in the Rabi amplitude with qubit frequency stems from poor matching of our transmission line at frequencies below few GHz. (b) Qubit relaxation rate $\Gamma_{1\rightarrow 0}$ as a function of the flux. Solid line shows the prediction for a constant quality factor. The positions of most prominent dips in the qubit $T_1$ coincide with the positions of the dips in the Rabi amplitude.}
    \label{fig:rabi_t1_vs_flux}
  \end{center}
\end{figure}

Figure~\ref{fig:rabi_t1_vs_flux}(b) shows how the qubit relaxation time depends on the qubit frequency. The relaxation time defined as the inverse rate of transitions from state $|1\rangle$ to state $|0\rangle$ measured in the absence of the drive. As is clear from the figure, the positions of the dips coincide with those in the measurement of the Rabi amplitude. At these frequencies the readout line is especially strongly coupled to the qubit due to accidental resonances. This leads to the enhancement of Purcell decay for the qubit.

%Figure~\ref{fig:rabi_t1_vs_flux}(b) shows the measure
%In the second, measurement we demonstrate how the relaxation time of our qubit depends on the qubit frequency. The results of this measurement are shown in Figure~\ref{fig:rabi_t1_vs_flux}(b). Notably, there is a dip in the relaxation time located at exactly the same frequency as the dip in the pulse amplitude. We believe that at this frequency our qubit is Purcell limited due to the coupling to the readout channel. This realization allows us to estimate the Purcell decay at all other frequencies using $\Gamma$ [cite sumeru]. The assumption behind this comparison is that the power delivered to the bottom of the fridge is only determined by the DAC amp of the pulse. This assumption is reasonable at frequencies below a gigahertz where attenuation is only weakly power-dependent. This shows that at the vast majority of qubit frequencies the lifetime of our qubit is not Purcell limited. 

\section{Alternative derivation of matrix element}

The method of deriving the rate of the Raman process ($|m\rangle\rightarrow |m+2\rangle$) presented above explicitly relied on the smallness of the anharmonicity of the transmon qubit. In this section, we provide an alternative derivation of the Raman rate suitable for describing the case when the anharmonicity is not small. We show that in the transmon regime the results of the two approaches coincide.

The main distinction of the present approach is that it does not rely on the normal mode decomposition of Eq.~\eqref{eq:phase_decomposition}. This decomposition is only well justified in the weakly non-linear case. Instead, we treat the coupling between the transmon and its environment as a perturbation. We compute the $T$-matrix to the lowest non-vanishing order in this perturbation. The transition rate then follows from the scattering theory as
\begin{gather}
\label{eq:VGV}
    \Gamma_{m \rightarrow m + 2} = \frac{2\pi}{\hbar} \nu_{out} |\mathcal{M}_{fi}|^2,\quad \mathcal{M}_{fi} = \langle \Psi_f | T | \Psi_i \rangle,\quad T = V \frac{1}{E_i - H_t - H_b} V.
\end{gather}
Here, as before, the initial and the final states are given by $|\Psi_{i}\rangle \propto (a_{\rm in}^\dagger)^{n_{\rm in}} |\Omega\rangle |m\rangle$ and $|\Psi_f\rangle \propto (a_{\rm in}^\dagger)^{n_{\rm in}-1}a_{\rm out}^\dagger |\Omega\rangle |m + 2\rangle$, respectively. Matrices $H_t$, $H_b$, and $V$ are defined in the main text. Equation~\eqref{eq:VGV} allow us to evaluate the transition rate regardless of whether the anharmonicity is small.

\subsubsection{Evaluation of the rate in the case of a weak anharmonicity}

To give an example of how the calculation via Eq.~\eqref{eq:VGV} proceeds, we focus on the case of weak anharmonicity. We show that this approach gives a result consistent with that obtained through normal mode decomposition in the main text of the paper.

The calculation proceeds as follows. First, we rewrite the matrix element using the decomposition of unity, $\hat{1}=\sum_p |\Psi_p\rangle\langle\Psi_p|$, where $|\Psi_p\rangle$ are the many-body states of the system. This yields
\begin{equation}
\label{eq:VGV_decomp}
\mathcal{M}_{fi} = \sum_p\frac{\langle \Psi_f | V| \Psi_p\rangle \langle \Psi_p| V |\Psi_i \rangle}{E_i - E_p},
\end{equation}
where $E_p$ is the energy of the state $|\Psi_p\rangle$ in the absence of coupling between the qubit and the bath.
There are three different contributions to the matrix element, $\mathcal{M}_{fi} = \mathcal{M}^{(1)}_{fi} + \mathcal{M}^{(2)}_{fi} + \mathcal{M}^{(3)}_{fi}$. They differ in the intermediate state of the qubit.  The first contribution $\mathcal{M}^{(1)}_{fi}$ corresponds to a transition through state $|m+1\rangle$. This gives rise to two intermediate states
\begin{gather}
|\Psi_{1}\rangle = (a_{\rm in}^\dagger)^{n_{\rm in}-1} |\Omega\rangle |m+1\rangle,\quad E_1 = (n_{\rm in} - 1)\hbar\omega_{\rm in} + \epsilon_{m+1},\\
|\Psi_{2}\rangle = (a_{\rm in}^\dagger)^{n_{\rm in}}a_{\rm out}^\dagger |\Omega\rangle |m+1\rangle,\quad E_2 = n_{\rm in}\hbar\omega_{\rm in} + \hbar\omega_{\rm out} + \epsilon_{m + 1}.
\end{gather}
They contribute to the matrix element as
\begin{gather}
\mathcal{M}^{(1)}_{fi} = \langle m+2|N|m+1\rangle \langle m+1|N|m\rangle \sqrt{n_{\rm in}} \lambda_{\rm in}\lambda_{\rm out}\frac{(2\epsilon_{m+1}-\epsilon_m-\epsilon_{m+2})}{(\hbar\win + \epsilon_{m+1} - \epsilon_{m+2})(\hbar\win + \epsilon_m - \epsilon_{m+1})}.
\end{gather}
Note that this contribution goes away for a linear oscillator. It relies explicitly on the presence of anharmonicity. In the limit of large $E_J/E_C$, we can approximate this term as
\begin{equation}
    \mathcal{M}^{(1)}_{fi} =  \sqrt{m+1}\sqrt{m+2}\nzpf^2 \sqrt{n_{\rm in}}\lambda_{\rm in}\lambda_{\rm out}\frac{E_C}{(\hbar\win - \hbar\wq)^2}.
\end{equation}
Contribution $\mathcal{M}_{fi}^{(2)}$ is mediated by different virtual processes, namely the transition goes through state $|m + 3\rangle$. The corresponding virtual states are
\begin{gather}
|\Psi_{3}\rangle = (a_{\rm in}^\dagger)^{n_{\rm in}-1} |\Omega\rangle |m+3\rangle,\quad E_3 = (n_{\rm in} - 1)\hbar\omega_{\rm in} + \epsilon_{m+3},\\
|\Psi_{4}\rangle = (a_{\rm in}^\dagger)^{n_{\rm in}}a_{\rm out}^\dagger |\Omega\rangle |m+3\rangle,\quad E_4 = n_{\rm in}\hbar\omega_{\rm in} + \hbar\omega_{\rm out} + \epsilon_{m+3}.
\end{gather}
The contribution to the matrix element reads
\begin{gather}
    \label{eq:M2}
    \mathcal{M}_{fi}^{(2)} = \langle m+2|N|m+3\rangle\langle m+3|N|m\rangle \sqrt{n_{in}} \lambda_{in}\lambda_{out}\frac{(2\epsilon_{m+3}-\epsilon_m-\epsilon_{m+2})}{(\hbar\win + \epsilon_{m+3} - \epsilon_{m+2})(\hbar\win + \epsilon_m - \epsilon_{m+3})}.
\end{gather}
Here, the matrix element of the charge operator is given by
\begin{equation}
    \label{eq:charge_by_3}
    \langle m+3|N|m\rangle = \frac{E_C}{4\hbar\omega_q}\sqrt{m+1}\sqrt{m+2}\sqrt{m+3}\nzpf.
\end{equation}
Substituting Eq.~\eqref{eq:charge_by_3} into Eq.~\eqref{eq:M2}, we obtain
\begin{equation}
        \mathcal{M}_{fi}^{(2)} = (m+3)\sqrt{m+1}\sqrt{m+2}N_\mathrm{zpf}^2 \sqrt{n_{in}} \lambda_{in}\lambda_{out} \frac{ E_C}{(\hbar\win + \hbar\wq)(\hbar\win - 3\hbar\wq)}.
\end{equation}
Finally, there is contribution $\mathcal{M}^(3)_{fi}$ that goes through a virtual states $|m-1\rangle$,
\begin{gather}
|\Psi_{5}\rangle = (a_{\rm in}^\dagger)^{n_{\rm in}-1} |\Omega\rangle |m-1\rangle,\quad E_5 = (n_{\rm in} - 1)\hbar\omega_{\rm in} + \epsilon_{m-1},\\
|\Psi_{6}\rangle = (a_{\rm in}^\dagger)^{n_{\rm in}}a_{\rm out}^\dagger |\Omega\rangle |m-1\rangle,\quad E_6 = n_{\rm in}\hbar\omega_{\rm in} + \hbar\omega_{\rm out} + \epsilon_{m-1}.
\end{gather}
Then corresponding contribution to the matix element reads
\begin{gather}
    \mathcal{M}_{fi}^{(3)} = \langle m+2|N|m-1\rangle\langle m-1|N|m\rangle\sqrt{n_{\rm in}} \lambda_{\rm in}\lambda_{\rm out}\frac{(2\epsilon_{m-1}-\epsilon_m-\epsilon_{m+2})}{(\hbar\win + \epsilon_{m-1} - \epsilon_{m+2})(\hbar\win + \epsilon_m - \epsilon_{m-1})}.
\end{gather}
Again, in the limit of weak anharmonicity we obtain
\begin{gather}
    \mathcal{M}_{fi}^{(3)} = -m \sqrt{m+1}\sqrt{m+2}\nzpf^2\sqrt{n_{\rm in}} \lambda_{\rm in}\lambda_{\rm out} \frac{E_C}{(\hbar\win + \hbar\wq)(\hbar\win-3\hbar\wq)}.
\end{gather}
Combining the three contributions to the matrix element we obtain the final result,
\begin{gather}
    \mathcal{M}_{fi} = \sqrt{m+1}\sqrt{m+2}\nzpf^2 \sqrt{n_{\rm in}}\lambda_{\rm in}\lambda_{\rm out}E_C \left(\frac{3}{(\hbar\win+\hbar\wq)(\hbar\win - 3\hbar\wq)}+\frac{1}{(\hbar\win - \hbar\wq)^2}\right)
\end{gather}
Combining the terms in the bracket and also using $E_C \nzpf^2 = \hbar\wq/16$, we recover
\begin{gather}
    \mathcal{M}_{fi} = \sqrt{m+1}\sqrt{m+2} \frac{\hbar\wq}{4} \sqrt{n_{\rm in}} \frac{\lambda_{\rm in}\hbar\win}{(\hbar\win)^2-(\hbar\wq)^2}\frac{\lambda_{\rm out }\hbar\wout}{(\hbar\wout)^2 - (\hbar\wq)^2}.
\end{gather}
Thus, in case of weak anharmonicity, the obtained matrix element coincides with that obtained through normal mode decomposition, Eq.~\eqref{eq:matrix_element_final}.

\bibliography{references.bib}